\documentclass{aa} 

\usepackage[T1]{fontenc}
\usepackage[utf8]{inputenc}
\usepackage{amssymb}
\usepackage[]{hyperref}
\usepackage{amsmath}
\usepackage{graphicx}
\usepackage[varg]{txfonts}
\usepackage[dvipsnames]{xcolor}
\usepackage{ulem}

\def\aap{A\&A}

\def\apjs{ApJSS}

\defcitealias{Mahlmann2020c}{Paper II}

\begin{document}
		
\title{Computational general relativistic force-free electrodynamics: I. Multi-coordinate implementation and testing}
\titlerunning{Computational GRFFE: Multi-coordinate implementation and testing}

\author{J. F. Mahlmann\inst{1}\thanks{jens.mahlmann@uv.es}, M. A. Aloy\inst{1}, V. Mewes\inst{2,3}, P. Cerdá-Durán\inst{1}}
\institute{Departament d'Astronomia i Astrofísica, Universitat de Val\`{e}ncia, 46100, Burjassot (Valencia), Spain
\and
National Center for Computational Sciences, Oak Ridge National Laboratory, P.O. Box 2008, Oak Ridge, TN 37831-6164, USA
\and
Physics Division, Oak Ridge National Laboratory, P.O. Box 2008, Oak Ridge, TN 37831-6354, USA}

\date{Received \textit{Month Day, Year}; accepted \textit{Month Day, Year}}

\abstract{General relativistic force-free electrodynamics is one possible plasma-limit employed to analyze energetic outflows in which strong magnetic fields are dominant over all inertial phenomena. The amazing images of black hole (BH) shadows from the Galactic Center and the M87 galaxy provide a first direct glimpse into the physics of accretion flows in the most extreme environments of the universe. The efficient extraction of energy in the form of collimated outflows or jets from a rotating BH is directly linked to the topology of the surrounding magnetic field. We aim at providing a tool to numerically model the dynamics of such fields in magnetospheres around compact objects, such as BHs and neutron stars. To do so, we probe their role in the formation of high energy phenomena such as magnetar flares and the highly variable teraelectronvolt emission of some active galactic nuclei. In this work, we present numerical strategies capable of modeling fully dynamical force-free magnetospheres of compact astrophysical objects. We provide implementation details and extensive testing of our implementation of general relativistic force-free electrodynamics in Cartesian and spherical coordinates using the infrastructure of the \textsc{Einstein Toolkit}. The employed hyperbolic/parabolic cleaning of numerical errors with full general relativistic compatibility allows for fast advection of numerical errors in dynamical spacetimes. Such fast advection of divergence errors significantly improves the stability of the general relativistic force-free electrodynamics modeling of BH magnetospheres.} 

\maketitle

\section{Introduction}
\label{sec:Introduction}

Relativistic, magnetically dominated plasma is a basic element of the environments around 
neutron stars \citep{Goldreich_Julian_1969ApJ...157..869,Michel_1973ApJ...180..207,
	Scharlemann_etal_1973ApJ...182..951,Contopoulos1999}, especially around magnetars \citep[][]
{Lamb_1982AIPC...77..249,Thompson_Duncan_1995MNRAS.275..255, Beloborodov_2007ApJ...657..967} and 
black holes (BHs) \citep[BHs;][]{Blandford1977}, including the accretion disks surrounding 
them \citep{MacDonald1982,Thorne1986,Beskin1997} and their outflows 
\citep{Takahashi1990,Lee2000,Punsly2001,Lyutikov_2003_jets}. Interest for the environments 
surrounding neutron stars and BHs has recently been sparked due to the overwhelming amount 
of new observations available, for example, in supermassive BHs \citep{ETH_2019a,ETH_2019b} and 
magnetars \citep{Turolla_2015RPPh...78k6901, Kaspi_2017ARAaA..55..261}. If magnetic fields 
dominate the dynamics, all inertial and thermal plasma contributions can be neglected. 
Thus, the only role of the plasma is to support the electromagnetic fields. Magnetic 
fields govern all plasma dynamics; the currents are not merely induced by the drift of 
the matter distribution but are completely determined by the electromagnetic fields. Under the aforementioned conditions, the plasma becomes force-free \citep[e.g.,][]{Uchida1997}. In this series of papers, we present comprehensive techniques for the modeling of force-free astrophysical plasma. We will point out various caveats of this regime with transparency. They are outweighed by the clear advantages of the force-free regime, such as numerical robustness in high magnetization. This is especially the case whenever it becomes important to know what the plasma is actually doing (on micro-scales), for example, the screening of nonideal electric fields or magnetic reconnection.

The correct interpretation of recent breakthrough observations requires building up a solid theoretical understanding of the astrophysical scenarios mentioned above. Due to their complexity and system size, they are well suited for numerical approaches. Two principal formulations for numerically modeling astrophysical plasma under force-free conditions have emerged in recent years \citep[see a detailed review in][]{Paschalidis2013}. \citet{Komissarov2004} suggests the time evolution of the full set of Maxwell's equations, where the magnetic induction and displacement encode the general relativistic spacetime geometry as non-vacuum effects. This formulation has also been employed by \citet{Petri2016} and in an implementation relying on spectral methods \citep[\textsc{Phaedra},][]{Parfrey2015,Parfrey2017}. 
Furthermore, \citet{Palenzuela2010} and \citet{Carrasco2017} carried out simulations in spherical geometries and higher-order finite difference approximations. \citet{McKinney2006} introduced a formulation that is based on an adaptation of general relativistic 
magnetohydrodynamics (GRMHD) to evolve the magnetic field as well as Poynting fluxes in time. As such, it was implemented, for example, in the GRMHD code \textsc{Harm} \citep{Gammie2003}. \citet{Paschalidis2013} improved upon the formulation introduced in \citet{McKinney2006} by explicitly accounting for the orthogonality relation between the magnetic field and the Poynting flux. A similar approach was implemented by \citet{Etienne2017} in the \textsc{GiRaFFE} code
provided in the
\textsc{Einstein Toolkit}\footnote{\url{http://www.einsteintoolkit.org}} \citep{Loeffler2012}. 
For this project, we have implemented the Maxwell equations 
evolution system in general relativity using the infrastructure of the \textsc{Einstein Toolkit}. 
To this end, we developed a new code for the numerical integration of the equations of general 
relativistic force-free electrodynamics (GRFEE) in dynamically evolving spacetimes. 
This tool has already been applied to the study of the dynamics in magnetospheres around 
compact objects \citep{Mahlmann2019,Mahlmann2020}. In this series of papers, we will 
extensively review implementation details and characterize the numerical properties of 
our new code.

The \textsc{Einstein Toolkit} infrastructure, which was originally designed for Cartesian coordinates,
has recently been adapted to support spherical coordinates \citep{Mewes2018,Mewes2020}. For certain 
applications in the realm of astrophysical compact objects, it is beneficial to exploit 
coordinates that reflect the approximate symmetries these systems possess. Spherical coordinates provide a coordinate system that enhances the accuracy of the employed method. Starting 
from the so-called reference metric approach, we write the evolution equations such that they 
correspond to conservation laws in a conformally related metric. We then use suitable 
finite volume discretizations \citep[in Cartesian and spherical coordinates,][]{CerdaDuran2008} 
for the integration of intercell fluxes in our time-marching scheme. Alternative approaches 
have been employed, for example, by \citet{Montero2014} and \citet{Mewes2020}, wherein all information about the 
underlying coordinate system is encoded in geometrical source terms.

This work is organized as a series of papers. This manuscript (Paper I) reviews the 
general theory of GRFFE and our implementation using the infrastructure of the 
\textsc{Einstein Toolkit}. The second paper \citep[][\citetalias{Mahlmann2020c}]{Mahlmann2020c} 
focuses on the characterization of the numerical diffusivity of our algorithm. 
Sections~\ref{sec:GRPrelims} and~\ref{sec:MaxConservative} lay out the theory background 
of GRFFE and introduce an augmented conservative system of partial differential 
equations (PDEs). We discuss the implementation of this system in our scientific code in 
the \textsc{Einstein Toolkit} in Sect.~\ref{sec:Code_Implementation}. 
Different finite volume integrators used for full support of both Cartesian and spherical 
coordinates are reviewed in Sects.~\ref{sec:FVCartesian} and~\ref{sec:FVSpherical}. We 
discuss two key ingredients for the successful numerical integration of GRFFE, namely 
the preservation of force-free conditions and the cleaning of numerical errors, 
in Sects.~\ref{sec:FFConstraints} and~\ref{sec:CleaningErrors}. 
Section~\ref{sec:Numerical_Tests} assembles a suite of tests for the numerical calibration 
and characterization of our code. We demonstrate its ability to reproduce the basic dynamics 
of force-free configurations (Sect.~\ref{sec:1D_Reconstruction}). We further demonstrate the code's 
potential for modeling astrophysical plasma in magnetar and BH magnetospheres in 
Sect.~\ref{sec:Astrophysical_Tests}. Finally, we outline distinct features of the 
presented methods as well as implications for GRFFE schemes in general in 
Sect.~\ref{sec:conclusions}.

\section{General relativistic force-free electrodynamics}
\label{sec:GRFFE}

The following sections as well as the code implementation in the \textsc{Einstein Toolkit} 
employ units where $M_\odot=G=c=1$, which sets the respective time and length scales to 
be $1M_\odot\equiv 4.93\times 10^{-6}\text{ s}\equiv 1477.98\text{ m}$. This unit system 
is a variation of the so-called system of geometrized units \citep[as introduced 
in appendix F of][]{Wald2010}, with the additional normalization of the mass to 
$1M_\odot$ \citep[see also][on unit conversion in the \textsc{Einstein Toolkit}]{Mahlmann2019}.
In the following, Latin indices denote spatial indices, running from 1 to 3; Greek indices denote 
spacetime indices, running from 0 to 3 (0 is the time coordinate). The Einstein 
summation convention is used.

\subsection{General relativity preliminaries}
\label{sec:GRPrelims}

To numerically solve the field equations of general relativity, a fully covariant formulation
is not optimal. 
Instead, to arrive at a Cauchy initial value problem that can be evolved forward in time, it is common to introduce a $3+1$ split of spacetime 
\citep[e.g.,][and references therein]{Darmois1927,Gourgoulhon2012,Tondeur2012}.
In doing so, the four-dimensional spacetime, characterized by the metric tensor $g_{\mu\nu}$, is foliated with a set of nonintersecting timelike 
hypersurfaces $\Sigma_t$, namely level surfaces of the scalar field $t$ 
(denoting the time coordinate). 
We denote the future-pointing, timelike normal on $\Sigma_t$ as $n_{\mu}$. It is 
defined through the constituting relation 
\begin{align}
	\alpha\, n^\mu \nabla_\mu t=1\,,
\end{align}
where $\nabla_{\mu}$ denotes the spacetime covariant derivative built from $g_{\mu \nu}$. The lapse function $\alpha$ indicates the separation in proper time between two hypersurfaces. The
spatial three-metric $\gamma_{ij}$ is the projection of the spacetime metric $g_{\mu \nu}$ onto
$\Sigma_t$: 
\begin{equation}
	\gamma_{i j} = (g^\mu_{\: i} + n^\mu n_i) (g^\nu_{\: j} + n^\nu n_j) g_{\mu\nu} = g_{i j}.
\end{equation}
Trajectories of constant spatial coordinates across different hypersurfaces $\Sigma_t$ define the time vector along them: 
\begin{align}
	t^\mu=\alpha n^\mu +\beta^\mu.
	\label{eq:tmu}
\end{align}
The component $\beta^\mu$ is the shift four-vector, which denotes the spacelike displacement 
in the direction perpendicular to $n^{\mu}$, required to reach the original base coordinate 
in a hypersurface $\Sigma_{t'}$ after leaving $\Sigma_t$. The shift vector satisfies $\beta^{\mu}n_{\mu}=0$ by 
definition, but it is otherwise arbitrary, as is the lapse function. 
Choosing the time coordinate such that $t^\mu=\left(1,0,0,0\right)$, the components of the 
normal vector $n^\mu$ and its (metric) dual $n_\mu$ (assuming the metric signature 
is $+1$) can be expressed in terms of lapse and shift as follows:
\begin{align}
	n^\mu=\left(\frac{1}{\alpha},-\frac{\beta^i}{\alpha}\right),\qquad\qquad n_\mu=\left(-\alpha,0,0,0\right).
\end{align}
The line element of the spacetime may be written in terms of the lapse, shift,
and spatial metric in the
3+1 formalism \citep{Lichnerowicz1944,Choquet-Bruhat1952,York1979} as
\begin{align}
	d s^{2}=-\alpha^{2} d t^{2}+\gamma_{ij}\left(d x^{i}+\beta^{i} d t\right)\left(d x^{j}+\beta^{j} d t\right).
	\label{eq:ADMmetric}
\end{align}
The spacetime metric $g_{\mu \nu}$ is given by:
\begin{equation}\label{3p1_metric}
	g_{\mu \nu} = \left( \begin{array}{cc}
		-\alpha^2 + \beta_i \beta^i & \beta_j \\
		\beta_i & \gamma_{i j}
	\end{array}\right).
\end{equation}
In this foliation, the Einstein equations can be cast into a set of evolution and constraint equations \citep[see, e.g.,][for textbook introductions]{Alcubierre2008,Baumgarte2010,Gourgoulhon2012}. One of the most widely used formulations to 
numerically solve the Einstein equations in $3+1$ form is the so-called Baumgarte-Shapiro-Shibata-Nakamura (BSSN) formulation \citep{Shibata1995,Baumgarte1999}. It evolves the conformally related metric $\bar{\gamma}_{ij}$ and the conformal factor $e^{4 \phi}$, which are related by
\begin{align}
	\bar{\gamma}_{ij}=e^{-4\phi}\gamma_{ij}\label{eq:confrelated}, 
\end{align}
where the conformal factor $e^{4 \phi}$ can be written as
\begin{equation}
	e^{4 \phi} = (\gamma/\bar{\gamma})^{\frac{1}{3}}.
\end{equation}
Here, $\gamma$ and $\bar{\gamma}$ are the determinants of the spatial and conformally related metric, 
respectively. The BSSN formalism also introduces the conformally related extrinsic curvature 
and the conformal connection functions as evolved variables. Throughout this work, we fix $\bar{\gamma}$ to be constant in time \citep[the so-called Lagrangian 
choice,][]{Brown_2005PhRvD..71j4011}:
\begin{align}
	\partial_t\bar{\gamma}=0. \label{eq:LagrangianGauge}
\end{align}
Thus, the time dependence of the spatial metric determinant is encoded only in the corformal factor, as $\sqrt{\gamma}=e^{6\phi}\sqrt{\bar{\gamma}}$. Keeping $\bar{\gamma}$ fixed to 
its initial value simplifies expressions in the GRFFE evolution equations. This choice is particularly 
useful when integrating GRFFE in spherical coordinates, as we elaborate below.
\subsection{Maxwell's equations in conservative form}
\label{sec:MaxConservative}

The evolution of the full set of Maxwell's equations is one possibility
\footnote{Another evolution scheme, developing energy fluxes rather than electric fields, 
	was employed by e.g., \citet{McKinney2006} or \citet{Etienne2017}.} to deal with 
electrodynamics in general relativity \citep{Komissarov2004}:
\begin{align}
	\nabla_\nu F^{\mu\nu}&=I^\mu, \label{eq:MaxI}\\
	\nabla_\nu\hspace{1pt} ^*\hspace{-2pt}F^{\mu\nu}&=0, \label{eq:MaxII} 
\end{align}
where $F^{\mu\nu}$ is the Maxwell tensor and
$^{\ast}F^{\mu\nu}$ is its dual, defined as:
\begin{align}
	^*\hspace{-2pt}F^{\mu\nu}&\equiv\frac{1}{2}\eta^{\mu\nu\lambda\zeta}F_{\lambda\zeta} ,\label{eq:Faradaytensor}
\end{align}
where 
\begin{align}
	\eta^{\mu\nu\lambda\zeta}=-(-g)^{-1/2}[\mu\nu\lambda\zeta],\qquad\qquad\eta_{\mu\nu\lambda\zeta}=(-g)^{1/2}[\mu\nu\lambda\zeta].
	\label{eq:LeviCivitatensor}
\end{align}
Here, $[\mu\nu\lambda\zeta]$ is the completely antisymmetric Levi-Civita symbol with $[0123]=+1$ and $g$ the determinant of the spacetime metric $g_{\mu \nu}$. 
$I^\mu$ is the electric current four-vector associated with the charge density 
$\rho=-n_{\mu}I^{\mu}=\alpha I^t$, and the current three vector $J^i=\alpha I^i$. A covariant definition of the current four-vector 
is \citep{Komissarov2004}
\begin{align}
	J^\mu=2 I^{[\nu}t^{\mu]}n_{\nu},
	\label{eq:Jmu}
\end{align}
where we use the standard terminology $I^{[\nu}t^{\mu]}=\frac{1}{2}(I^\nu t^\mu - I^\mu t^\nu)$.
We note that $\rho$ is the charge density as measured by the normal
(Eulerian) observer defined by $n^\mu$. The current density 3-vector as measured by the Eulerian observer is the projection 
of $I^\mu$ onto the hypersurface $\Sigma_t$:
\begin{equation}
	j^{\, i} = (g^i_{\: \mu} + n^i n_\mu) I^\mu = \alpha^{-1} (J^{\, i} + \rho \beta^i).
\end{equation}
\citet{Komissarov2004} introduces a set of vector fields, which are analogous to the electric 
and magnetic fields, $\mathbf{E}$ and $\mathbf{B}$, and electric displacement and magnetic 
induction, $\mathbf{D}$ and $\mathbf{B}$, of the electrodynamic theory of continuous media 
\citep[see, e.g.,][\S I.4]{jackson1999}. They have the following spatial components in a 
$3+1$ decomposition of spacetime:
\begin{align}
	E_i =&\:F_{it},\label{eq:MacroscopicE}\\
	B^i =&\:\frac{1}{2}e^{ijk}F_{jk}, \label{eq:MacroscopicB}\\
	D^i =&\:\frac{1}{2}e^{ijk} {^*}\hspace{-2pt}F_{jk}, \label{eq:MacroscopicD}\\
	H_i =&\:{^*}\hspace{-2pt}F_{ti}.\label{eq:MacroscopicH}
\end{align}
Following the convention in, for example, \citet{Baumgarte2003}, the four-dimensional volume 
element induces a volume element on the hypersurfaces of the foliation:
\begin{align}
	e^{a b c}=n_\mu \eta^{\mu a b c}=-\alpha \eta^{0 a b c}.
	\label{eq:LeviCivitatensor3d}
\end{align}
In the previous expression, $e^{abc}\ne 0$ only for spatial indices, thus, we can write 
$e^{ijk}=-\alpha\eta^{0ijk}=[ijk]/\sqrt{\gamma}$. 

Using the definitions \eqref{eq:MacroscopicB} and \eqref{eq:MacroscopicD} in the time 
components of Eqs.~\eqref{eq:MaxI} and~\eqref{eq:MaxII}, one obtains the familiar 
constraints
\begin{align}
	\text{div}\mathbf{D}=\rho,\label{eq:ellipticD3}\\
	\text{div}\mathbf{B}=0 . \label{eq:ellipticB3}
\end{align}
We separately evolve the charge continuity equation, which is obtained from the covariant derivative of Eq.~\eqref{eq:MaxI},
\begin{align}
	\nabla_\nu I^\nu=\:0\label{eq:ChargeConservation}\,,
\end{align}
to ensure the conservation of the (total) electric charge in the computational domain, as 
well as the compatibility of the charge distribution obtained numerically, with the currents 
that they generate \citepalias[see also the detailed discussion in][]{Mahlmann2020c}. If this is not done, the difference $|\text{div}\,\mathbf{D}-\rho|$ 
may grow unbounded with time due to the accumulation of numerical errors 
\citep[][]{Munz_1999CRASM.328..431}.

Embodied in the 
definitions~(\ref{eq:MacroscopicE}) to~(\ref{eq:MacroscopicH}) one finds the following vacuum 
constitutive relations \citep{Komissarov2004}:
\begin{align}
	\mathbf{E}=&\alpha \mathbf{D}+\boldsymbol{\beta}\times\mathbf{B},\label{eq:EvecCorot}\\
	\mathbf{H}=&\alpha \mathbf{B}-\boldsymbol{\beta}\times\mathbf{D}.\label{eq:HvecCorot}
\end{align} 
We may now write the Maxwell tensor as measured by the Eulerian observer
in terms of the electric, $D^\mu$, and magnetic, $B^\mu$, field four-vectors 
\citep[cf.][]{McKinney2006,Anton2006}:
\begin{align}
	F^{\mu\nu}&=\: n^\mu D^\nu-D^\mu n^\nu-\eta^{\mu\nu\lambda\zeta}B_\lambda n_\zeta\label{eq:MaxTensorI}\,,\\
	{^*}\hspace{-2pt}F^{\mu\nu} &=\: -n^\mu B^\nu+B^\mu n^\nu-\eta^{\mu\nu\lambda\zeta}D_\lambda n_\zeta\,, \label{eq:MaxTensorII}
\end{align}
which satisfy
\begin{align}
	D^\mu &= F^{\mu\nu}n_\nu \,= \left(0, D^i\right)\,, \label{eq:Displacement4vector}\\
	B^\mu &= ^{*}\hspace{-2pt}F^{\nu\mu}n_\nu \,= \left(0, B^i\right)\,.\label{eq:MagInduction4vector}
\end{align}
For later reference, we provide two Lorentz invariants of the Faraday tensor, namely:
\begin{align}
	^{*}\hspace{-2pt}F^{\mu\nu}F_{\mu\nu}=&\: 4D^\mu B_\mu, \label{eq:*FxF} \\
	F^{\mu\nu}F_{\mu\nu} =&\: 2(\mathbf{B}^2 - \mathbf{D}^2). \label{eq:FxF} 
\end{align}
In order to build up a stationary magnetic configuration (as, e.g., in the magnetosphere 
around a compact object), it is necessary to guarantee that there are either no forces 
acting on the system or, more generally, that the forces of the system are in equilibrium. 
Except along current sheets the latter condition implies that the electric four-current 
$I^{\mu}$ satisfies the force-free condition \citep{Blandford1977}:
\begin{align}
	F_{\mu\nu}I^\nu=0.
	\label{eq:ForceFree}
\end{align}
Eq.~\eqref{eq:ForceFree} is equivalent to a vanishing
Lorentz force density $f_\mu$ on the charges measured by the Eulerian observer:
\begin{align}
	\nabla_\nu\,T^\nu_{\hspace{4pt}\mu}=-F_{\mu\nu}I^\nu=-f_\mu\equiv 0. \label{eq:LorentzCov}
\end{align}
Also, Eq.~(\ref{eq:ForceFree}) can be seen as a system of linear equations, the 
nontrivial solution \citep[i.e., non-electrovacuum or equivalently, $I^\nu\ne0$, cf. the discussion in][]{Paschalidis2013} of which demands that the determinant of $F_{\mu\nu}$ vanishes. 
Since $\det F_{\mu\nu}=({^*}\hspace{-2pt}F_{\mu\nu}F^{\mu\nu})^2/16=(D^\mu B_\mu)^2$, 
the force-free condition (\ref{eq:ForceFree}) reduces to
\begin{align}
	{^*}\hspace{-2pt}F_{\mu\nu}F^{\mu\nu}=0,
	\label{eq:FFCondICova}
\end{align}
or, equivalently (see Eq.~\ref{eq:*FxF})
\begin{align}
	D^\mu B_\mu=0.
	\label{eq:FFCondICov}
\end{align}
Hence, the component of the electric field parallel to the magnetic always vanishes. 
Since $\det F_{\mu\nu}=0$, the rank of $F_{\mu\nu}$ (regarded as a $4\times4$ matrix) is two, provided $F_{\mu\nu}$ has 
nonvanishing components. If $a^\mu$ is a zero eigenvector of $F_{\mu\nu}$, i.e., 
$F_{\mu\nu}a^\mu=0$, then another null eigenvector orthogonal to $a^\mu$ is 
$b^\mu={^*}\hspace{-2pt}F^{\mu\nu}a_\nu$, and the Faraday tensor can be expressed as 
$F_{\mu\nu}=\eta_{\mu\nu\lambda\delta}a^\lambda b^\delta$ \citep[cf.][]{Komissarov2002}. 
Hence, it admits a two-dimensional space of eigenvectors associated with the null eigenvalue 
\citep[cf.][]{Uchida1997}. These zero eigenvectors are time-like if the Lorentz invariant 
$F_{\mu\nu}F^{\mu\nu}$ is positive \citep{Uchida1997}.
The sign of the invariant $F_{\mu\nu}F^{\mu\nu}$ is not unanimously defined for generic 
electromagnetic four-vectors $B^\mu$ and $D^{\mu}$. To chose the sign of the invariant, 
it is useful to consider the force-free approximation as a low inertia limit of relativistic 
MHD. This means that a physical force-free electromagnetic field should be compatible with 
the existence of a velocity field of the plasma. Recalling that the plasma four-velocity 
$u^\mu$ is a unit time-like vector ($u^\mu u_\mu=-1$), and that the Lorentz force is 
$f_\mu\propto F_{\mu\nu}u^\nu$, a physical force-free electromagnetic field ($f_\mu=0$) must
satisfy $F_{\mu\nu}u^\nu=0$ (we note that this is also required by the ideal MHD condition). 
Hence, the sign of the Lorentz invariant $F_{\mu\nu}F^{\mu\nu}$ (see Eq.~\ref{eq:FxF}) 
should consistently be positive, namely
\begin{align}
	F_{\mu\nu}F^{\mu\nu} = 2(B^2 - D^2)>0.
	\label{eq:FFCondIICov}
\end{align}
In the introduced language of the full system of Maxwell's equations in $3+1$ decomposition, (\ref{eq:FFCondICov}) and (\ref{eq:FFCondIICov}) read 
\begin{align}
	\mathbf{D}\cdot\mathbf{B}=0\label{eq:FFCondI}\, ,\\
	\mathbf{B}^2-\mathbf{D}^2> 0\label{eq:FFCondII}.
\end{align}
Condition~(\ref{eq:FFCondII}) implies that the magnetic field is always stronger than the 
electric field. Equivalently, one can classify the degenerate electromagnetic tensor as 
magnetic since condition~(\ref{eq:FFCondIICov}) guarantees that there exists a frame in 
which an observer at rest measures zero electric field \citep[cf.][]{Uchida1997}. This observer 
is the comoving observer with four-velocity $u^\mu$ in the ideal MHD limit. In GRFFE, such a frame exists but is not unique \citep{McKinney2006,Paschalidis2013}.

The challenge of maintaining the physical constraints of $\text{div}\,\mathbf{B}=0$ and 
$\text{div}\,\mathbf{D}=\rho$ in numerical simulations has been reviewed throughout the 
literature \citep[e.g.,][]{Dedner2002,Mignone2010}, and applied to GRFFE, for example, by 
\citet{Komissarov2004} and the relativistic MHD regime, e.g., by \citet{Palenzuela2009} and \citet{Miranda-Aranguren2018}.
Following \citet{Palenzuela2009}, \citet{Palenzuela2010} as well as \citet{Mignone2010}, we suggest 
to modify the system of Maxwell's equations (Eqs.~\ref{eq:MaxI} and \ref{eq:MaxII}) in the 
following way \citep[cf.][]{Alic2012}:
\begin{align}
	\nabla_\nu\left(F^{\mu\nu}+g^{\mu\nu}\Phi\right)&=I^\mu+t^\mu\kappa_{\Phi}\Phi ,\label{eq:MaxAugI}\\
	\nabla_\nu\left(^*\hspace{-2pt}F^{\mu\nu}+s^{\mu\nu}\Psi\right)&=t^\mu\kappa_{\Psi}\Psi .\label{eq:MaxAugII}
\end{align}
Here, the definition of $t^\mu$ is given in Eq.~(\ref{eq:tmu}), and we define $s^{\mu\nu}\equiv c_h^2\gamma^{\mu\nu}-n^\mu n^\nu$. $c_h$ corresponds to a speed of propagation of the divergence cleaning errors (see below); $\kappa_{\Phi}$ and $\kappa_{\Psi}$ are adjustable constants that control the parabolic damping of the aforementioned numerical errors. The scalar potentials $\Psi$ and $\Phi$ are ancillary variables employed to control the errors in the elliptic constrains $\text{div}\,\mathbf{B}=0$ and $\text{div}\,\mathbf{D}=\rho$, respectively. This is implemented in practice by  augmenting the system of Maxwell's equations with extra evolution equations for $\Phi$ and $\Psi$.
Contracting Eq.~(\ref{eq:MaxAugII}) with $\nabla_\mu$ yields 
for the simplified case of stationary spacetimes \citep[cf.][]{Komissarov2004}:
\begin{align}
	\begin{split}
		-\kappa_\Psi\nabla_t\Psi=\:&\nabla_\mu\nabla_\nu\left({^*}\hspace{-2pt}F^{\mu\nu}+\left(c_h^2\gamma^{\mu\nu}-n^\mu n^\nu\right)\Psi\right)\\
		=\:&\nabla_\mu\nabla_\nu\left(c_h^2\gamma^{\mu\nu}-n^\mu n^\nu\right)\Psi\\
		=\:&c_h^2\nabla_i\nabla^i\Psi-\nabla_\mu\nabla_\nu n^\mu g^{\nu \alpha}n_\alpha\Psi\\
		=\:&c_h^2\nabla_i\nabla^i\Psi+\nabla_t\nabla^t \Psi.
	\end{split}
\end{align}
This compares to telegrapher equations, used for example to describe signal propagation in lossy wires. In this analogy, $\kappa_{\Psi}$ and $c_h$ are the parameters controlling the 
damping and advection of numerical errors \citep{Mignone2010}. We stress the correspondence 
of $c_h$ with a finite propagation speed for divergence errors \citep{Mignone2010} and their 
decay according to the damping factor $\kappa_\Psi$. For $c_h$ chosen equal to the speed of 
light, Eq.~(\ref{eq:MaxII}) reduces to the evolution system given in \citet{Palenzuela2009}.

The augmented system of Maxwell Equations (Eqs.~\ref{eq:MaxAugI} and~\ref{eq:MaxAugII}), can be 
written as a system of balance laws of the form
\begin{align}
	\partial_t\:\mathcal{C}+\bar{\nabla}_j\:\mathcal{F}^j=\mathcal{S}_{\rm n}+\mathcal{S}_{\rm s}\label{eq:ConsGeneral}\,,
\end{align}
where $\bar{\nabla}$ is the covariant derivative with respect to the conformally related 
metric, $\bar{\gamma}$ (Eq.~\ref{eq:confrelated}). $\mathcal{C}$ denotes the vector of conserved 
variables, $\mathcal{F}^j$ the flux vectors, $\mathcal{S}_{\rm n}$ the geometrical and 
current-induced source terms, and finally $\mathcal{S}_{\rm s}$ are the potentially stiff 
source terms \citep[cf.][App. C2]{Komissarov2004}. We note that each of these quantities consists 
of elements in a multidimensional space. In general, the conserved variables are derived 
from the so-called primitive variables; primitive variables are usually the 
quantities measured by the Eulerian observer, namely $\rho$, $\mathbf{B}$, and $\mathbf{D}$, as well as the numerical cleaning 
potentials $\Psi$ and $\Phi$. Adapting the notation used by \citet{CerdaDuran2008} and 
\citet{Montero2014}, we specify the components of Eq.~(\ref{eq:ConsGeneral}) in terms 
of the determinant of a reference metric $\hat{\gamma}$. We define the conserved variables as
\begin{align}
	\arraycolsep=1.4pt\def\arraystretch{1.5}
	\mathcal{C}\equiv\left(\begin{array}{c}
		\mathcal{L}\\
		\mathcal{Q}\\ 
		\mathcal{P}\\
		b^i\\ 
		d^i
	\end{array}\right)=\: e^{6\phi}\sqrt{\frac{\bar{\gamma}}{\hat{\gamma}}}\:\left(\begin{array}{c}
		\rho \\
		\frac{\Psi}{\alpha}\\ 
		\frac{\Phi}{\alpha}\\
		B^i+\frac{\Psi}{\alpha}\beta^i\\ 
		D^i-\frac{\Phi}{\alpha}\beta^i
	\end{array}\right)\label{eq:KCons}\,,
\end{align}
with their corresponding fluxes
\begin{align}
	\arraycolsep=1.4pt\def\arraystretch{1.5}
	\mathcal{F}^j\equiv e^{6\phi}\sqrt{\frac{\bar{\gamma}}{\hat{\gamma}}}\:\left(\begin{array}{c}
		\alpha J^j\\
		B^j-\frac{\Psi}{\alpha}\beta^j\\ 
		-\left(D^j+\frac{\Phi}{\alpha}\beta^j\right)\\
		e^{ijk}E_k+\alpha\left(c_h^2\gamma^{ij}-n^i n^j\right)\Psi\\ 
		-\left(e^{ijk}H_k+\alpha g^{ij}\Phi\right)
	\end{array}\right)\label{eq:KFluxes}\,.
\end{align}
For the source terms, the split according to equation (\ref{eq:ConsGeneral}) yields the 
source terms $\mathcal{S}_{\rm n}$, and the potentially stiff source terms $\mathcal{S}_{\rm s}$:
\begin{align}
	\arraycolsep=1.4pt\def\arraystretch{1.5}
	\mathcal{S}_{\rm n}\equiv e^{6\phi}\sqrt{\frac{\bar{\gamma}}{\hat{\gamma}}}\:\left(\begin{array}{c}
		0\\
		\alpha\Psi\Gamma^t_{\mu\nu}s^{\mu\nu}\\ 
		\alpha\Phi\Gamma^t_{\mu\nu}g^{\mu\nu}-\rho\\
		-\alpha\Psi\Gamma^i_{\mu\nu}s^{\mu\nu}\\ 
		\alpha\Phi\Gamma^i_{\mu\nu}g^{\mu\nu}- J^i
	\end{array}\right)\,,
\end{align}
\begin{align}
	\arraycolsep=1.4pt\def\arraystretch{1.5}
	\mathcal{S}_{\rm s}\equiv e^{6\phi}\sqrt{\frac{\bar{\gamma}}{\hat{\gamma}}}\:\left(\begin{array}{c}
		0\\
		-\alpha\kappa_\Psi\Psi\\ 
		-\alpha\kappa_\Phi\Phi\\
		0\\ 
		0
	\end{array}\right)\label{eq:KSources}.
\end{align}
In the previous expressions, $\Gamma^\alpha_{\beta\gamma}$ are the Christoffel symbols
of the Levi-Civita connection associated with the spacetime metic $g_{\mu \nu}$.
In both Cartesian and spherical coordinates, we always make the initial choice 
$\bar{\gamma}(t=0) = \hat{\gamma}$, so that, due to the Eq.~(\ref{eq:LagrangianGauge}), the ratio $\bar{\gamma}/\hat{\gamma} =1$
throughout the evolution. This is an algebraic constraint for the components of the 
conformally related metric $\bar{\gamma}_{ij}$ and is continuously enforced in the spacetime evolution by making the replacement
\begin{equation}
	\bar{\gamma}_{ij} \to \left(\hat{\gamma}/\bar{\gamma} \right)^{\frac{1}{3}} \bar{\gamma}_{ij}
\end{equation}
at every sub-step of the time integration.

\subsection{The force-free current}
\label{sec:FFCurrent}

In force-free electrodynamics there is no uniquely defined rest frame for the fluid motion 
\citep[e.g.,][]{Uchida1997,McKinney2006,Paschalidis2013,Shibata2015}; the electromagnetic 
current $I^\mu$ cannot be determined by tracking the velocity of charges throughout the 
domain. Rather, the enforcement of the force-free conditions~(\ref{eq:FFCondI}), and 
(\ref{eq:FFCondII}) determines a suitable current. The conservation condition (implicitly embodied in Maxwell's 
equations) 
\begin{align}
	\mathcal{L}_{n} \left(\mathbf{D}\cdot\mathbf{B}\right) = n^\mu \nabla_\mu \left(\mathbf{D}\cdot\mathbf{B}\right)=0,
	\label{eq:DBcons}
\end{align}
where $\mathcal{L}_{n} $ is the Lie derivative with respect to $n^\mu$ is equivalent to $\partial_t\left(\mathbf{D}\cdot\mathbf{B}\right)=0$. Together with conditions (\ref{eq:FFCondI}) and 
(\ref{eq:FFCondII}), it
can be combined to obtain an explicit expression for the so-called force-free current 
$I^\mu_{\textsc{ff}}$ \citep[cf.][]{McKinney2006,Komissarov2011,Parfrey2017}:
\begin{align}
	\begin{split}
		I^\mu_\textsc{ff}=&\:\rho n^\mu+\frac{\rho}{\mathbf{B}^2}\eta^{\nu\mu\alpha\beta}n_\nu D_\alpha B_\beta\\
		&+\frac{B^\mu}{\mathbf{B}^2} \:\eta^{\alpha\beta\lambda\sigma}n_\sigma\left(B_{\lambda; \, \beta}B_\alpha-D_{\lambda; \, \beta}D_\alpha\right).
		\label{eq:FFCurrent}
	\end{split}
\end{align}
The current is one important closure relation for GRFFE. In this form, Eq.~(\ref{eq:FFCurrent}) induces primitive variable derivatives to the source terms. Such nonconservative splitting - chipping off parts of the flux terms - requires diligent attention and is prone to have a significant impact on the quality of the numerical evolution. We dedicate section \ref{sec:ParallelCurrent} and a large part of Paper II \citep{Mahlmann2020c} to the implementation of current closure.

In practice, the combination of the force-free current (\ref{eq:FFCurrent}) as a 
source-term to Eq.~(\ref{eq:MaxI}) - or Eq.~(\ref{eq:MaxAugI}) if we consider 
the augmented system of equations - with numerically enforcing conditions (\ref{eq:FFCondI}) 
and (\ref{eq:FFCondII}) restricts the evolution to the force-free regime. 
The discussion of techniques to ensure a physical \citep[cf.][]{McKinney2006} 
evolution of numerical force-free codes is a recurrent topic that can be found throughout 
the literature \citep[e.g.,][]{Lyutikov2003,Komissarov2004,Palenzuela2010,Alic2012,Paschalidis2013,Carrasco2017,Parfrey2017,Mahlmann2019}. 
We review one of these techniques in Sect.~\ref{sec:FFConstraints}.

\section{Numerical methodology}
\label{sec:Code_Implementation}

Our GRFFE method uses the framework of the \textsc{Einstein Toolkit} \citep{Loeffler2012}. 
The \textsc{Einstein Toolkit} is an open-source software package utilizing the modularity 
of the \textsc{Cactus}\footnote{\url{http://www.cactuscode.org}} code \citep{Goodale2002a}, 
which enables the user to specify so-called \textsc{thorns} in order to set up customized 
simulations and provides (basic) adaptive mesh refinement (AMR) via the 
\textsc{Carpet}\footnote{\url{https://bitbucket.org/eschnett/carpet/src/master/}}
driver \citep{Goodale2002a,Schnetter2004}. The spacetime evolution is performed using the 
\textsc{MacLachlan}\footnote{\url{http://www.cct.lsu.edu/~eschnett/McLachlan/}} thorn \citep{Brown2009} as an implementation of the BSSN formalism. 
Recently, numerical relativity in spherical grids has been successfully enabled on the 
traditionally Cartesian \textsc{Einstein Toolkit} by the new implementation of \textsc{SphericalNR} 
\citep{Mewes2018,Mewes2020}, which is built upon a reference-metric formulation of the 
BSSN equations \citep{BrownCovariant2009,Montero2012b,Baumgarte2013}. 
We make use of a variety of open-source thorns within the 
\textsc{Einstein Toolkit}, such as the apparent horizon finder \textsc{AHFinderDirect} 
\citep{Thornburg2004}, the extraction of quasilocal quantities \textsc{QuasiLocalMeasures} 
\citep{Dreyer2003}, and the efficient \textsc{SummationByParts} thorn \citep{Diener2007}.

In our code, the time update of the system of PDEs (see Eq.~\ref{eq:ConsGeneral}) is done applying the method-of-lines 
\citep[e.g.,][]{LeVeque2007} implemented in the \textsc{Einstein Toolkit} 
thorn \textsc{MoL}. For the numerical test shown in this paper we make use of the 
fourth-order accurate (not strictly TVD) Runge-Kutta method implemented in the thorn \textsc{MoL}.

To ensure the conservation properties of the algorithm, it is critical to employ 
refluxing techniques correcting numerical fluxes across different levels of mesh 
refinement \citep[e.g.,][]{Collins2010}. Specifically, we make use of the thorn 
\textsc{Refluxing}\footnote{Refluxing at mesh refinement interfaces by Erik Schnetter: 
	\url{https://svn.cct.lsu.edu/repos/numrel/LSUThorns/Refluxing/trunk}} in combination with 
a cell-centered refinement structure \citep[cf.][]{Shibata2015}. We highlight the fact that 
employing the refluxing algorithm makes the numerical code $2-4$ times slower for the 
benefit of enforcing the conservation properties of the numerical method (especially of 
the charge). Refluxing also reduces the numerical instabilities, which tend to develop at 
mesh refinement boundaries \citep{Mahlmann2019,Mahlmann2020}.

This section reviews in detail techniques that are inherently important components of GRFFE. Apart from these, we use a wide range of numerical recipes, such as higher-order 
monotonicity preserving (MP) reconstruction at cell interfaces \citep[][]{Suresh1997} and 
the cleaning of numerically induced divergence and charges.

\subsection{Finite volume integration}
\label{sec:FVIntegration}

We solve Eq.~(\ref{eq:ConsGeneral}) by discretizing its integral over the volume $V$ 
of a cell of our numerical mesh \citep[cf.][]{LeVeque2007,Mignone2014,Marti2003},
\begin{align}
	\partial_t\left\langle\mathcal{C}\right\rangle+\frac{1}{V}\int_{\partial V}\text{d}\mathbf{A}\cdot\mathbf{F}=\left\langle\mathcal{S}_{\rm n}\right\rangle+\left\langle\mathcal{S}_{\rm s}\right\rangle.\label{eq:FV_Integration}
\end{align}
Here, $\left\langle\right\rangle$ denotes the volume average of a quantity. The divergence 
term $\bar{\nabla}_j\:\mathcal{F}^j$ appearing in Eq.~(\ref{eq:ConsGeneral}) is 
integrated by applying Stoke's theorem and summing up the reconstructed fluxes $\mathbf{F}$ 
through the cell interfaces with respective area elements $\text{d}\mathbf{A}$.

In practice, we approximate volume averages by cell-centered values for each grid element. We identify each of these elements by the indices $(i,j,k)$ that correspond 
to the locations $x_i=x_0+i\Delta x$, $y_j=y_0+j\Delta y$, and $z_k=z_0+k\Delta x$. 
$\Delta x$, $\Delta y$, and $\Delta z$ represent the (uniform) numerical grid spacing in 
each coordinate direction. The quantities $(x_0, y_0,z_0)$ denote the coordinates of an 
arbitrary reference point in 3D. Face-centered quantities are indicated by the subscript 
of a half-step added to the respective index. For example, subscript $i+1/2$ denotes the value 
located at the face between the two elements $(i,j,k)$ and $(i+1,j,k)$. If no subscript 
is provided, we refer to the cell-centered values.

\subsubsection{Cartesian coordinates}
\label{sec:FVCartesian}

The system of Eqs.~(\ref{eq:KCons}) to~(\ref{eq:KSources}) is specified to its 
application in Cartesian coordinate systems ($x,y,z$) by setting $\sqrt{\hat{\gamma}}=1$. 
In this case, the cell volume is
\begin{align}
	V=\Delta x\times\Delta y\times\Delta z\,,
\end{align}
and the area elements are denoted by
\begin{align}
	\text{d}\mathbf{A}=\left(\Delta y\times\Delta z,\Delta x\times\Delta z,\Delta x\times\Delta y\right).
\end{align}
Eq.~(\ref{eq:FV_Integration}) is approximated by evaluating the fluxes $\mathbf{F}$ 
as reconstructed averages at cell interfaces: 
\begin{align}
	\begin{split}
		\frac{1}{V}\int_{\partial V}\text{d}\mathbf{A}\cdot\mathbf{F}\approx&\frac{F^x_{i+1/2}-F^x_{i-1/2}}{\Delta x}+\frac{F^y_{j+1/2}-F^y_{j-1/2}}{\Delta y}\\
		&+\frac{F^z_{k+1/2}-F^z_{k-1/2}}{\Delta z}.
	\end{split}
\end{align}

\subsubsection{Spherical coordinates}
\label{sec:FVSpherical}

In spherical coordinates ($r,\theta,\phi$), $\sqrt{\hat{\gamma}}=r^2\sin\theta$, and the 
cell volume is
\begin{align}
	V=-\frac{\Delta r^3}{3}\times\Delta\cos\theta\times\Delta\phi,
\end{align}
where $\Delta r^3=r_{i+1/2}^3-r_{i-1/2}^3$ and $\Delta\cos\theta=\cos\theta_{j+1/2}-\cos\theta_{j-1/2}$. 
The numerical stability of the spacetime integral in Eq.~(\ref{eq:FV_Integration}) 
critically depends on the balancing of coordinate singularities, such as the polar 
axis and the origin of the spherical coordinate system. 
We guarantee an exact evaluation of metric contributions at the location of the 
cell-interfaces by transforming the reconstructed fluxes $\mathbf{F}$ to an orthonormal 
basis. The area elements in an orthonormal basis are denoted by
\begin{align}
	\text{d}\hat{\mathbf{A}}=\left(r^2\sin\theta\times\Delta\theta\times\Delta\phi,r\sin\theta\times\Delta r\times\Delta\phi,r\times\Delta r\times\Delta\theta\right).
\end{align}
Eq.~(\ref{eq:FV_Integration}) is approximated by evaluating the fluxes $\mathbf{F}$ as 
reconstructed averages in an orthonormal basis at cell interfaces: 
\begin{align}
	\begin{split}
		\frac{1}{V}\int_{\partial V}\text{d}\hat{\mathbf{A}}\cdot\hat{F}\approx\:&\, 3\,\frac{r^2_{i+1/2}\,{\hat{F}_{i+1/2}^r}-r^2_{i-1/2}\,{\hat{F}_{i-1/2}^r}}{\Delta r^3}\\
		&-\frac{3\Delta r^2}{2\Delta r^3}\,\frac{\sin\theta_{j+1/2}\,{\hat{F}_{j+1/2}^\theta}-\sin\theta_{j-1/2}\,{\hat{F}_{j-1/2}^\theta}}{\Delta\cos\theta}\\
		&-\frac{3\Delta r^2}{2\Delta r^3}\,\frac{\Delta\theta}{\Delta\cos\theta}\,\frac{{\hat{F}_{k+1/2}^\phi}-{\hat{F}_{k-1/2}^\phi}}{\Delta\phi}.
	\end{split}
\end{align}
In analogy to the above, we use $\Delta r^2=r_{i+1/2}^2-r_{i-1/2}^2$. The reconstructed 
fluxes $\mathbf{F}$ (coordinate basis) are related to their orthonormal counterparts 
$\hat{\mathbf{F}}$ by the following relations:
\begin{align}
	\hat{F}^r=\,F^r,\qquad\quad
	\hat{F}^\theta=\,r\times F^\theta,\qquad\quad
	\hat{F}^\phi=\,r\times\sin\theta\times F^\phi.
\end{align}

\subsection{Numerical fluxes across cell Interfaces}
\label{sec:NumericalFluxes}

We employ an approximate (HLL) Riemann solver \citep[][]{Harten1997} to derive the 
numerical fluxes at the cell interfaces:
\begin{align}
	\mathbf{F}^j=\frac{\lambda_+\mathcal{F}^j\left(\mathbf{U}^-\right)-\lambda_-\mathcal{F}^j\left(\mathbf{U}^+\right)+\lambda_+\lambda_-\left(\mathbf{U}^+-\mathbf{U}^-\right)}{\lambda_+-\lambda_-}.
\end{align}
Here, $\mathbf{U}^+$ and $\mathbf{U}^-$ correspond to the reconstructed (conserved) variables at 
the cell interfaces. $\lambda_\pm$ are given by the minimal or maximal wave speeds:
\begin{align}
	\lambda_+=\max\left(0,\mathbf{w}\right),\qquad\qquad\lambda_-=\min\left(0,\mathbf{w}\right).
\end{align}
In flat space, the propagation speeds for the conservative scheme derived from 
Eqs.~(\ref{eq:MaxI}) and~(\ref{eq:MaxII}) are $\lambda_+=1$ and $\lambda_-=-1$. Characteristic speeds of the force-free electrodynamics equations have been obtained, 
e.g., by \citet{Komissarov2002} and \citet{Carrasco2016}. For the GRFFE system of 
Eqs.~(\ref{eq:KCons}) to~(\ref{eq:KSources}), the characteristic speeds $\mathbf{w}$ are
\begin{align}
	\arraycolsep=11pt\def\arraystretch{2.0}
	\mathbf{w} =\:\left[\begin{array}{ccc}
		-\beta^i\pm\alpha\sqrt{\gamma^{ii}} & m=3 & (\text{EVI})\\ 
		-\beta^i\pm c_h\alpha\sqrt{\gamma^{ii}} & m=1 & (\text{EVII})\\
		w_q & m=1 & (\text{EVIII})\\ 
	\end{array}\right].
	\label{eq:wavespeeds}
\end{align}
Here, we do not employ the summation convention; by $m$ we denote the multiplicity of the 
respective eigenvalues. The speeds EVI correspond to the coordinate velocity of light as 
defined by \citet{CorderoCarrion2008}. The other two eigenspeeds (EVII) account for the 
propagation of the divergence cleaning potentials at speed $c_h$. Finally, EVIII corresponds 
to the wavespeed induced by the continuity equation of charge conservation, which is at 
most the coordinate velocity of light (EVI). 

\subsection{Preservation of force-free conditions}
\label{sec:FFConstraints}

Across the literature \citep[e.g.,][]{Komissarov2004,Alic2012,Parfrey2017} we find various 
modifications in the definition of $I^\mu$ to drive the numerical solution of the system of 
PDEs (Eqs.~\ref{eq:MaxI} and~\ref{eq:MaxII}) toward a state that fulfills the magnetic 
dominance condition~(\ref{eq:FFCondII}) by introducing a suitable cross-field conductivity. 
This effectively augments condition (\ref{eq:DBcons}) used to determine expression 
(\ref{eq:FFCurrent}) by a recipe to avoid (numerically) building up violations of the 
perpendicularity condition~\eqref{eq:FFCondI}.

A straightforward way to guarantee the preservation of the $\mathbf{D}\cdot\mathbf{B}=0$ 
constraint is the introduction of a numerical correction to the electric field after every 
time step of the evolution. In practice, this correction is also applied after every 
intermediate step of the employed time-integration method. To this end, the electric 
field ($\mathbf{D}$) is projected onto the direction perpendicular to the magnetic field 
($\mathbf{B}$) in every point of the numerical mesh \citep[e.g.,][]{Palenzuela2010}:
\begin{align}
	D^i\rightarrow D^k\left(\delta^i_{\hspace{4pt}k}-B_k\frac{B^i}{\mathbf{B}^2}\right).
	\label{eq:DBcutback}
\end{align}
Alternatively, dissipative currents (induced by so-called driver terms) may ensure the 
evolution of the electromagnetic fields toward physically allowed (force-free) configurations. 
Using driver terms, the numerical evolution does not guarantee that the electromagnetic 
fields are exactly force-free after every time-step. However, violations of force-free conditions are significantly reduced. While \citet{Komissarov2004}, \citet{Komissarov2011}, and 
\citet{Alic2012} introduce a modified Ohm's law with a suitably chosen cross-field dissipation, 
\citet{Parfrey2017} modify the force-free currents in Eq.~(\ref{eq:FFCurrent}) with additional 
dissipation driving the evolution toward a target ($\mathbf{D}\cdot\mathbf{B}=0$ in ideal FFE) 
configuration without further models for cross-field dissipation. They generalize 
the conservation of Eq.~(\ref{eq:FFCondI}) by introducing a target current fulfilling 
the following equation:
\begin{align}
	\mathcal{L}_{n}  (\mathbf{B}\cdot\mathbf{D})& = \kappa_I\left(\alpha^{-1}  \eta\, \mathbf{J} \cdot \mathbf{B}
	- \mathbf{D} \cdot \mathbf{B} \right).	\label{eq:CurrentDriver}
\end{align}
Here, $\kappa_I$ is the decay rate driving the left-hand side of Eq.~(\ref{eq:DBcons}) toward 
the target value and $\eta$ is a dissipation coefficient for the electric field, which is 
parallel to the current.

As for the preservation of the $\mathbf{B}^2-\mathbf{D}^2> 0$ condition (\ref{eq:FFCondII}), 
one can also employ an algebraic correction of the electric field after every step of the 
time evolution. Following \citet{Palenzuela2010}, the electric field ($\mathbf{D}$) is 
rescaled in every point of the numerical mesh to the length of the magnetic field 
($\mathbf{B}$) in a qualitatively similar manner as in Eq.~\eqref{eq:DBcutback}:
\begin{align}
	D^i\rightarrow D^i\left(1-\Theta\left(\chi\right)+\frac{\left|\mathbf{B}\right|}{\left|\mathbf{D}\right|}\Theta\left(\chi\right)\right), \label{eq:DsqBsqRescale}
\end{align}
where $\Theta$ is the Heaviside function, 
$\chi=\:\mathbf{D}^2-\mathbf{B}^2$ \citep[other choices of the functional dependence $\Theta$ that maintain strict magnetic dominance have been employed, e.g., by][]{Paschalidis2013}. Again, an 
alternative is employed by \citet{Komissarov2011}, and \citet{Alic2012}, introducing driver 
terms for additional dissipative currents, also for the conservation of the 
$\mathbf{B}^2-\mathbf{D}^2> 0$ condition. 

Our GRFFE scheme employs, by default, the algebraic correction of electric fields in every 
(intermediate) step of the time evolution as given by Eqs.~(\ref{eq:DBcutback}) 
and~(\ref{eq:DsqBsqRescale}). However, in \citet{Mahlmann2019} we resorted to a suitably 
chosen resistivity model \citep[in analogy to][]{Komissarov2004} replacing the instant 
algebraic cutback of the electric displacement field by a gradual relaxation of force-free 
violations. For a review on the interpretation of constraint violations in GRFFE, we refer 
to \citet{Mahlmann2019}.

\subsection{Treatment of the parallel current}
\label{sec:ParallelCurrent}
The last term in Eq.~\eqref{eq:FFCurrent} is the component of the current parallel to 
the magnetic field, with the spatial projection
\begin{align}
	\mathbf{j}_{||}=\:&
	\frac{\mathbf{B} \cdot(\nabla \times \mathbf{B})
		-\mathbf{D} \cdot(\nabla \times \mathbf{D})}{B^2} \mathbf{B}.
	\label{eq:CurrentParallel}
\end{align}
We have empirically found (see Paper II) that the discretization of the parallel current 
is one of the main sources of numerical diffusivity in our code in certain tests. Indeed, the presence of 
derivatives of conserved quantities in the parallel current (Eq.~\ref{eq:FFCurrent}) makes the 
practical evaluation of this term cumbersome in numerical simulations. This has brought 
some authors \citep[e.g.,][]{Yu2011} to ignore it completely (i.e., assuming 
$\mathbf{j}_{||}=\mathbf{0}$), and removing the accumulated parallel component of the 
electric field employing an algebraic procedure akin to that of Eq.~\eqref{eq:DBcutback}. 
However, this specific strategy of \citet{Yu2011} did not yield satisfactory results when 
employed with the numerical framework described in this paper.

The order of the discretization of the derivatives has to be comparable with the order of 
accuracy of the spatial reconstruction. Otherwise, the global order of accuracy of the 
scheme decreases (see Sect.~2.1 of Paper II). In the previous applications of our code 
\citep{Mahlmann2019,Mahlmann2020} and independently of the order of the spatial reconstruction, 
we employed a fourth-order accurate discretization of the partial derivatives in 
Eq.~\eqref{eq:FFCurrent}. In case of the (exemplary) sweep in the $x$ direction, the finite 
difference discretization is of the following form (for uniform grids):
\begin{align}
	\left[\frac{\partial A}{\partial x}\right]_i^{4th}\approx\,\frac{A_{i-2}-8A_{i-1}+8A_{i+1}-A_{i+2}}{12\times\Delta x}, 
	\label{eq:4thorderderivative}
\end{align}
where $A$ denotes a quantity on the numerical mesh (e.g., $\mathbf{D}$ or $\mathbf{B}$) and 
the respective locations are labeled as in Sect.~\ref{sec:FVIntegration}. In Paper II, we will evaluate the improvements by changing the discretization of $\mathbf{j}_{||}$ according to the sixth-order and eighth-order 
accurate formulae
\begin{align}
	\left[\frac{\partial A}{\partial x}\right]_i^{6th}\approx\,&\frac{-A_{i-3}+9A_{i-2}-45A_{i-1}+45A_{i+1}-9A_{i+2}+A_{i+3}}{60\times \Delta x}, \label{eq:6thorderderivative}\\
	\begin{split}
		\left[\frac{\partial A}{\partial x}\right]_i^{8th}\approx\,&\frac{3A_{i-4}-32A_{i-3}+168A_{i-2}-672A_{i-1}}{840\times\Delta x}\\
		&+\frac{672A_{i+1}-168A_{i+2}+32A_{i+3}-3A_{i+4}}{840\times \Delta x}. \label{eq:8thorderderivative}
	\end{split}
\end{align}

\subsection{Cleaning of numerical errors}
\label{sec:CleaningErrors}

We extend the augmented evolution equations by a hyperbolic/parabolic divergence error 
cleaning with the possibility of having a hyperbolic advection speed, $c_h>1$ (see below), 
as suggested by \citet{Mignone2010}. In order to minimize violations of $\text{div}\mathbf{B}=0$ in spacetimes containing BHs, 
we find it beneficial to employ $1\le c_h\le 2$. In practice, a propagation speed within 
this interval does not limit the time step strongly, since the numerical evolution of the BSSN equations usually demands Courant–Friedrichs–Lewy (CFL) factors significantly smaller than unity 
(say $f_\textsc{cfl}\sim 0.1-0.3$) and, often, 
choosing $c_h>1$ allows somewhat larger values of the same. Hence, we choose to advect 
numerical errors of this constraint with a speed faster than the speed of light 
(typically, $c_h=2$) to significantly reduce the numerical noise. We employ the same 
scheme with $c_h=1$ for the cleaning of numerically induced errors in charge conservation 
by the scalar potential $\Psi$.

The variables $\kappa_{\Psi}$ and $\kappa_{\Phi}$ are damping rates, introducing time scales that act 
in addition to the advection time scales of the hyperbolic conservation laws of the augmented 
GRFFE system (Eqs.~\ref{eq:KCons} to~\ref{eq:KSources}). In order to deal with the 
potential stiffness introduced by the parabolic damping numerically, we employ a time step 
splitting technique \citep[Strang splitting, see, e.g.,][]{LeVeque2007}, which has been 
applied previously to GRFFE by \citet{Komissarov2004}. Prior to and after the method-of-lines 
time integration\footnote{Specifically, we add the exact evaluation of stiff source terms 
	before the scheduling bin \textsc{MoL\_Step} and before \textsc{MoL\_PostStep}. The latter 
	has to be restricted to the last intermediate step of the method-of-lines integration.}, 
we evaluate the equations
\begin{align}
	\mathcal{P}\left(t\right)&=\mathcal{P}_0\exp\left[-\alpha^2\kappa_\Phi c_h t\right]\,,\\
	\mathcal{Q}\left(t\right)&=\mathcal{Q}_0\exp\left[-\alpha^2\kappa_\Psi t\right]\,,
\end{align}
for a time $t=\Delta t/2$. We find it beneficial to choose a large value for $\kappa_\Phi$, 
in some cases $\sim 200$, effectively dissipating charge conservation errors on very short 
time scales (and justifying the time-splitting approach). As for the divergence cleaning, 
we conducted a series of tests, optimizing $\kappa_\Psi$ to yield stable and converging 
evolution for all shown resolutions, ultimately resorting to $\kappa_\Psi = 0.125-0.25$ 
\citep[see also][]{Mahlmann2019}.

\section{Numerical tests}\label{sec:Numerical_Tests}

We present several tests with results that specifically depend on the various numerical methods 
(e.g., reconstruction, cleaning of numerical errors) available in our new code. Since the 
code is genuinely 3D, in 1D and 2D simulations, the surplus dimensions are condensed to the 
extension of one cell by applying appropriate boundary conditions to them. If not stated otherwise, all plasma fields at the remaining boundaries are either extrapolated linearly or the (outer) boundary itself is located sufficiently far away from the grid-region of interest, so that a simple copy boundary is enough. For dynamical spacetime evolutions, we use radiation (Sommerfeld) boundary conditions \citep[see, e.g.,][]{2003PhRvD..67h4023A} for all evolved spacetime fields at the outer boundary of the domain.
Section~\ref{sec:1D_Reconstruction} reviews the 1D tests of signal propagation and stability 
in GRFFE following the work by \citet{Komissarov2004} and \citet{Yu2011} closely. In 
Sect.~\ref{sec:FFE_Interaction} we probe the correct representation of force-free plasma 
wave interactions by reproducing key results by \citet{Li2019}. 
All tests in this section are performed in a fixed background Minkowski spacetime.
The initial value of the charge density is computed as $\rho = \text{div} \mathbf{D}$ 
and the cleaning potentials are set to $\Psi=\Phi=0$.

\subsection{Testing the 1D reconstruction methods}
\label{sec:1D_Reconstruction}

GRFFE allows two modes of plasma waves \citep{Komissarov2002,Punsly2003,Li2015,Li2019}: 
Alfv\'{e}n waves that transport energy, charge, and current along magnetic field lines 
and fast waves that correspond to the linearly polarized waves of vacuum electrodynamics 
(see also Sect.~\ref{sec:NumericalFluxes}). The following set of 1D problems is selected 
to demonstrate (a) the correct propagation of fast waves, (b) the formation of a 
current-sheet when magnetic dominance breaks down and (c) the correct modeling of stationary 
Alfv\'{e}n waves that do not transport energy across magnetic field lines \citep[cf.][]{Li2019}. 
In problem (c), the wave can only diffuse due to a finite numerical resistivity if the force-free conditions 
are not preserved (see Paper II). The numerical solution to all these problems critically 
depends on the employed reconstruction algorithms. Since our code employs numerical 
reconstruction in 1D sweeps across all dimensions, we consider the following suite of 1D 
tests a fundamental measure for the performance of our GRFFE scheme. We verify 
\citep[in the sense of][]{Roache1997} the correct implementation of the reconstruction 
methods evaluating the convergence order from several data-sets with increasing 
resolution. Specifically, we evaluate the (global) difference measure 
\citep[cf.][]{Anton2010}
\begin{align}
	\epsilon^{ab}=\frac{1}{N}\times\sum_{i}\left|u^a_i-u^b_i\right|,
\end{align}
where $u^a$ and $u^b$ are the one-dimensional vectors (of $N$ elements) of the considered 
evolved quantity at different levels of resolution, $a,b\in\left[1,2,3\right]$. We denote 
the resolution on each of these levels by $\Delta x_1$, $\Delta x_2$, $\Delta x_3$, 
where $\Delta x_3/\Delta x_2=\Delta x_2/\Delta x_1$. With level $1$ being the level with 
the finest resolution, the (empirical) order of convergence is then defined as
\citep[see also][]{BonaConvergence1998}:
\begin{align}
	\mathcal{O}=\frac{\log\left(\epsilon^{23}/\epsilon^{12}\right)}{\log\left(\Delta x_3/\Delta x_2\right)}\label{eq;ConvergenceOrder}.
\end{align}

Unless stated otherwise, in the following tests we employ a fourth-order accurate 
discretization of the parallel current $\mathbf{j}_{||}$.

\subsubsection{(Degenerate) current sheet test}
\label{sec:Deg_CurrentSheet}

\begin{figure}
	\centering
	\includegraphics[width=0.48\textwidth]{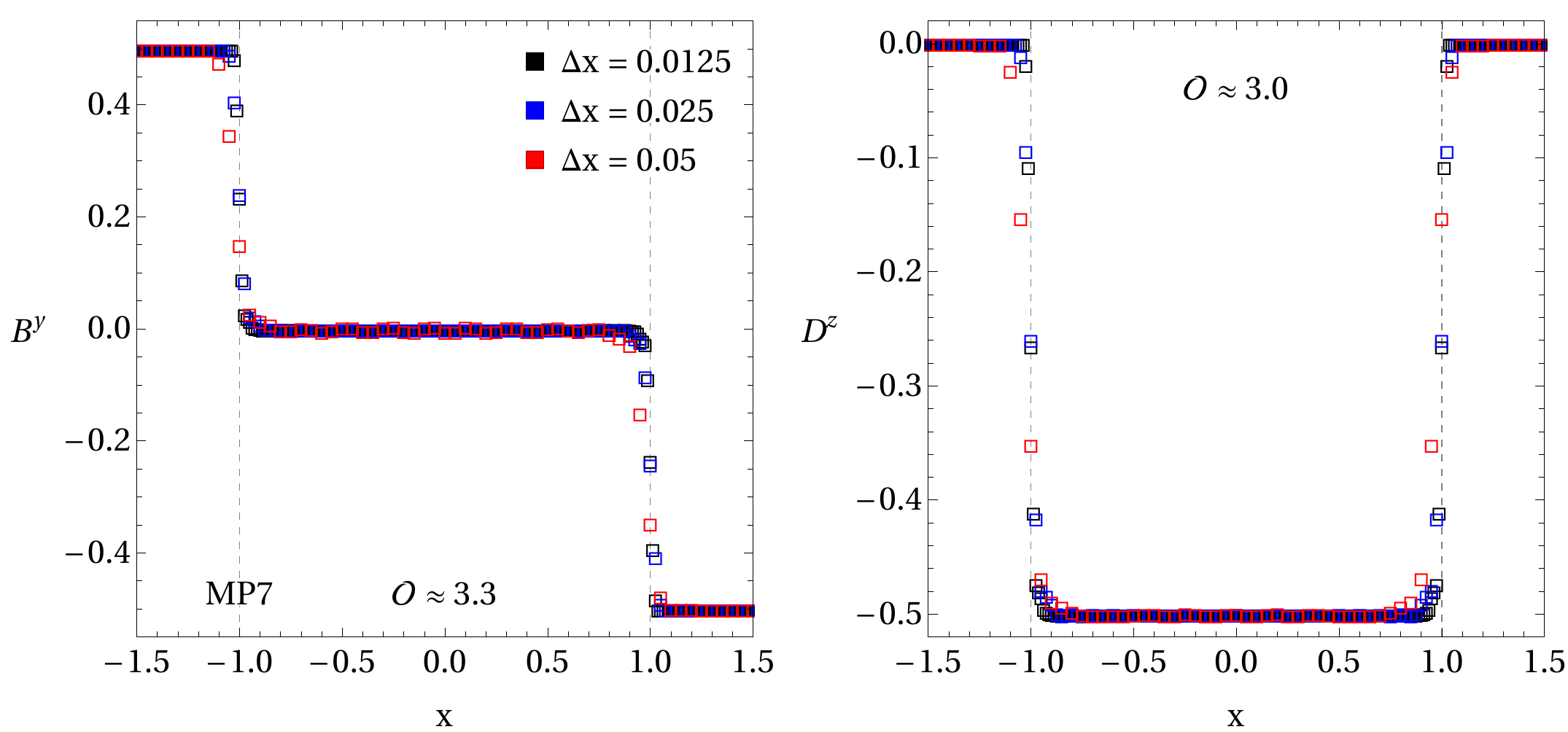}\\
	\includegraphics[width=0.48\textwidth]{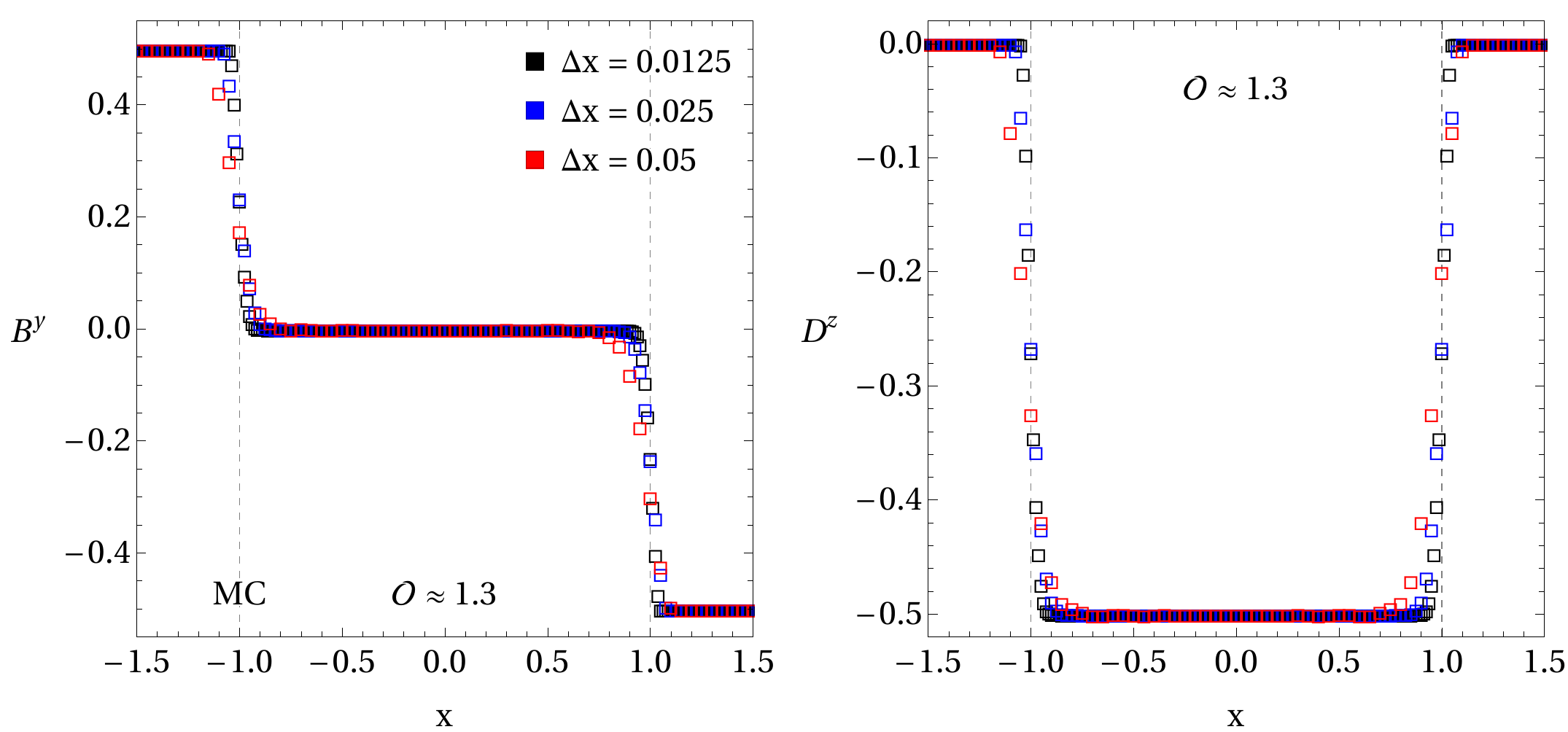}
	\caption{Current sheet test \citep{Komissarov2004,Yu2011} as described by the initial 
		data in Eq.~(\ref{eq:currensheetinitialdata}) on a $x\in\left[-2,2\right]$ 
		grid ($f_\textsc{cfl}=0.25$) at $t=1.0$ for $B_0=0.5$ and different resolutions. 
		Two fast waves emerge from the original discontinuity and propagate outward with the 
		speed of light (analytical position of the waves are indicated by dashed vertical lines). 
		The order of convergence, $\mathcal{O}$ is indicated according to Eq.~(\ref{eq;ConvergenceOrder}). \textit{Top:} MP7 reconstruction. \textit{Bottom:} 
		MC reconstruction.}
	\label{fig:CurrentSheet}
\end{figure}

\begin{figure}
	\centering
	\includegraphics[width=0.48\textwidth]{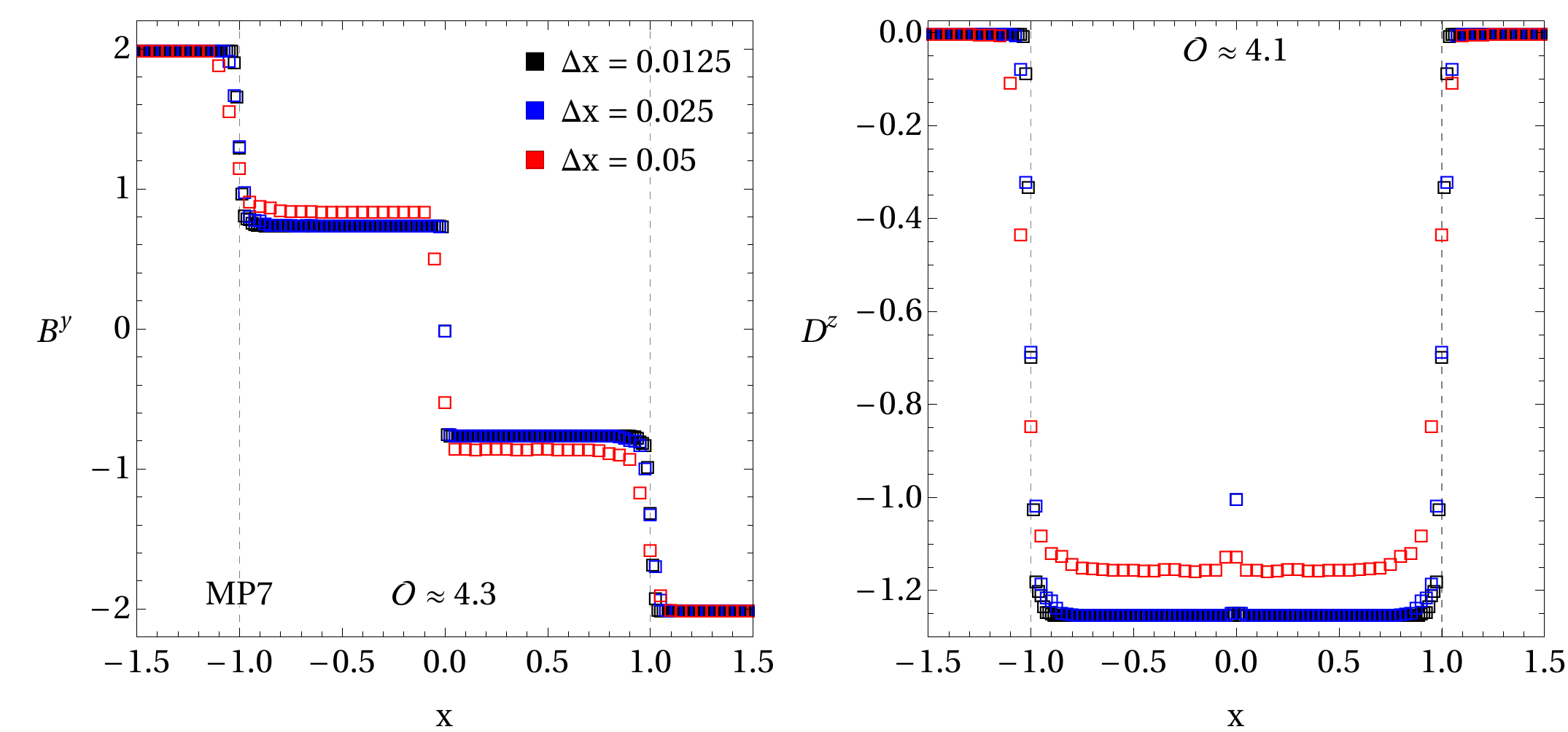}\\
	\includegraphics[width=0.48\textwidth]{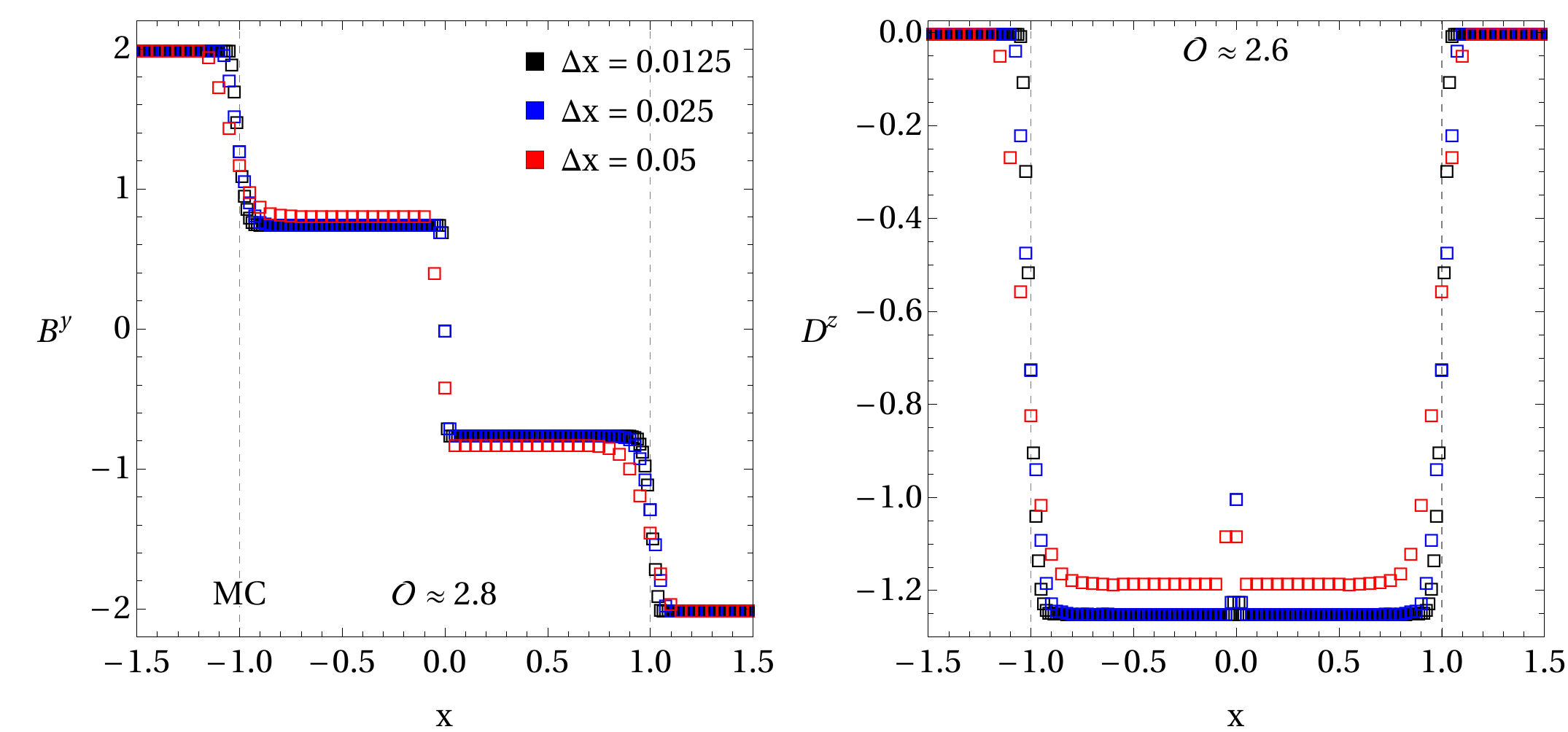}
	\caption{Degenerate current sheet test \citep{Komissarov2004,Yu2011} as as described 
		by the initial data (Eq.~\ref{eq:currensheetinitialdata}) with $B_0=2.0$. Two fast waves 
		emerge from the original discontinuity and propagate outward with the speed of light. 
		The cross field conductivity (induced by conserving conditions~\ref{eq:FFCondI}, 
		and~\ref{eq:FFCondII}) terminates the fast waves in the breakdown-zone. \textit{Top:} 
		MP7 reconstruction. \textit{Bottom:} MC reconstruction.}
	\label{fig:DegCurrentSheet}
\end{figure}

\citet{Komissarov2004} examines two variations of a current sheet problem, one of which has 
a solution in force-free electrodynamics, while the other violates the force-free condition 
(Eqs.~\ref{eq:FFCondI}, and \ref{eq:FFCondII}). The tests for physical current sheets 
(Fig.~\ref{fig:CurrentSheet}) and degenerate current sheets (Fig.~\ref{fig:DegCurrentSheet}) 
are initialized by the following set of data:
\begin{align}
	\begin{split}
		\mathbf{D}=0,\qquad \mathbf{B}=\left(1.0,B_y,0.0\right),\qquad
		B_y&=\left\{\begin{array}{cc}
			B_0& x<0\\
			-B_0& x>0
		\end{array}\right.
	\end{split}. \label{eq:currensheetinitialdata}
\end{align}
If $B_0<1$, there exists a force-free solution given by two fast waves traveling at the 
speed of light \citep[see Fig.~\ref{fig:CurrentSheet}, also Fig.~C2 in][]{Komissarov2004}. 
For $B_0>1$ the solution is dominated by an increasing cross-field conductivity that 
locks $\mathbf{B}^2-\mathbf{D}^2$ to zero in a current sheet located at $x=0$. At this 
location, the preservation of the force-free conditions (Sect. \ref{sec:FFConstraints}) 
becomes important for the field dynamics, namely it changes the structure of the propagating 
waves. The states bounded by the fast waves are terminated at the current sheet and a 
standing field reversal remains \citep[see Fig.~\ref{fig:DegCurrentSheet}, 
cf. Fig.~C2 in][]{Komissarov2004}. We take advantage of this test to compare the performance 
of two different reconstruction schemes: The second-order accurate, slope limited TVD reconstruction 
with a monotonized central (MC) limiter \citep{vanLeer1977}, and the seventh-order accurate 
monotonicity preserving \citep[MP7,][]{Suresh1997} reconstruction. 

From the results of the presented tests 
(Figs.~\ref{fig:CurrentSheet} and~\ref{fig:DegCurrentSheet}), we draw the following 
conclusions. Fast electromagnetic waves propagate correctly at the speed of light.  For a resolution similar to the one employed in \citet{Komissarov2004}, where $\Delta x=0.015$, the time evolution of the (degenerate) current sheet is in good qualitative agreement with the literature \citep{Komissarov2004,Yu2011}. For resolutions below the lowest presented resolution (i.e., for $\Delta x>0.05$) the wave structure of the presented test quickly smears out.	

\begin{figure}
	\centering
	\includegraphics[width=0.48\textwidth]{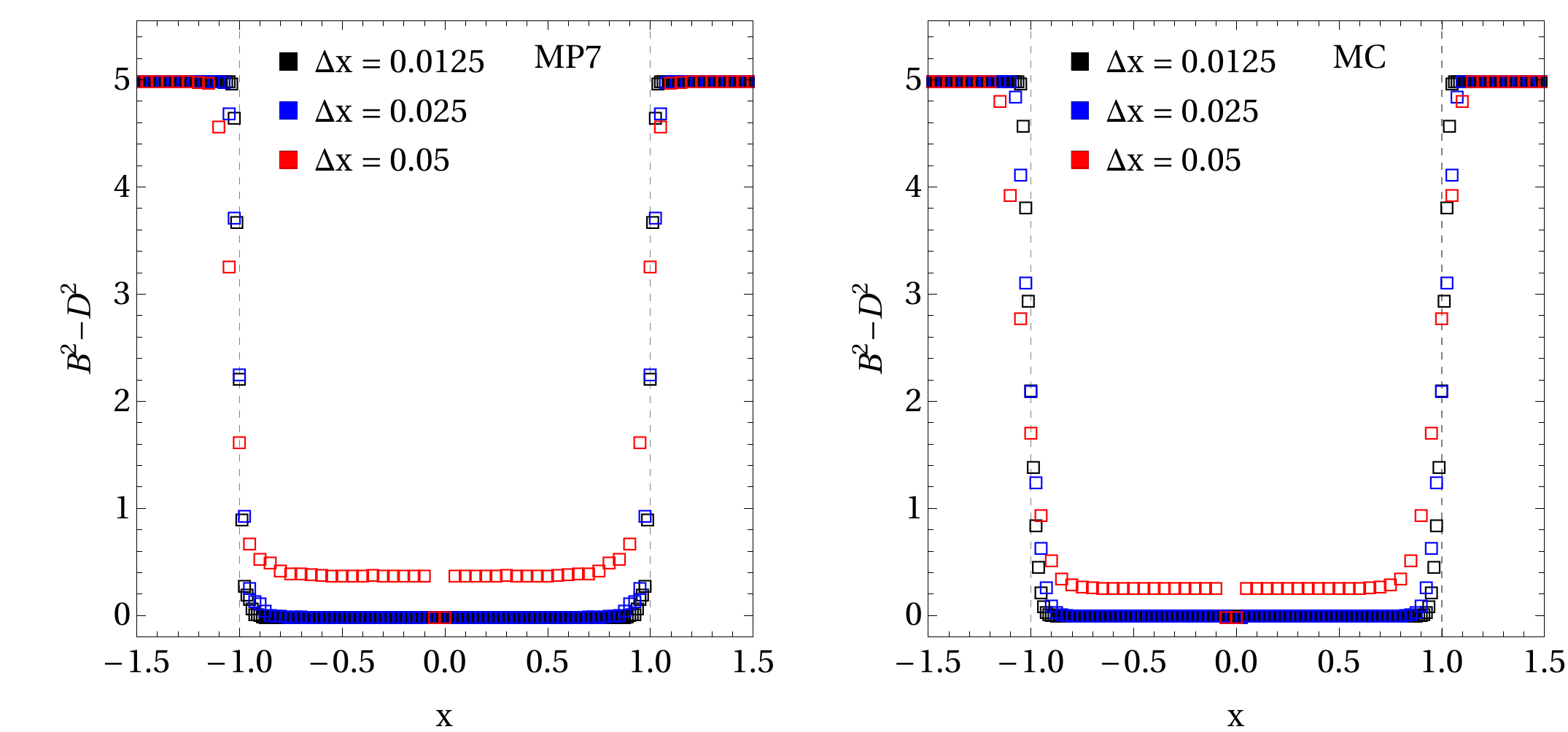}
	\caption{Same scenario as Figure \ref{fig:DegCurrentSheet}. The panels show the magnetic dominance state \citep[cf. Fig. C2,][]{Komissarov2004}. $\mathbf{D}^2>\mathbf{B}^2$ develops at the location of the central current sheet, such that the electric field is altered by the algebraic adjustment maintaining condition (\ref{eq:FFCondII}). Our numerical implementation of the adjustment (\ref{eq:DsqBsqRescale}) drives $\mathbf{D}^2-\mathbf{B}^2\rightarrow 0$ instantaneously, restoring magnetic dominance.}
	\label{fig:BDsqDegCurrentSheetMP}
\end{figure}

Conservation of force-free conditions in the degenerate current sheet test is working well and agrees with similar tests throughout the literature (see Fig. \ref{fig:BDsqDegCurrentSheetMP}). While monotonicity preserving reconstruction is slightly more oscillatory than, for example, monotonized central flux limiters, the higher-order schemes allow a steeper resolution of wave-fronts and current sheets. While the order of convergence of the (more diffusive) MC reconstruction approaches the formal theoretical order of convergence ($\mathcal{O}=2$),  the order of convergence degrades below its theoretical value for MP7.

Although some degradation of the order of convergence is expected in non-smooth regions of the flow (e.g., the discontinuities associated with fast or Alfv\'{e}n waves), the algebraic enforcement of the violated force-free conditions seems to have a large impact on the computed value of $\mathcal{O}$. Very likely, such algebraic enforcement is the main source of deviation from the theoretical expectations regarding the order of convergence.

Given the previous statements, the developed GRFFE code passes the 1D (degenerate) current sheet test.


\subsubsection{Three-wave and stationary alfv\'{e}n wave test}
\label{sec:ThreeWave}

\begin{figure}
	\centering
	\includegraphics[width=0.48\textwidth]{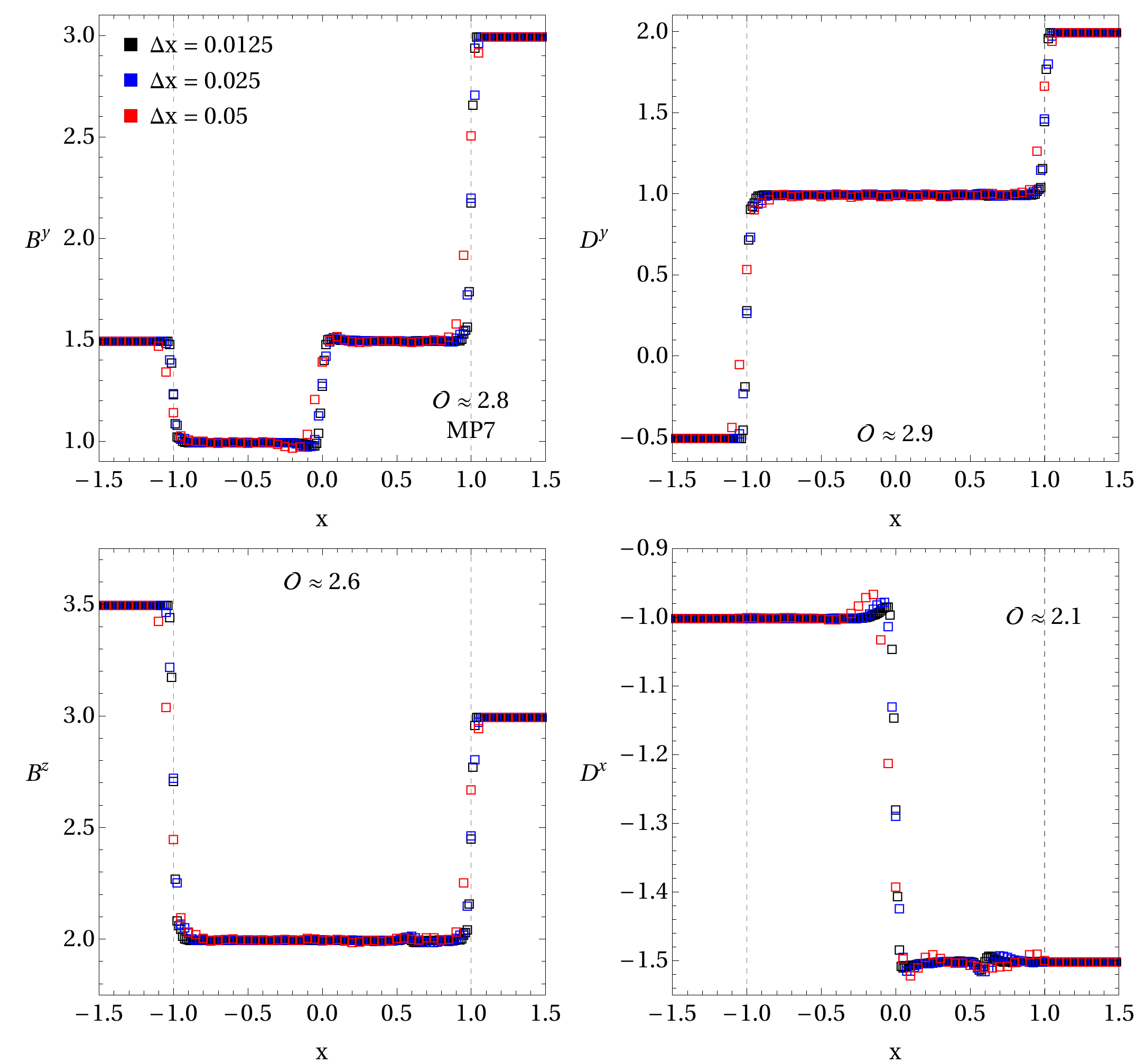}
	\caption{Three-wave problem \citep{Paschalidis2013} as described by the setup in 
		Eq.~(\ref{eq:3waves}) and employing MP7 reconstruction. Numerical setup and labels are the same as in 
		Fig.~\ref{fig:CurrentSheet}. The initial discontinuity at $x=0$ splits into two fast 
		discontinuities and one stationary Alfv\'{e}n wave.}
	\label{fig:Threewave}
\end{figure}
\begin{figure}
	\centering
	\includegraphics[width=0.48\textwidth]{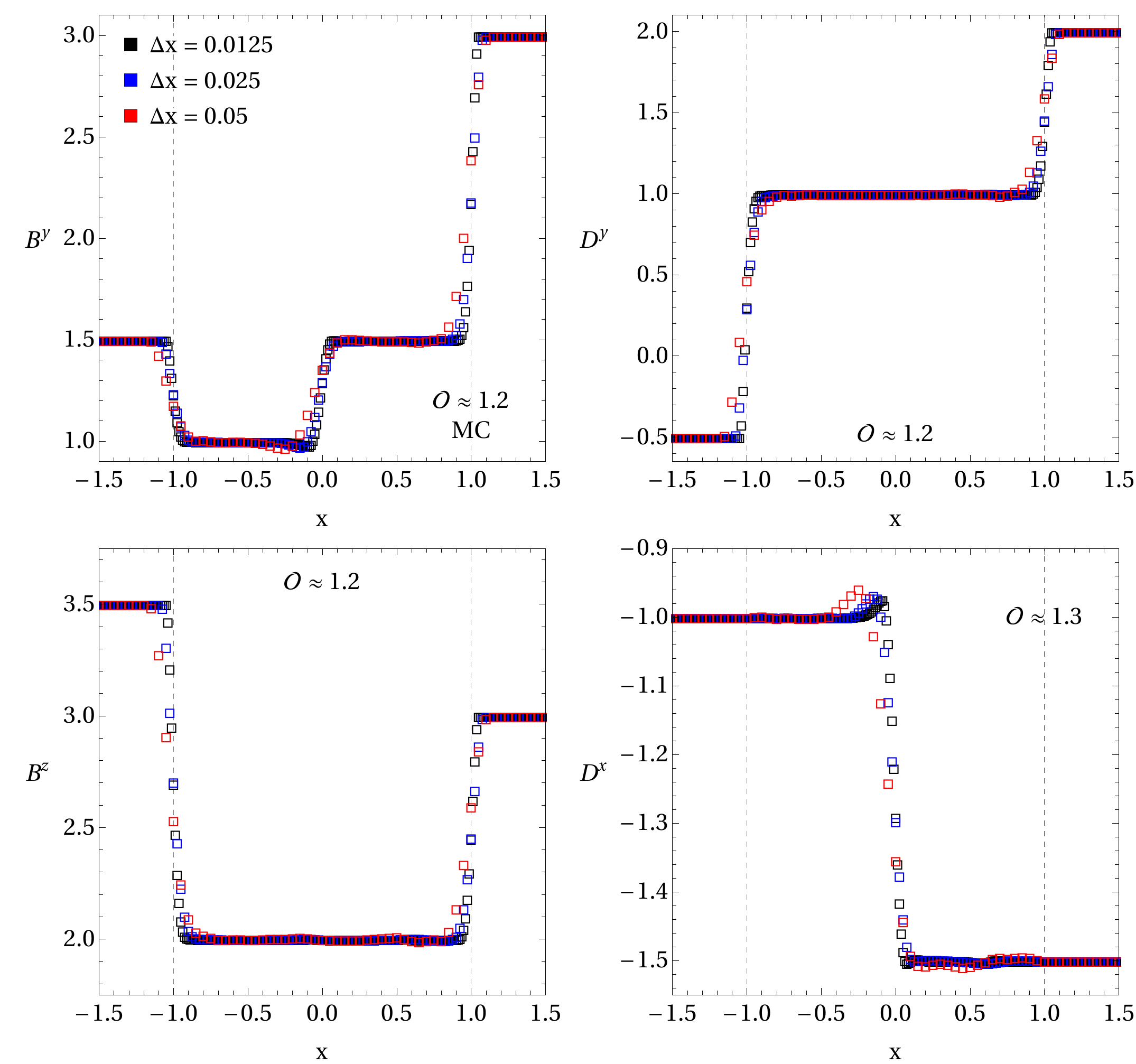}
	\caption{Same as Fig.~\ref{fig:Threewave} but employing MC reconstruction.}
	\label{fig:ThreewaveMC}
\end{figure}
\begin{figure}
	\centering
	\includegraphics[width=0.48\textwidth]{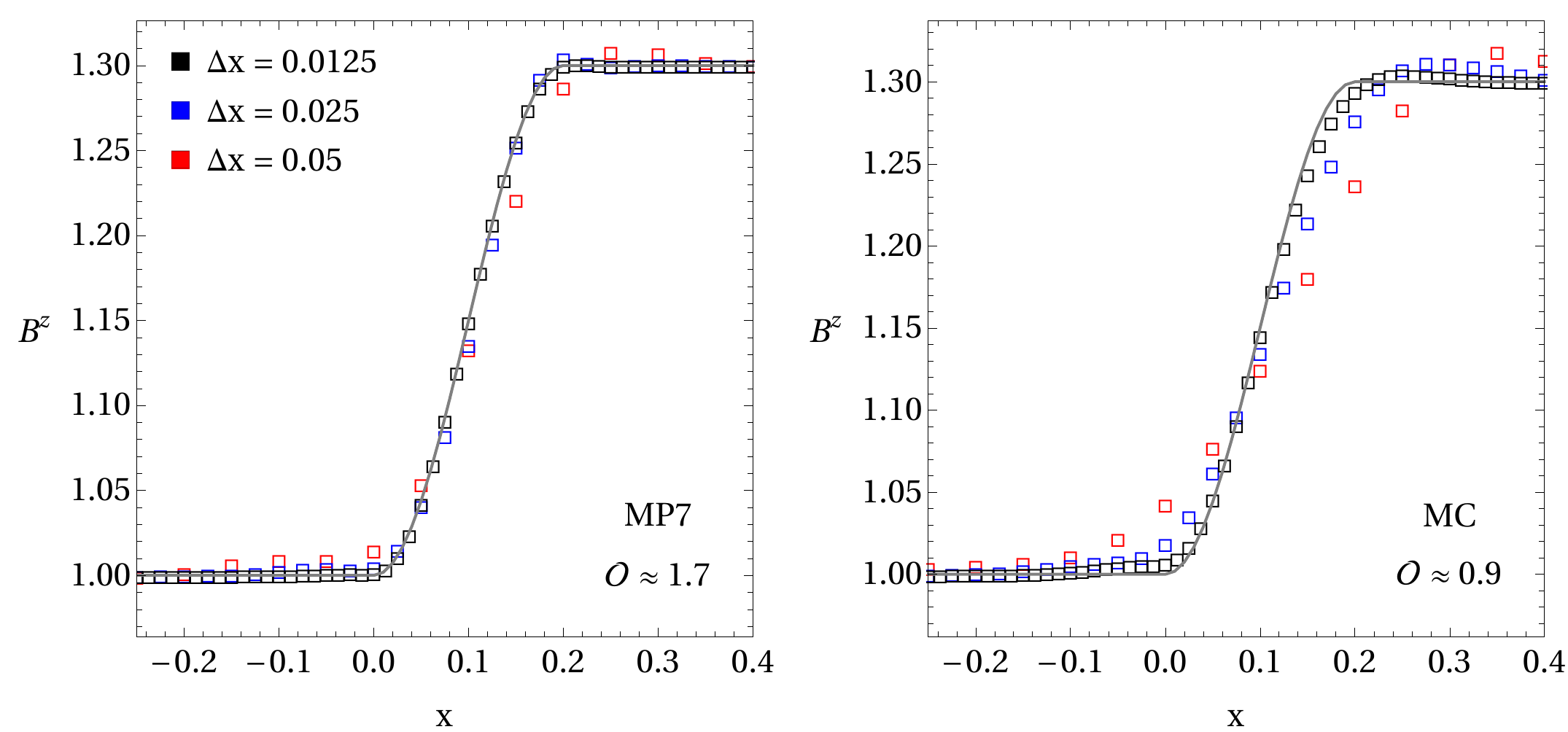}
	\caption{Stationary Alfv\'{e}n wave problem \citep{Komissarov2004}, same numerical setup as in 
		Fig.~\ref{fig:CurrentSheet}. The analytic solution (Eq.~\ref{eq:standingAlfven}) is indicated by a gray line.}
	\label{fig:Stationary}
\end{figure}

\citet{Komissarov2002}, \citet{Yu2011} and \citet{Paschalidis2013} suggest the three-wave 
problem (or a variant thereof, see Fig.~\ref{fig:Threewave}) as a test for force-free 
electrodynamics. The initial discontinuity at $x=0$ splits into two fast discontinuities and 
one stationary Alfv\'{e}n wave. This effectively combines the previously introduced test of 
Sect.~\ref{sec:Deg_CurrentSheet} with the standing Alfv\'{e}n wave test thats was also 
employed by \citet{Komissarov2004}. The initial electromagnetic field is given by \citep{Paschalidis2013}:
\begin{align}
	\begin{split}
		\mathbf{B}&=\left(1.0,1.5,3.5\right),\qquad \mathbf{D}=\left(-1.0,-0.5,0.5\right),\qquad \text{if } x<0;\\
		\mathbf{B}&=\left(1.0,3.0,3.0\right),\qquad \mathbf{D}=\left(-1.5,2.0,-1.5\right),\qquad \text{if } x>0.
	\end{split}
	\label{eq:3waves}
\end{align}

We evolve this setup in time and present the results for different resolutions in 
Figs.~\ref{fig:Threewave} and~\ref{fig:ThreewaveMC}. Around the standing Alfv\'{e}n wave at 
$x=0$, high-order reconstructions tend to develop small-scale oscillations, especially 
visible in the plots of $D^x$, restricted to the region delimited by the fast waves 
(at $x=\pm 1$ for $t=1$). Oscillations around this discontinuity can also be observed 
(for higher resolutions) in part of the literature \citep[specifically, Fig.~4 in][]{Yu2011}. 
The order of convergence is slightly reduced when compared to the results shown in the previous 
section, probably due to the specific challenges of resolving stationary Alfv\'{e}n waves, not 
only in GRFFE but also in relativistic MHD \citep[see, e.g.,][]{Anton2010}. 

The empirical order of convergence grows employing MP7 in combination with a sixth-order 
accurate discretization of the parallel current \eqref{eq:6thorderderivative}. This growth manifests, for example, in an increase from $\mathcal{O}\approx 2.1$ to $\mathcal{O}\approx 2.8$ for $D^x$, but 
is negligible in other variables. In any case, the overall numerical solution does not 
significantly change modifying the order of accuracy of the calculation of $\mathbf{j}_{||}$ 
in the 1D problems involving discontinuities.

\citet{Komissarov2004} achieves high accuracy maintaining a single standing Alfv\'{e}n wave 
stationary during evolution for resolutions comparable to the highest one shown in 
Figs.~\ref{fig:Threewave} and~\ref{fig:ThreewaveMC}. The numerical techniques in \citet{Komissarov2004} are slightly different from ours, employing, for example, a linear Riemann solver, which makes use of the full spectral decomposition of the FFE equations. 
As such, it distinguishes all physical, and nonphysical wave speeds and may provide additional 
accuracy at critical locations (in the context of GRFFE, e.g., at current sheets). 
Additionally, \citet{Komissarov2004} employs a different form of the current in Faraday's 
Eq.~(\ref{eq:MaxI}) based on a specific (numerical) resistivity model to drive electromagnetic 
fields toward a force-free state throughout the evolution. Although one could suspect that 
this different treatment of the currents may alter the numerical solution significantly in 
this test (which is dominated by the numerical diffusivity of the standing wave), we 
find that our results are quite similar to the ones of \citet{Komissarov2004}; a more detailed analysis is presented in Paper II.

Next, we consider the analytical solution of 
a standing Alfv\'{e}n wave as initial data in the following:
\begin{align}
	\begin{split}
		\mathbf{B}&=\left(1,1,B^z\right),\qquad \mathbf{D}=\left(-B^z,0,1\right),\\
		B^z&=\left\{\begin{array}{cc}
			1& x\leq 0\\
			1+0.15\left[1+\sin\left[5\pi\left(x-0.1\right)\right]\right]& 0<x\leq 0.2\\
			1.3 & x>0.2
		\end{array}\right.
	\end{split}. \label{eq:standingAlfven}
\end{align}
We present the results of the Alfv\'{e}n stationarity test in Fig.~\ref{fig:Stationary}. 
With resolutions comparable to the one employed in \citet{Komissarov2004}, namely 
$\Delta x\approx 0.015$, the numerical solution converges to the analytic one with an order of convergence of $\approx 2$ for MP7 reconstruction. This order of convergence is dominated by the numerical errors around the transition layer $0\lesssim x\lesssim 0.2$. As mentioned in the previous tests, 
standing Alfv\'{e}n waves seem to introduce severe degradation of the order of convergence 
in MP methods (we also verified these results with MP5). This is very likely related 
to the preservation of the $\mathbf{D}\cdot\mathbf{B}=0$ condition, in the extended region 
$0\le x\le 0.2$ where $B^z$ is not uniform. In that region, the 
cutback of the electric displacement generates numerical errors that accumulate mostly 
close to its lower boundary (see the behavior of $D^x$ in $-0.5\lesssim x \lesssim 0$ in 
Fig.\,\ref{fig:ThreewaveMC}).

\subsection{FFE wave interaction (2D/3D)}
\label{sec:FFE_Interaction}
\begin{figure}
	\centering
	\includegraphics[width=0.48\textwidth]{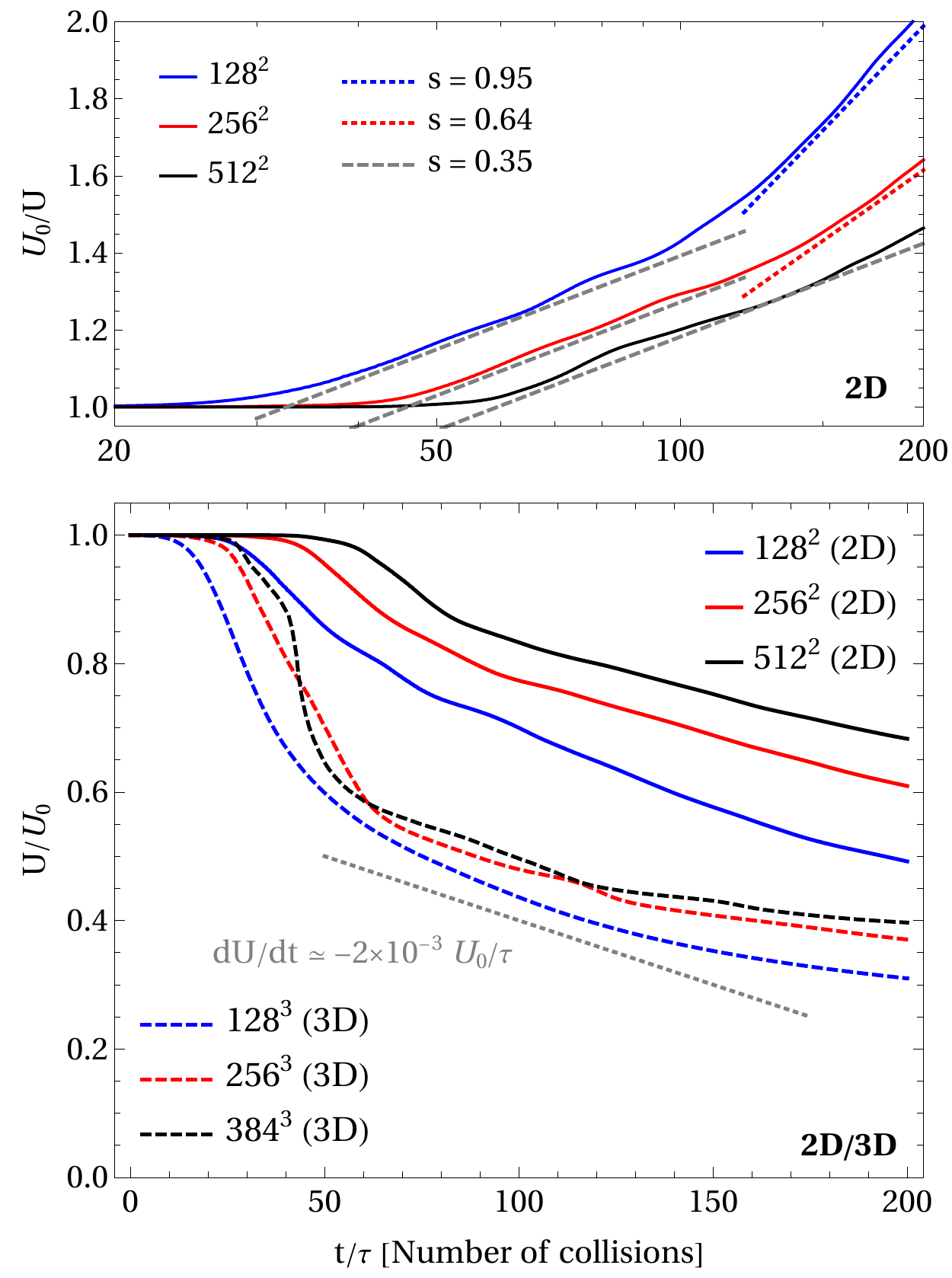}
	\caption{Free-energy $U$ (normalized to its initial value $U_0$) during the collision 
		of Alfv\'{e}n wave packets on numerical meshes (2D/3D) of various resolutions 
		(indicated by different line styles). \textit{Top:} Evolution of $U_0/U$ for the set of 
		2D models. Slopes for the asymptotic linear relation between $U_0/U$ and $\text{ln}\:t$ 
		are indicated by dashed/dotted lines, comparing to Fig.~7 of \citet{Li2019}.
		\textit{Bottom:} Free-energy evolution normalized to $U_0$ comparing to Figs.~2 
		and~5 from \citet{Li2019}. The asymptotic slope for 3D models found by \citet{Li2019} 
		is indicated by a gray dashed line for reference.}
	\label{fig:AlfvenDissipation}
\end{figure}
\begin{figure}
	\centering
	\includegraphics[width=0.48\textwidth]{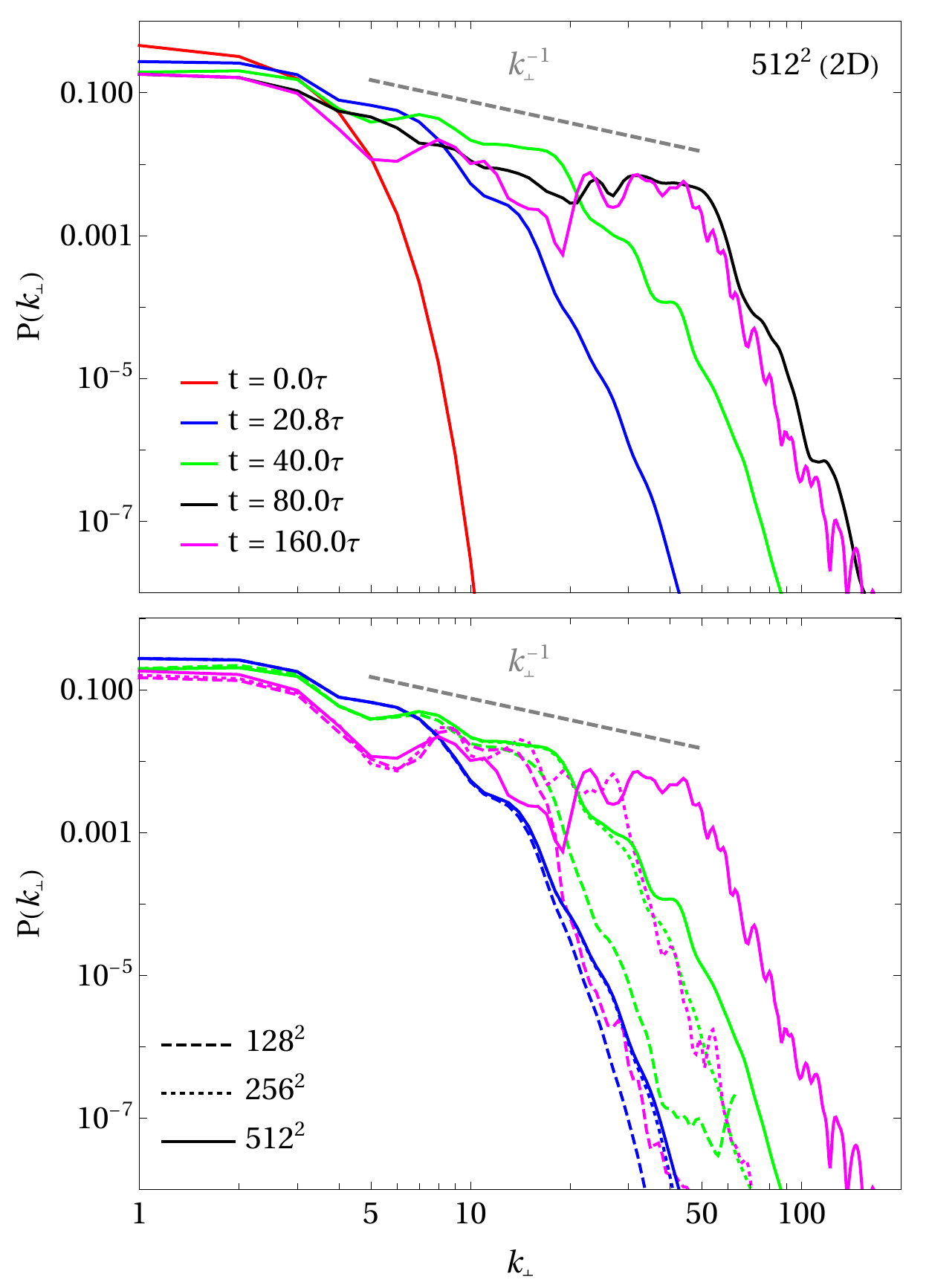}
	\caption{Spectrum evolution for the 2D simulation of Alfv\'{e}n interactions initialized 
		according to Eq.~(\ref{eq:AlfvenPackage2D}). We show the spectral energy distribution 
		for wavenumbers $k_\perp$ (perpendicular to the guide field) in analogy to Fig.~6, 
		\citet{Li2019}. \textit{Top:} Spectral energy distribution at different times for the 
		resolution $512^2$. \textit{Bottom:} Spectral energy distribution for selected times 
		and wavenumbers (color code as in \textit{top} panel) and different resolutions, 
		indicated by dashed ($128^2$), dotted ($256^2)$ and solid ($512^2$) lines. No visible 
		convergence is reached.}
	\label{fig:Alfven2DSpectrum}
\end{figure}
\begin{figure}
	\centering
	\includegraphics[width=0.48\textwidth]{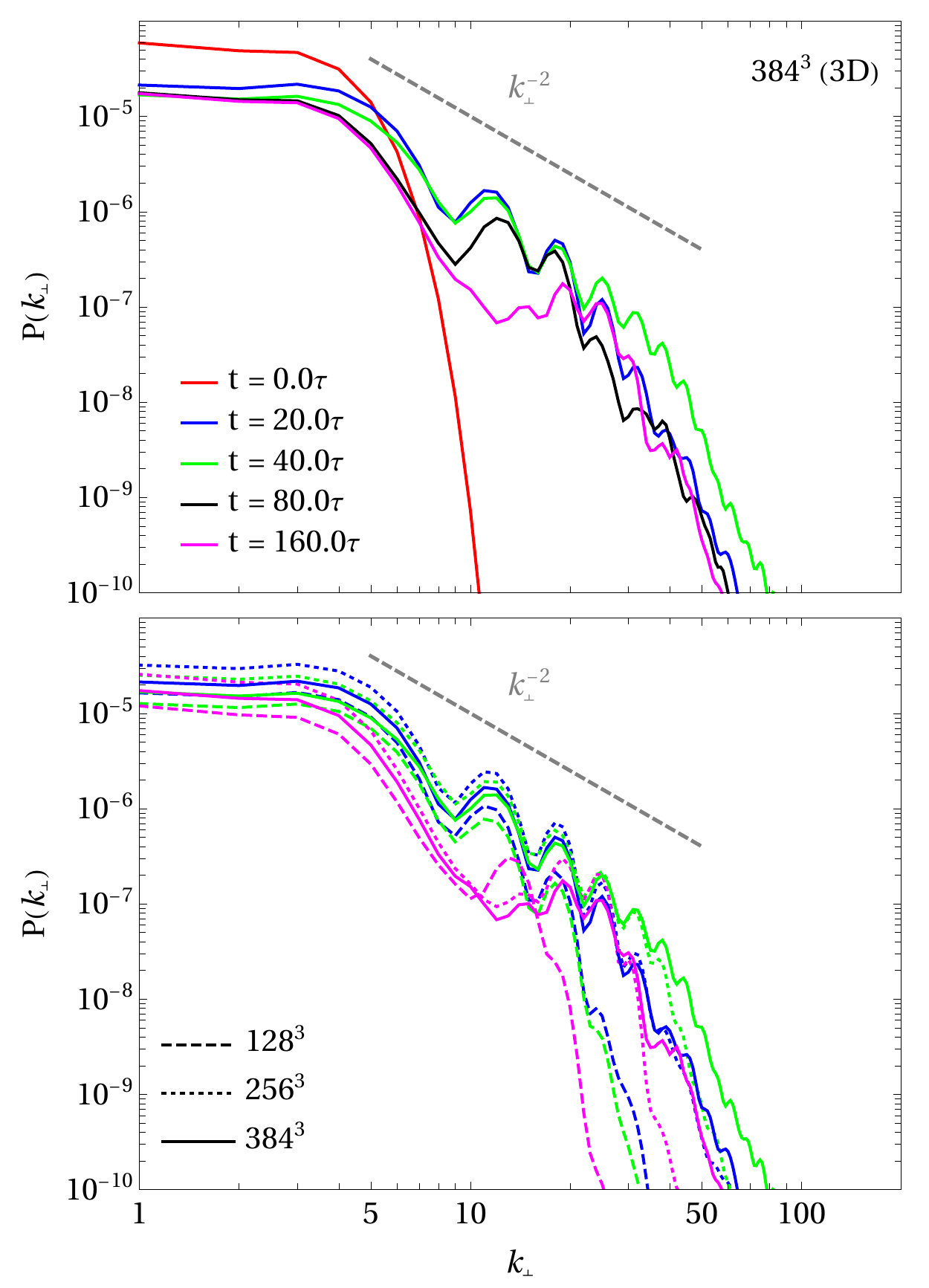}
	\caption{Spectrum evolution for the 3D simulation of Alfv\'{e}n interactions initialized 
		according to Eq.~\eqref{eq:Bfield-collisions3D}. We show the spectral energy 
		distribution for wavenumbers $k_\perp$ (perpendicular to the guide field) in analogy 
		to Fig.~2, \citet{Li2019}. \textit{Top:} Spectral energy distribution at different 
		times for the resolution $384^3$. \textit{Bottom:} Spectral energy distribution for 
		selected times and wavenumbers (color code as in \textit{top} panel) and different 
		resolutions, indicated by dashed ($128^3$), dotted ($256^3)$ and solid ($384^3$) lines. 
		For wavenumbers $k_\perp\lesssim 40$, convergence is reached for the shown high-resolution 
		cases.}
	\label{fig:Alfven3DSpectrum}
\end{figure}

We perform a test \citep[explored in extensive detail and high resolution by][]{Li2019} of 
the interaction between colliding Alfv\'{e}n modes in suitably chosen 2D and 3D computational 
boxes. In this section, we intend to reproduce the most basic results of energy cascades 
from Alfv\'{e}n wave interactions to show our GRFFE scheme's ability to explore such 
phenomena in further detail in the future.

On the respective numerical meshes, one initializes counter-propagating Gaussian 2D or 3D 
wave packets traveling along a uniform guide field $B^y=B_0$. Periodic boundary conditions 
facilitate the recurring superpositions and interaction of the wave packets, eventually 
triggering an energy cascade of rapid dissipation. The 3D Gaussian wave packets are initialized as 
\begin{align}
	\mathbf{B}=B_0\mathbf{\hat{y}}+B_0\nabla\times\left(\phi\mathbf{\hat{y}}\right),
	\label{eq:Bfield-collisions3D}
\end{align}
where $\mathbf{\hat{y}}$ is the unit vector in the $y$-direction and the scalar field
\begin{align}
	\phi(\mathbf{r})=\xi\, l\sum_{i=1,2}\exp\left(-\frac{\left|\mathbf{r}-\mathbf{r}_i\right|^2}{l^2}\right).\label{eq:AlfvenPackage3D}
\end{align}
In this section, $\xi$ denotes the perturbation strength, $l$ the width of the wave packet, 
with centers are located at $\mathbf{r}_1$ and $\mathbf{r}_2$. We follow \citet{Li2019} 
in choosing $\xi=0.5$, $l=0.1$, $\mathbf{r}_1=\left(0.5,0.25,0.5\right)$ and $\mathbf{r}_2=\left(0.5,0.75,0.5\right)$ 
for the 3D wave packets. With this setup, the field perturbation is purely azimuthal with 
respect to the $y$-axis. On a reduced 2D mesh, we initialize Gaussian wave packets with 
magnetic fields
\begin{align}
	\mathbf{B}=B_0\mathbf{\hat{y}}+B_z\mathbf{\hat{z}},
	\label{eq:Bfield-collisions2D}
\end{align}
where $\mathbf{\hat{z}}$ is the unit vector in the $z$-direction and
\begin{align}
	B_{z}=B_0\xi\sum_{i=1,2}\exp\left(-\frac{\left|\mathbf{r}-\mathbf{r}_i\right|^2}{l^2}\right).\label{eq:AlfvenPackage2D}
\end{align}
We employ $\xi=0.4$, $l=0.1$, $\mathbf{r}_1=\left(0.5,0.25\right)$ and 
$\mathbf{r}_2=\left(0.5,0.75\right)$ for the 2D setup. The motion of the wave packets is 
induced with a drift speed $\mathbf{D}\times\mathbf{B}/B^2$, that results form an initial electric field
\begin{align}
	\mathbf{D}=\pm \mathbf{\hat{y}}\times \mathbf{B},
	\label{eq:Dfield-collisions}
\end{align}
with opposite signs for each Gaussian wave packet. 

After the initialization of the electromagnetic 
fields, the bounding box of length $L=1$ and periodic boundaries are left to evolve for 
$200\tau$ ($f_\textsc{cfl}=0.2$). $\tau=0.5$ is the interval between two subsequent collisions of the wave packets, and $t/\tau$ the number of collisions. For these tests, we employ MP7 
reconstruction (with a fourth-order discretization of $\mathbf{j}_\parallel$). Following \citet{Li2019}, 
we employ the free-energy $U$ as the measure of total electromagnetic energy of the system 
$e_{\rm tot}$ under removal of the background magnetic field $B_0$:
\begin{align}
	U=e_{\rm tot}-\frac{1}{2}\int\text{d}V B_0^2.
\end{align}

Fig.~\ref{fig:AlfvenDissipation} shows the free-energy for the collision of the wave packets defined in Eqs.~(\ref{eq:AlfvenPackage3D}) 
and~(\ref{eq:AlfvenPackage2D}). The wave packets are spherical and - due to their curvature - 
prone to redistribute energy across wave modes and rapid dissipation in cascade-like processes 
\citep[e.g.,][]{Howes2013,Nielson2013}. Such processes are likely to be found along curved 
guide-fields, for example in magnetar magnetospheres. Collisions excite waves of higher 
frequency than initially setup and eventually trigger the rapid decay of the wave free-energy. 
In order to make a more quantitative comparison of the results, we also compute the spectral 
distribution of free-energy according to components of the propagation wave vector which 
are parallel ($k_{||}$) and perpendicular ($k_\perp$) to the guide field, following the 
same prescription as in \citet{Li2019} (see Figs.~\ref{fig:Alfven3DSpectrum} and 
\ref{fig:Alfven2DSpectrum}). {We compare our 2D and 3D results with the reference work 
	of \citet{Li2019} below.
	
	\subsubsection{2D models}
	Our GRFFE code is able to reproduce the dissipation patterns of free electromagnetic energy 
	presented in Fig.~5 of \citet{Li2019} for the 2D setup of Eq.~(\ref{eq:AlfvenPackage2D}). 
	We note that our tests correspond to the lowest three mesh resolutions employed by 
	\citet{Li2019}.} In the top panel of Fig.~\ref{fig:AlfvenDissipation}, we display the 
evolution of the inverted free-energy ($U_0/U$) along with the slopes for their decay in 
2D. Contrasting the findings by \citet{Li2019}, the decay of $U$ initially proceeds at the 
same rate ($s\approx 0.35$) for all of the analyzed 2D models, independent of the chosen 
resolution. Only at later times, the slopes deviate and (roughly) approach the numerical 
values given in Fig.~7 of \citet{Li2019}. The redistribution of spectral energy happens 
at all times from the smaller to the larger values of $k_\perp$ (Fig.~\ref{fig:Alfven2DSpectrum}), 
and an approximate $k_\perp^{-2}$ spectral dependence is observed for intermediate values 
$10\lesssim k_\perp\lesssim 70$. The maximum value of $k_\perp$, $k_{\perp,{\rm max}}$, 
is resolution dependent (the finer the resolution, the larger $k_{\perp,{\rm max}}$). Hence, 
there is no evidence of spectral convergence in 2D \citep[in agreement with][see 
Fig.~\ref{fig:Alfven2DSpectrum} lower panel]{Li2019}. For the respective resolution, the 
decay of $U$ begins later than in \citet{Li2019}, suggesting that the numerical diffusivity 
in our method (combining MP7 reconstruction, a fourth-order accurate Runge-Kutta time-integrator 
and no additional driving terms in $\mathbf{j}_{||}$) is smaller (compared to a fifth-order 
spatial WENO reconstruction, a third order accurate Runge-Kutta time-integrator, and an 
extra dissipation term in $\mathbf{j}_{||}$). As a result, our models with a grid of $512^2$ 
zones display an evolution of $U$ trending roughly in between the curves corresponding to 
$\sim 1024^2$ and $2048^2$ in \citet{Li2019}. A more thorough characterization of 
the (numerical) dissipation of our algorithm is considered in Paper II.

\subsubsection{3D models}

For tests in 3D, we are limited to the two lower resolutions of the corresponding model in \citet{Li2019} to 
stay within the computational costs that are reasonable for a test setup. In spite of the 
reduced resolution we employ compared to the literature, we find a remarkable agreement with 
the reference results. For instance, we observe a faster onset of the energy cascade than 
in 2D, and the same asymptotic value of the free-energy \citep[$U/U_0\approx 0.4$ for the 
best resolved 3D model; Fig.~\ref{fig:AlfvenDissipation} lower panel; cf. with Fig.~2 
of][]{Li2019}. We also find comparable (although slightly shallower) asymptotic slopes for 
the free-energy decay; the slope found by \citet{Li2019} is indicated in 
Fig.~\ref{fig:AlfvenDissipation} with a dashed gray line. Another important key feature 
obtained by \citet{Li2019} is the existence of a characteristic time, $t_{\rm onset}$ after 
which dissipation commences in 3D models. After this time there is a relatively sharp drop 
of the free-energy, which tends to level off for sufficiently long times. However, the 
feature that unanimously sets $t_{\rm onset}$ \citep[according to the definition of][]{Li2019} 
is found in the evolution of the spectral energy distribution. Before $t_{\rm onset}$, the 
colliding Gaussian packets shuffle energy toward smaller scales (larger values of the wave 
number $k$) until a maximum $k=k_{\rm max}$ is reached. However, after $t=t_{\rm onset}$ 
there exists a redistribution of energy from the smaller to the larger scales, which 
manifest itself as an increasing spectral power at small values of $k$. For \citet{Li2019}, 
$t_{\rm onset}\approx 24\tau$. The conclusions of \citet{Li2019} are based upon 3D models 
with finer resolution than the ones employed in this section (e.g., models with $512^3$ 
and $768^3$ zones). With a smaller resolution, we also find a time after which there is 
a steep drop of the free-energy, which begins at $t\gtrsim 30\tau$ (Fig.~\ref{fig:Alfven3DSpectrum} 
top panel). Nevertheless, we do not clearly see an increase in the spectral power at small 
values of $k_\perp$, but we observe a decrease of power for $k_\perp\gtrsim 30$ for a 
time $30\tau \lesssim t_{\rm onset}\lesssim 40\tau$. At about this time, the fast drop off 
in $U$ takes place. Since we have evolved our models longer in time, we note that at 
$t=160\tau$ there is a decrease of spectral energy at intermediate values 
$10\lesssim k_\perp\lesssim 25$ (Fig.~\ref{fig:Alfven3DSpectrum} top panel). The spectral 
evolution of the modes parallel to the guide field proceeds qualitatively as in \citet{Li2019}, 
with the only difference that we observe some (small) excess of power in the range 
$10\lesssim k_{||}\lesssim 50$. As in 2D, most of the (small) quantitative differences 
observed in the comparison with \citet{Li2019} results can be attributed to the different 
order of accuracy of our codes and the different terms entering in the current parallel to 
the magnetic field.

\section{Astrophysically motivated tests}
\label{sec:Astrophysical_Tests}

\subsection{Magnetar magnetospheres}
\label{sec:Magnetar_Magnetospheres}

\subsubsection{Grid aligned magnetar magnetospheres}
\label{sec:MagnetarsAligned}

\begin{figure}
	\centering
	\includegraphics[width=0.48\textwidth]{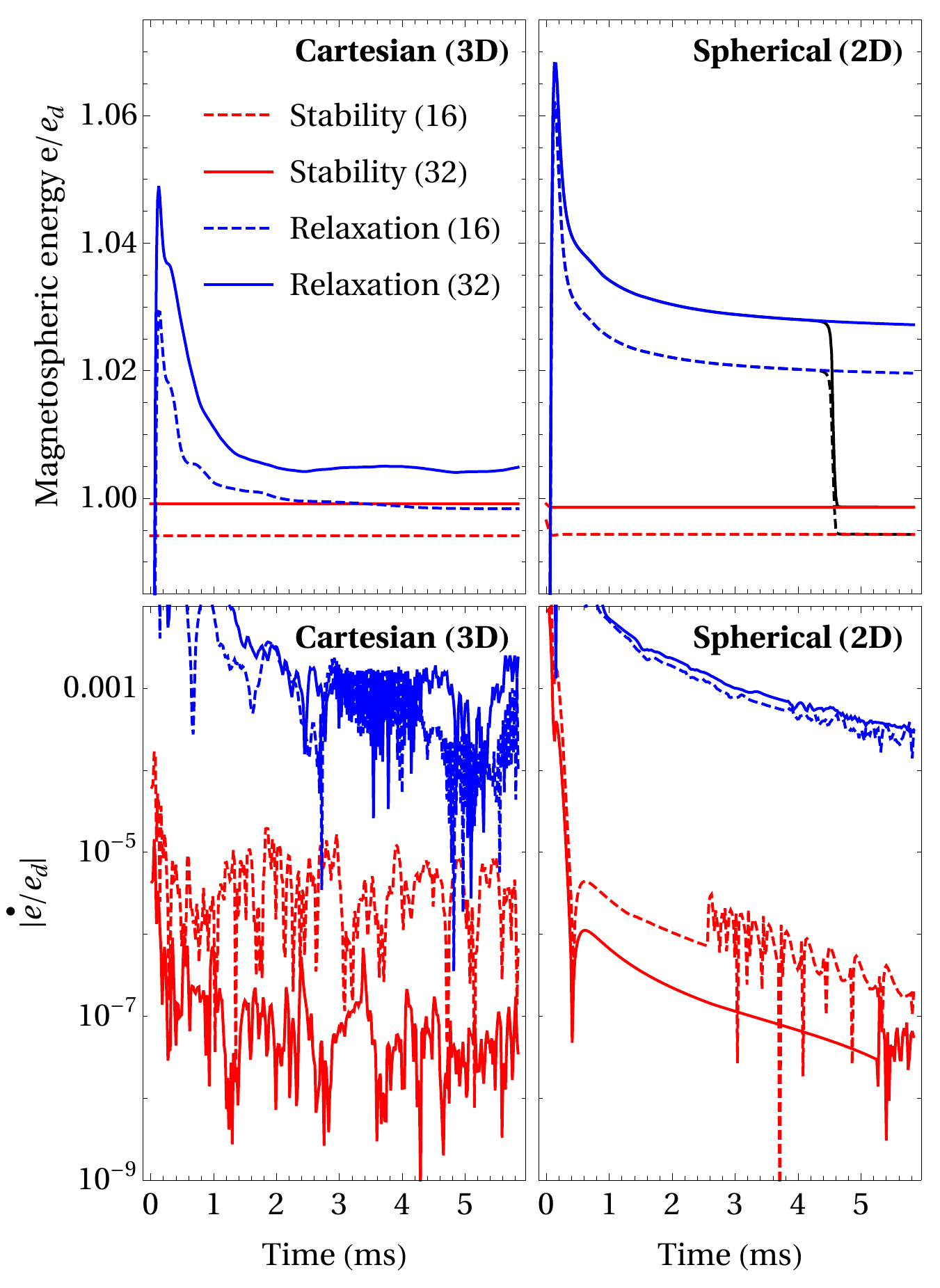}
	\caption{Stability and relaxation test of magnetar magnetospheres endowed with an 
		analytic dipole field structure for different resolutions. 
		Upper panels show the total magnetospheric energy normalized
		to the energy of the corresponding dipole. The lower panels show the time derivative of the 
		energy normalized to the dipole. The resolutions of
		16 and 32 points per stellar 
		radius in a 3D Cartesian \textsc{Carpet} grid (\textit{left}) correspond to the setup 
		in \citet{Mahlmann2019}. Axially symmetric simulations in spherical coordinates 
		(\textit{right}) are set up with the indicated resolution at the stellar surface. 
		Black (solid and dashed) lines in the upper-right panel correspond to a simulation on a 
		smaller domain, extending up to $r=935.26M_\odot-6\Delta r$, in which the pulses of the initial relaxation 
		have sufficient time to leave the domain.}
	\label{fig:MagnetarEnergy}
\end{figure}

The magnetospheres of magnetars are a well suited laboratory for numerical methods dealing 
with force-free plasma, and we have explored their dynamics in \citet{Mahlmann2019}. Prior 
to those numerical simulations of a potentially very dynamic scenario, we performed numerical 
tests to assess the ability of our GRFFE code to maintain the structural stability of a 
magnetosphere around a spherical, nonrotating neutron star with a dipolar magnetic field. 
We use a spherical mask to cut out the neutron star interior in order to avoid dealing with 
the equation of state of nuclear matter, the different phases of matter that may occur 
inside of the neutron star, and the solid structure of the stellar crust. This is achieved 
by setting an internal boundary in a 3D Cartesian grid (i.e., stair-stepping along the 
spherical boundary mask) inside of which the evolution is frozen (see below). We note that 
3D Cartesian coordinates are neither adapted to the spherical shape of the neutron star nor 
the axial symmetry of the magnetospheric dipole. Therefore, we expect significantly improved 
results when employing GRFFE in a spherical coordinate system. We compare the results of 
simulations in 3D Cartesian coordinates to axisymmetric (2D) tests in spherical coordinates, with the magnetic axis aligned to the symmetry axis of the initial data.

It is straightforward to specify the employed initial data in spherical coordinates 
$\left(r,\theta,\phi\right)$ and, subsequently, map it to the computational grid. The analytically 
derived equilibrium dipolar magnetic field in (the coordinate basis of) spherical coordinates 
reads:
\begin{align}
	\begin{split}
		\mathbf{B}=\left(\frac{2\cos\theta}{r^3},\frac{\sin\theta}{r^4},0\right),\qquad\qquad\mathbf{D}=\left(0,0,0\right).\label{eq:magnetar_initial}
	\end{split}
\end{align}
The Cartesian simulations are conducted in a 3D box with dimensions 
$\left[4741.12 M_\odot\times4741.12 M_\odot\times4741.12 M_\odot\right]$ with a grid spacing 
of $\Delta_{x,y,z}=74.08M_\odot$ on the coarsest grid level. For the chosen magnetar model 
of radius $R_*=9.26M_\odot$, this corresponds to a $\left[512R_*\times 512R_*\times 512R_*\right]$ 
box with a grid spacing of $\Delta_{x,y,z}=8R_*$. For the low-resolution and high-resolution tests, we 
employed seven and eight additional levels of mesh refinement, respectively, each of which increase the resolution by a factor of two and encompass the central object. This means that the 
finest resolution of our models (close to the magnetar surface) are 
$\Delta_{x,y,z}^{\rm min}=0.0625\times R_* = 0.5787M_\odot$ and $\Delta_{x,y,z}^{\rm min}=0.03125\times R_*= 0.2894M_\odot$ 
for the low and high-resolution models. This corresponds to 16 and 32 points per $R_*$, respectively. The spherical simulations are conducted in axial symmetry enclosing the volume $\left[9.26M_\odot,2555.76M_\odot-6\Delta r\right]\times\left[0,\pi\right]$. In order to issue a resolution that is comparable to the Cartesian setup, we employ 
$\Delta r\in\left[R_*/16,R_*/32\right]$ and $\Delta\theta\in\left[\pi/50,\pi/100\right]$. 
The setup is evolved for a period of $t=1185.28M_\odot\simeq 5.84\,$ms. We provide extensive 
details on the internal boundary conditions (frozen electromagnetic fields but balanced 
radial current) in \citet{Mahlmann2019}. For this section we employ $f_{\rm CFL}=0.2$, the 
MP7 reconstruction and a fourth-order accurate discretization of $\mathbf{j}_{||}$. We note that in this test we assume a flat spacetime. More general tests including a general relativistic, dynamically evolving background have been considered, for example, in \citet{Ruiz_2014PhRvD..89h4045}. However, performing such tests here is beyond the scope of this paper.

The stability test initializes the dipole structure throughout the entire computational 
domain and tracks the stability during a dynamical evolution. The relaxation test is even 
more challenging than the stability test since it requires the time evolution toward the 
physical topology set by the boundary conditions. Precisely, in a relaxation test we fix 
the dipolar structure inside of the star, but fill the magnetosphere with a purely radial 
field at the start of the simulation. In order to track the changes in the magnetosphere, we define its total energy as the volume integral
\begin{align}
	e=\frac{1}{8\pi}\int \left(\mathbf{D}^2 + \mathbf{B}^2\right) \sqrt{-g}\, d^{3}x .
\end{align}
Once initialized, the energy of the dipole magnetosphere
\citep[cf.][]{Mahlmann2019} is well conserved (stability) or else gradually approaches the 
dipole energy (relaxation) once all initially introduced perturbations leave the domain. 
Figure~\ref{fig:MagnetarEnergy} shows stability and relaxation tests of the dipole magnetosphere for different resolutions in both, Cartesian (3D) and spherical (2D, axial 
symmetry) coordinates.

The initial spike of the relaxation model can be attributed to a surge of electromagnetic 
energy during a rapid rearrangement in the early phase. The excited energy pulses propagate 
as plasma waves through the magnetosphere. A part of these pulses is confined to closed 
field lines in the vicinity of the central object. The rest of this energy propagates 
outward through the domain. As the dissipation of electromagnetic energy in 
collisions of force-free waves strongly depends on the employed resolution (see 
Sec.~\ref{sec:FFE_Interaction}, and Paper II) and grid geometry, the confined energy 
pulses remain within the domain longer for higher resolution and spherical coordinates. 
This is visible best in the relaxation test presented in Fig.~\ref{fig:MagnetarEnergy}. 
For different resolutions, the asymptotic energy differs by $<1\%$. Complete relaxation of 
this energy will require longer simulation times, such that waves emerging from the initial 
relaxation can leave the domain. Also, accurate treatment of the interior boundary will be necessary so that plasma 
waves in the region of closed field lines dissipate physically. The first of these effects, 
associated with the total simulated time, can be partly addressed by considering a computational 
domain with a reduced outer radial boundary. In the top right panel of Fig.~\ref{fig:MagnetarEnergy}, 
we show the time evolution in a reduced computational domain with black lines. The abrupt 
drop-off the magnetospheric energy is due to the desertion of the initial perturbation through 
the outer radial boundary. Remarkably, the energy level to which these models evolve is the 
same as the corresponding stability tests with the corresponding numerical resolution.

\subsubsection{Tilted magnetar magnetospheres}

\begin{figure*}
	\centering
	\includegraphics[width=0.99\textwidth]{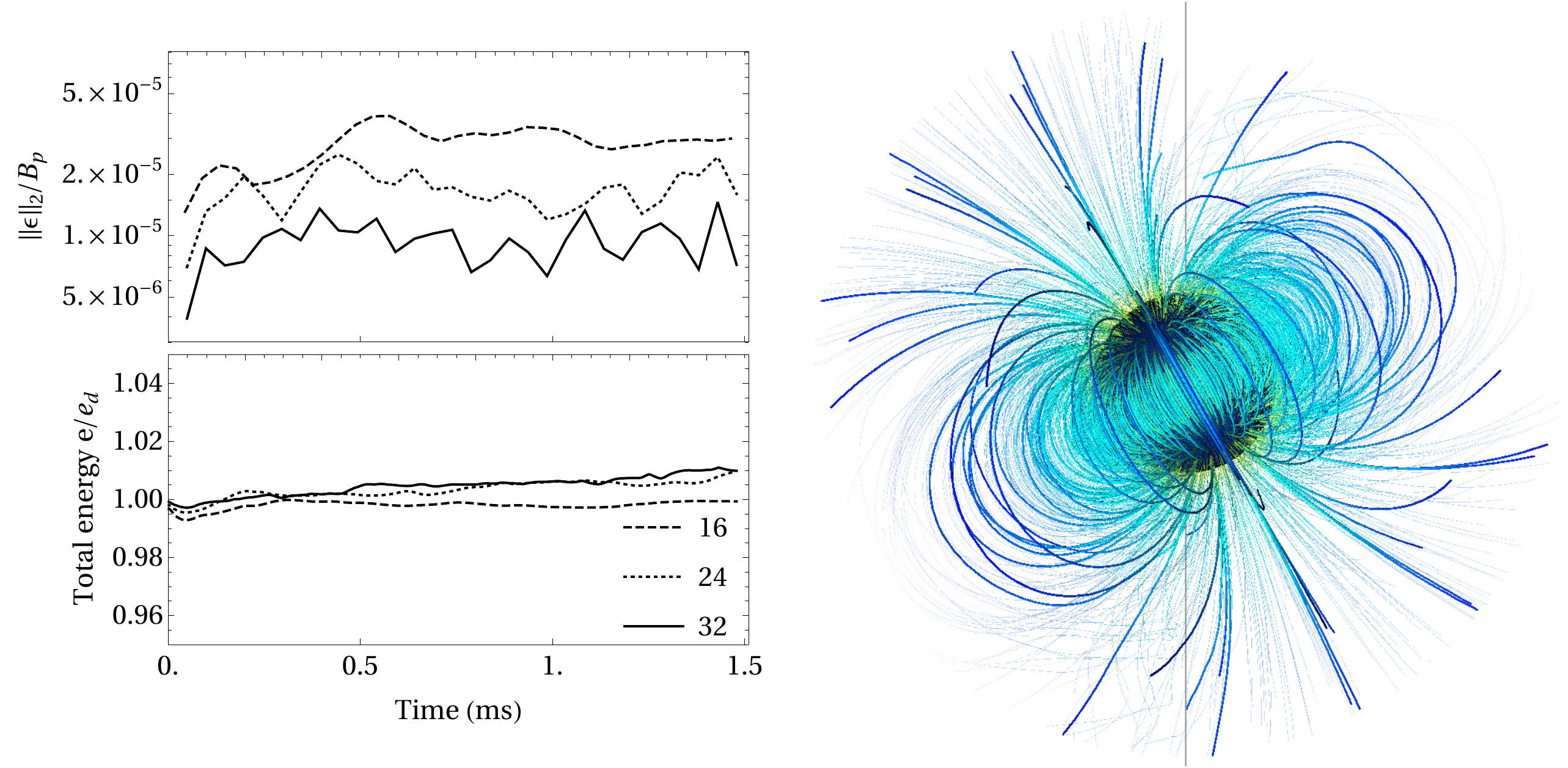}
	\caption{Stability test of magnetar magnetospheres endowed with a tilted (30 degrees) 
		analytic dipole field structure for different resolutions. \textit{Left:} Evolution of 
		the $l^2$-norm of the error with respect to the analytical (initial) configuration 
		normalized to the magnetic field strength $B_p$ at the magnetar pole (\textit{top}), 
		and of the total magnetospheric energy content (\textit{bottom}). \textit{Right:} 3D 
		impression of the final state ($t\approx 1.5\text{ms}$) of the intermediate resolution 
		simulation ($\Delta r=R_*/24$). We display the magnetic field lines colorized by their norm.}
	\label{fig:TiltedMagnetar}
\end{figure*}

In this section, we explore the full 3D capabilities of our newly developed code in 
spherical coordinates $\left(r,\theta,\phi\right)$ by considering the stability test from 
the previous section, but tilting the magnetic dipole axis by an angle $\alpha$ with respect to 
the spherical polar axis along $\theta=0$. For this, we carry out the transformation 
$\mathbf{B}_{\rm t} = \mathbf{R}_{-\alpha}\mathbf{B}\left(\mathbf{\tilde{r}}\right)$, 
where $\mathbf{\tilde{r}}=\mathbf{R}_{\alpha}\mathbf{r}$, $\mathbf{B}$ corresponds to the initial data given in 
Eq.~(\ref{eq:magnetar_initial}), and
\begin{align}
	\begin{split}
		\mathbf{R}_{-\alpha}&=\left[\begin{array}{ccc}
			1 & 0 & 0\\ 
			0 & \frac{\cos\alpha\sin\theta-\sin\alpha\cos\theta\sin\phi}{\chi} & \sin\alpha\cos\phi\\
			0 & -\frac{\sin\alpha\csc\theta\cos\phi}{\chi} & \cos\alpha-\sin\alpha\cot\theta\sin\phi\\ 
		\end{array}\right],\\
		\chi&=\sqrt{\sin^2\theta\cos^2\phi+\left(\sin\alpha\cos\theta-\cos\alpha\sin\theta\sin\phi\right)^2}.
	\end{split}
\end{align}
We chose $\alpha=30^\circ$ for simulations that are exclusively conducted in a spherical 
domain with dimensions $\left[9.26M_\odot,611.16M_\odot-6\Delta r\right]\times\left[0,\pi\right]\times\left[0,2\pi\right]$. 
In order to use a resolution that is comparable to Sect.~\ref{sec:MagnetarsAligned}, we 
employ $\Delta r\in\left[R_*/16,\,R_*/24,\,R_*/32\right]$ and 
$\Delta\phi=\Delta\theta\in\left[\pi/50,\, \pi/75,\, \pi/100\right]$. The setup was evolved 
for a period of $t=300M_\odot\simeq 1.48\,$ms with $f_{\rm CFL}=0.25$ using MP7 spatial 
reconstruction and the default fourth-order discretization of $\mathbf{j}_{||}$.

Besides the measurement of the total energy in the magnetosphere, in order to quantify the 
deviation of the numerical solution $\mathbf{B}$ from the analytical (initial) configuration 
$\mathbf{B}_0$, we define the $l^2$ error norm of the magnetic field as
\begin{align}
	\lVert\epsilon\rVert_2=\frac{1}{N}\sqrt{\sum_{i,\,j,\,k}\left[\mathbf{B}(r_i,\theta_j,\phi_k)-\mathbf{B}_0(r_i,\theta_j,\phi_k)\right]^2},\label{eq:magnetarerror}
\end{align}
where indices $i$, $j$, and $k$ extend over all the computational cells. Fig.~\ref{fig:TiltedMagnetar} 
shows the evolution in time of the error norm in Eq.~(\ref{eq:magnetarerror}) normalized to the 
magnetic field strength at the pole of the neutron star, $B_p$. Increasing the mesh resolution 
by a factor of two (i.e., from $\Delta r=R_*/16$ to $R_*/32$) reduces the error by roughly the 
same factor. At the same time, the total magnetospheric energy slightly increases throughout 
the simulation time ($\lesssim 1\%$). Such a (continuous) increase in energy does not occur 
in the aligned (axially symmetric) setups, as we show in Fig.~\ref{fig:MagnetarEnergy}. 
However, the global structure of the 3D tilted dipole is conserved throughout the simulation 
(Fig.~\ref{fig:TiltedMagnetar}), with only slight kinks arising around the polar axis of the spherical coordinates.

Solving hyperbolic PDEs in spherical coordinates suffer from very small timesteps due to the 
fact that cell volumes are not constant in space and get smaller as the polar axis or the origin of
the coordinate system are approached. The timestep is proportional to 
$r \times \sin (\Delta \theta/2) \times \Delta \phi$ and hence becomes prohibitively small 
for full 3D simulations with high angular resolutions, this is the reason for the shorter simulation 
time compared to the axially symmetric models in Sect.~\ref{sec:MagnetarsAligned}). 
In order to mitigate this limitation for simulations in spherical coordinates, additional 
grid coarsening or filtering approaches will be necessary when considering 
computationally feasible long-term 3D evolution \citep[cf.][]{Obergaulinger_2017MNRAS.469L..43,Mewes2020,Zlochower2020,Aloy_2020arXiv200803779,Obergaulinger_2020MNRAS.492.4613}. The impact of 
the tilt across the axis on the magnetospheric energy (as observed in Fig.~\ref{fig:TiltedMagnetar}) 
may be further diminished by such techniques.

In this test we employ the full 3D capacities of our GRFFE method in spherical coordinates. 
The tilted magnetar magnetospheres maintain their topology stable for $\sim 32$ light-crossing 
times of the central object. Extrapolating the deviation of the magnetospheric energy to 
longer times is uncertain due to the non-monotonic evolution. However, we foresee that stationary tilted magnetospheres may be maintained approximately stable sufficiently for more than a few hundred light-crossing times of the central object. These longer periods of evolution may suffice to address numerically 
dynamical phenomenae in the magnetosphere \citep[e.g.][]{Carrasco_etal_2019_10.1093/mnrasl/slz016, Mahlmann2019}. 
Besides, our results show that increasing the resolution decreases the global error, hence, 
if needed, finer grids may be used to address longer evolutionary times.

\subsection{The force-free aligned Rotator}
\label{sec:ffaligned}
\begin{figure*}
	\centering
	\includegraphics[width=0.95\textwidth]{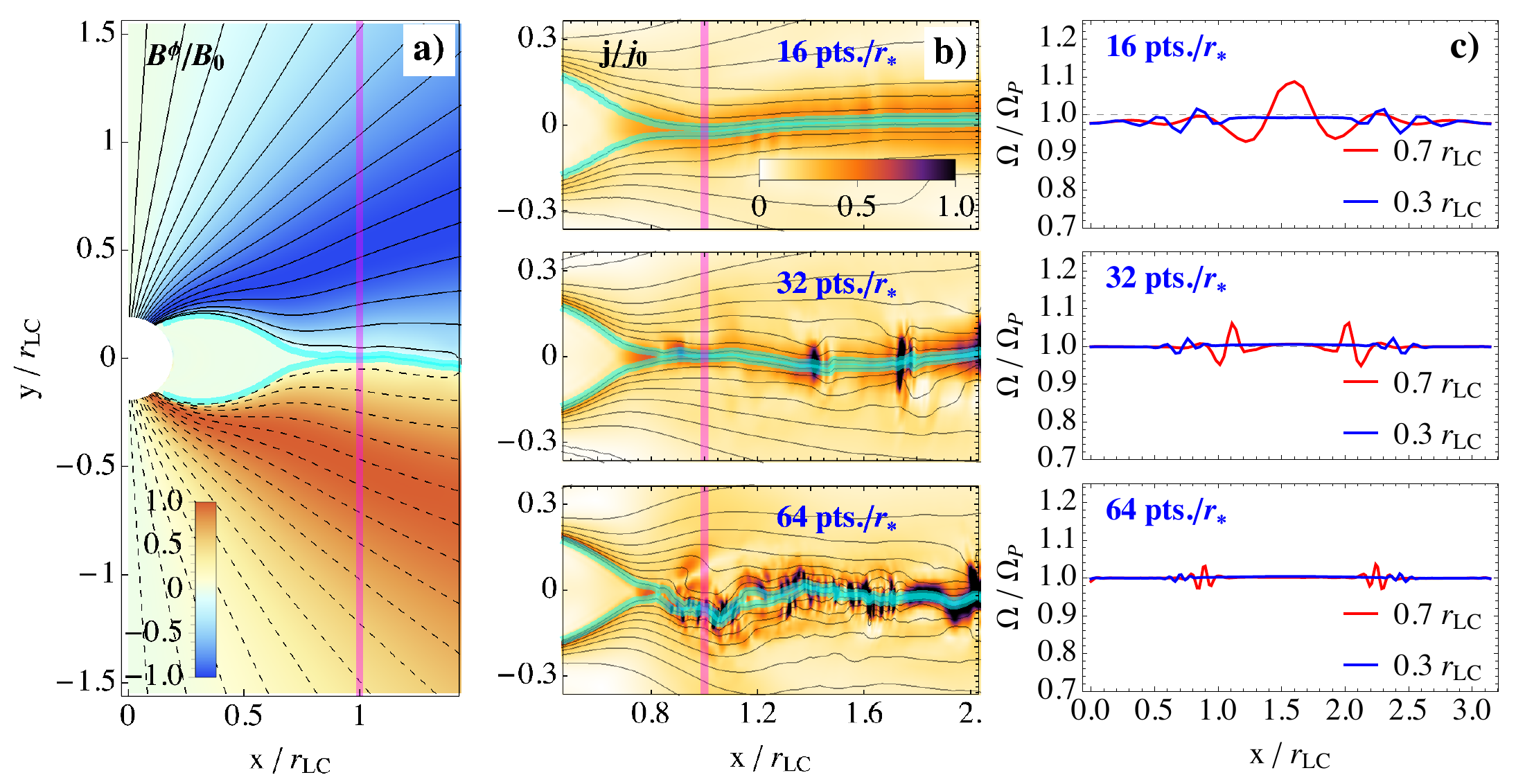}
	\caption{Force-free aligned rotator test in spherical ideal FFE for different resolutions. \textit{Left} panel (a): Global structure of the magnetosphere for intermediate resolution (32 pts./$r_*$). The poloidal fieldlines are indicated by black solid and dashed lines, depending on the magnetic field direction. Blue and red colors represent direction and magnitude of the toroidal magnetic field (normalized to its maximum value). \textit{Middle} panel (b): Zoom on the Y-point and equatorial current sheet (alternating magnetic field highlighted by cyan field lines) for different resolutions. The color scale represents the current density (normalized to the current density of the low-resolution data). \textit{Left} panel (c): Field line angular velocity measured at different radii. }
	\label{fig:AlignedRotator}
\end{figure*}
An astrophysically motivated test that deliberately breaches the limits of FFE is the aligned rotator test. It sets up an initially dipolar magnetic field on a star rotating with angular velocity $\Omega_p$. Equilibrium solutions of the axisymmetric pulsar magnetospheres, as solutions to the so-called pulsar equation, have been studied, for example, in \citet{Contopoulos1999} and \citet{Timokhin2006}. Time-dependent solutions to the aligned rotator magnetosphere have been presented; for instance, by \citet[][resistive MHD and resistive FFE]{Komissarov3006}, \citet[][resistive FFE]{Spitkovsky2006}, \citet[][dissipative FFE]{Paschalidis2013}, \citet[][MHD]{Tchekhovskoy2013}, \citet[][dissipative FFE]{Etienne2017}. All of these schemes make sure that the equatorial current sheet can be resolved in FFE by either adding additional dissipation or a resistivity model to FFE, or by combining FFE and MHD in order to capture such genuinely non-force-free regions.

In this section, we present results of the aligned rotator test in ideal FFE conducted on a spherical, axisymmetric mesh. We demonstrate that our code correctly reproduces the features of the force-free magnetosphere inside of the pulsar light cylinder $r_{\rm LC}=c/\Omega_p$. We transparently point out the consequences of the emerging equatorial current sheet in ideal FFE, to which we ultimately dedicate paper II of this series \citep{Mahlmann2020c}. This test sets up the initial data of Eq.~(\ref{eq:magnetar_initial}) in a domain with dimensions $\left[9.26M_\odot,6954.26M_\odot-6\Delta r\right]\times\left[0,\pi\right]$. 
We employ $\Delta r\in\left[R_*/16,\,R_*/32,\,R_*/64\right]$ and 
$\Delta\phi=\Delta\theta\in\left[\pi/50,\, \pi/100,\, \pi/200\right]$. The setup is evolved 
for a period of $t=3150M_\odot$ (10 revolutions of the central object) with $f_{\rm CFL}=0.25$, using MP5 spatial 
reconstruction, and the default fourth-order discretization of $\mathbf{j}_{||}$. We implement the pulsar boundary condition by following closely the interior boundary conditions described by \citet{Parfrey2012} and setting $\Omega_p=0.02$.

Fig.~\ref{fig:AlignedRotator} visualizes the results of this resolution study. Our code correctly reproduces the expected magnetospheric features for $r<r_{\rm LC}$. A twist-free, closed region emerges for $r\lesssim 0.8 r_{\rm LC}$. There, field lines co-rotate with the central object in good accordance with the imposed pulsar boundary conditions. Slight deviations from perfect co-rotation (of a few percent, decreasing with higher resolution) occur at the sheath of the twist-free region. The location of the Y-point $r\sim 0.9 r_{\rm LC}$ is stable for the employed resolution. As reasoned in \citet{Spitkovsky2006}, reconnection is the key factor setting the location of the Y-point, $r_{\rm Y}$ relative to $r_{\rm LC}$, as well as the rate at which $r_{\rm Y}$ approaches $r_{\rm LC}$ from its initial location. The effects of a finite (numerical) resistivity induce an oscillatory behavior of the Y-point as well as \citep[in the case of][]{Spitkovsky2006} the ejection of (small) plasmoids. Episodically, non-axisymmetric, small-scale structures emerge out of the equatorial current sheet. The scale and temporal wavelength of the variations in the current sheet decrease with increasing resolution  (\textit{middle} panels Fig.~\ref{fig:AlignedRotator}b).  The unsteadiness of the current sheet is a direct consequence of the very low numerical resistivity of our high-order schemes applied to \textit{ideal} FFE. They significantly reduce the numerically driven reconnection \citep[cf.][]{Mahlmann2020c}. The morphology of the current sheet results from localized (numerical) reconnection events that happen as a result of the enforcement of the magnetic dominance condition ($\mathbf{B}^2-\mathbf{E}^2\ge0$) at low latitudes, where the magnetic field strength passes through zero as it crosses the equator.  In the absence of a more elaborated physical description of the layer where the plasma pressure becomes larger than the magnetic pressure (e.g., using MHD), recipes to handle the numerical resistivity completely determine the spatial smoothness and time variability of the equatorial current sheet.  

The Poynting flux of outflowing energy agrees well with the expectations collected in the literature, $L\approx\left(1.0 + 0.1\right)L_0$ (for the highest-resolution run), where $L_0=\mu^2\Omega_p^4/c^3$ for a magnetic moment $\mu$. The value of $L$ larger than $L_0$ is directly related to the fact that $r_{\rm Y}<r_{\rm LC}$ during the computed time. On longer time-scales (computationally prohibitive for a single code test) the luminosity decreases as the closed zone expands \citep{Spitkovsky2006}, something that generically happens for all resolutions considered here (but at significantly different rates; faster at lower resolution). For the lowest employed resolution (16 pts./$r_*$), we find that the luminosity gradually decreases by $\sim10\%$ for $r<5 r_{\rm LC}$. For higher resolutions, the luminosity level is stable but showing variations due to small structures emerging in the current sheet (\textit{middle} panels Fig.~\ref{fig:AlignedRotator}b). Our rigid confinement to only enforcing the FFE conditions (Sect.~\ref{sec:FFConstraints}) without adding additional dissipation mechanisms or resistivity models is the main difference to the aforementioned array of literature available for this particular setup.

The results presented in this section are sensitive to both the boundary conditions and the modeling of the equatorial current sheet. As general features of the force-free magnetosphere ($r<r_{\rm LC}$) are reproduced as expected (twist-free region, field line angular velocity and luminosity), our code passes the ideal FFE aligned rotator test. We stress, however, that a physical modeling of the equatorial current sheet beyond the Y-point requires suitable techniques to develop such genuinely resistive regions in time. In \citep{Mahlmann2020c}, we outline the difficulties and some possible remedies of this task. We are convinced that a physical modeling of equilibrium solutions containing both force-free regions and current sheets either requires exceptional fine-tuning of the employed resistivity models, or a mixing of different plasma regimes \citep[such as FFE and resistive MHD; see, e.g.,][]{Ruiz_2014PhRvD..89h4045}. In any case, such modeling is beyond the scope of a test for ideal FFE.

\subsection{Black hole magnetospheres}
\label{sec:BH_Magnetospheres}

\subsubsection{Black hole monopole tests}
\begin{figure*}
	\centering
	\includegraphics[width=0.98\textwidth]{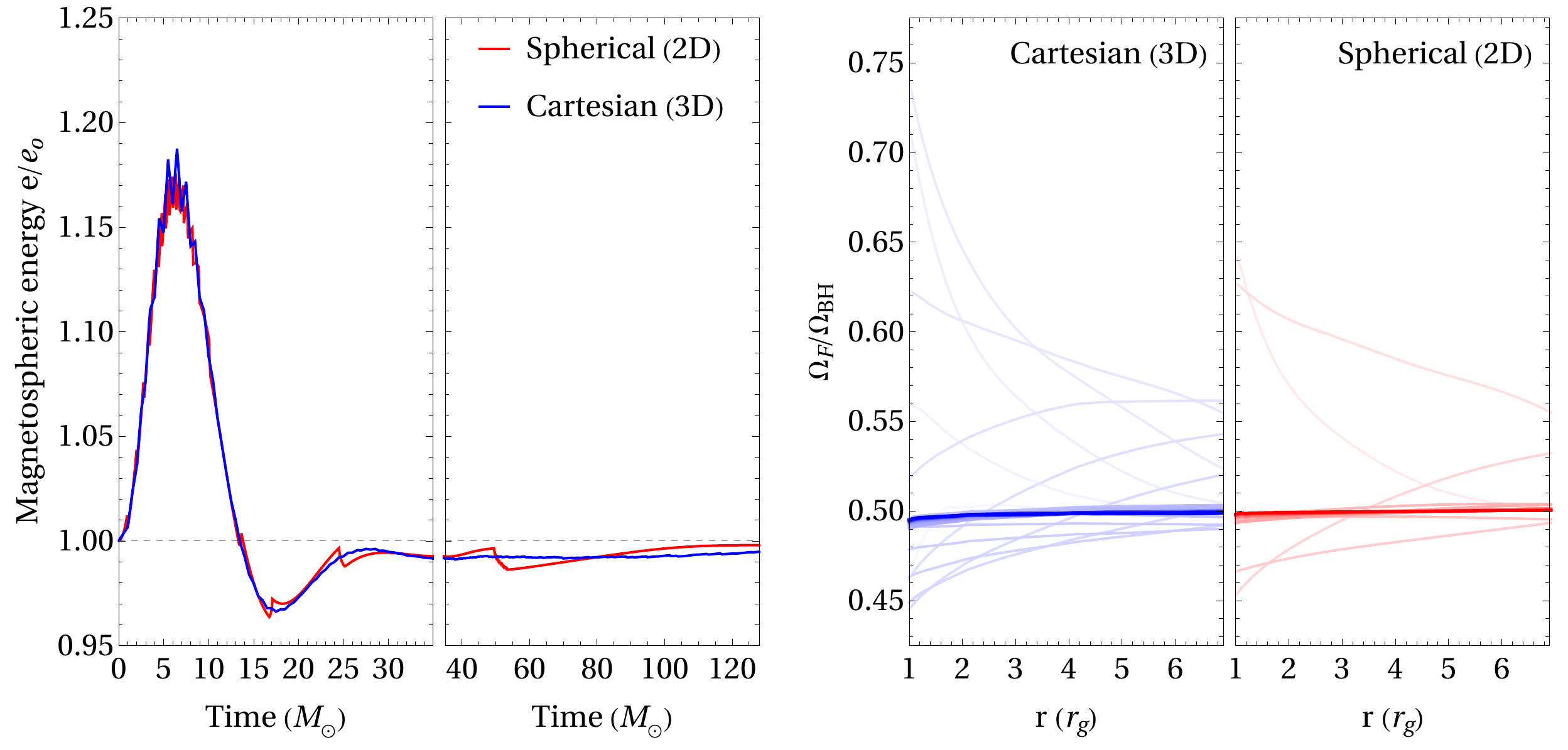}
	\caption{Time evolution of the Schwarzschild monopole ($\Omega_{\rm F}=\Omega_{\rm BH}/2$) 
		of a slowly spinning Kerr BH ($a^*=0.1$,$M=1$) in Cartesian (3D \textsc{Carpet} grid with 
		nine refinement levels, with the highest resolution of $0.03125M_\odot$ completely 
		enclosing the central object) and spherical (2D, axsymmetric) coordinates. The spacetime 
		metric is dynamically evolving. \textit{Left:} Evolution of the total magnetospheric 
		(electromagnetic) energy normalized to the initial value, $e_0$. \textit{Right:} 
		Field line angular velocity along the equatorial plane. The final value is shown in 
		a strong color. Intermediate states throughout the simulation are depicted by lighter 
		colored lines (the strength of the color increasing with simulation time).}
	\label{fig:Schwarzschildmono}
\end{figure*}
\begin{figure}
	\centering
	\includegraphics[width=0.49\textwidth]{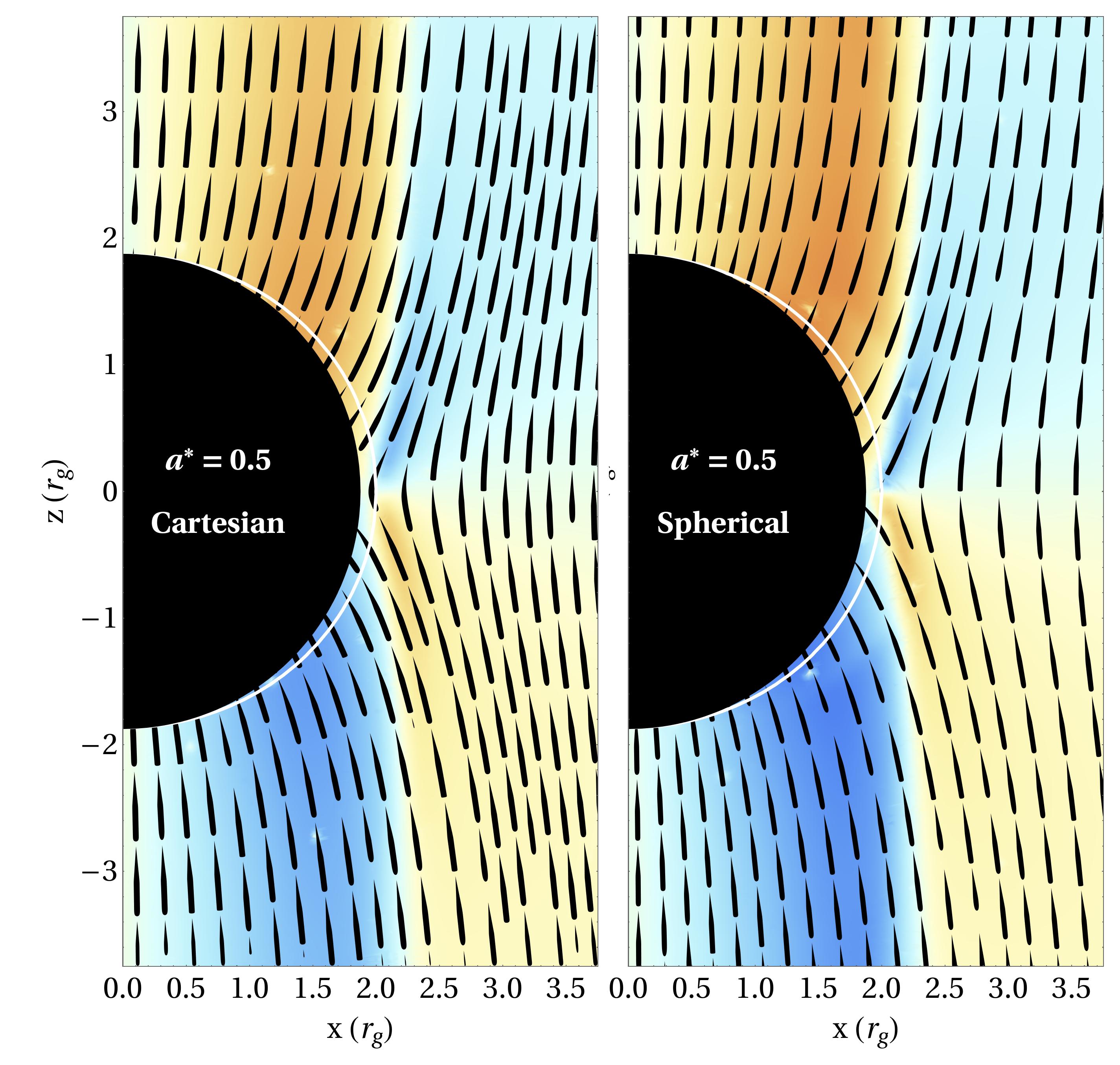}
	\caption{Simulations of Wald magnetospheres for $a^*=0.5$ and $M=1$ at $t=128M_\odot$. 
		\textit{Left:} 3D Cartesian \textsc{Carpet} grid (vacuum spacetime modeled by 
		\textsc{MacLachlan}). \textit{Right:} 2D (axially symmetric) spherical grid 
		(vacuum spacetime modeled by \textsc{SphericalBSSN}). The poloidal field is indicated 
		by streamlines, the toroidal field by red and blue colors (color scale coincides for 
		all panels) indicating whether the toroidal field leaves or enters into the displayed 
		plane, respectively. The BH ergosphere is denoted by a solid white line.}
	\label{fig:WaldA05}
\end{figure}
\begin{figure}
	\centering
	\includegraphics[width=0.49\textwidth]{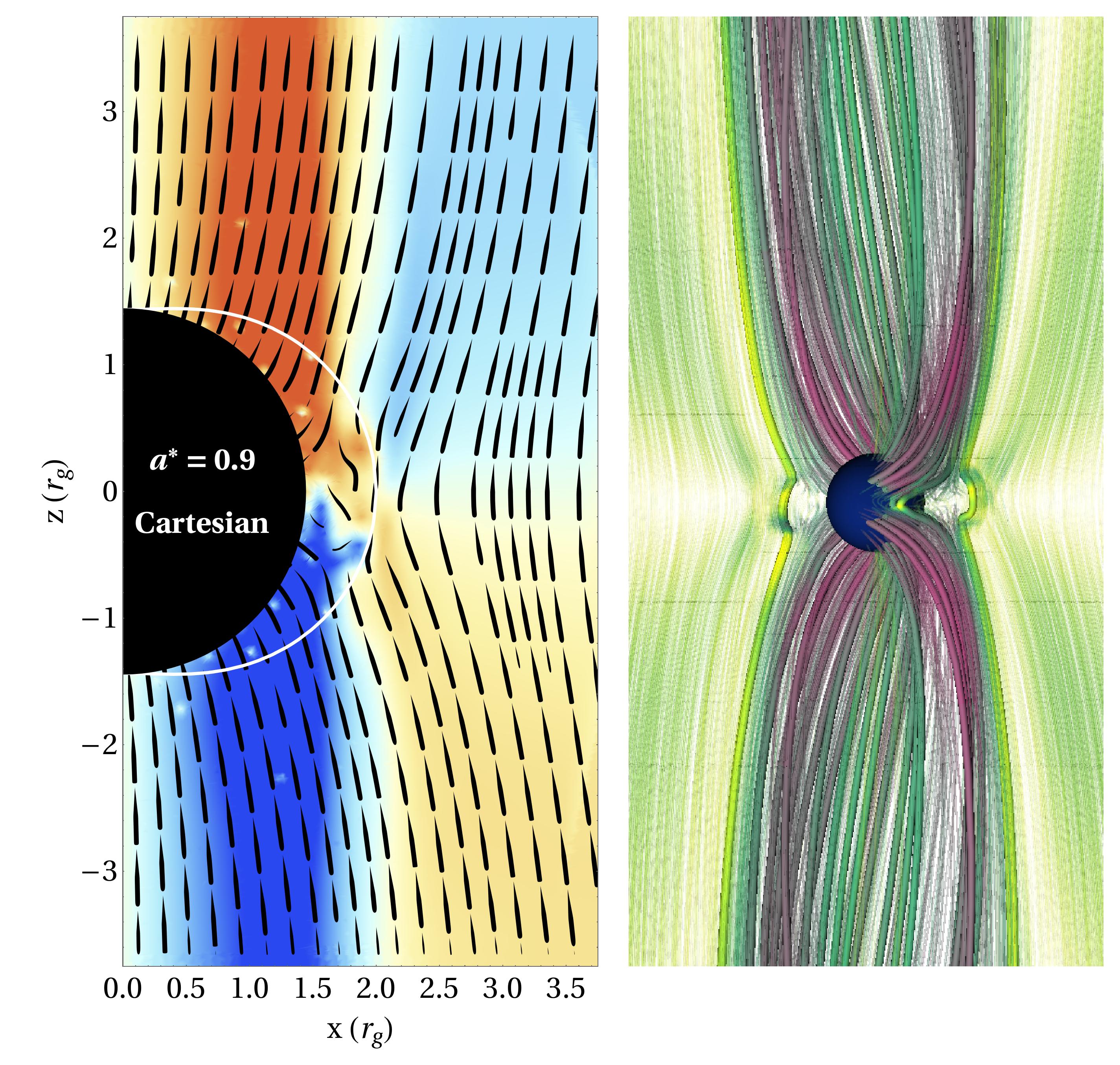}
	\caption{Simulations of a Wald magnetosphere for $a^*=0.9$ in Cartesian coordinates 
		for $t=256M_\odot$. \textit{Left:} Same format and color scale as in Fig. ~\ref{fig:WaldA05}. \textit{Right:} 3D magnetic field line 
		impression of the Wald magnetosphere. Darker colors indicate a stronger twist of the 
		magnetic field.}
	\label{fig:WaldA09}
\end{figure}
\begin{figure}
	\centering
	\includegraphics[width=0.475\textwidth]{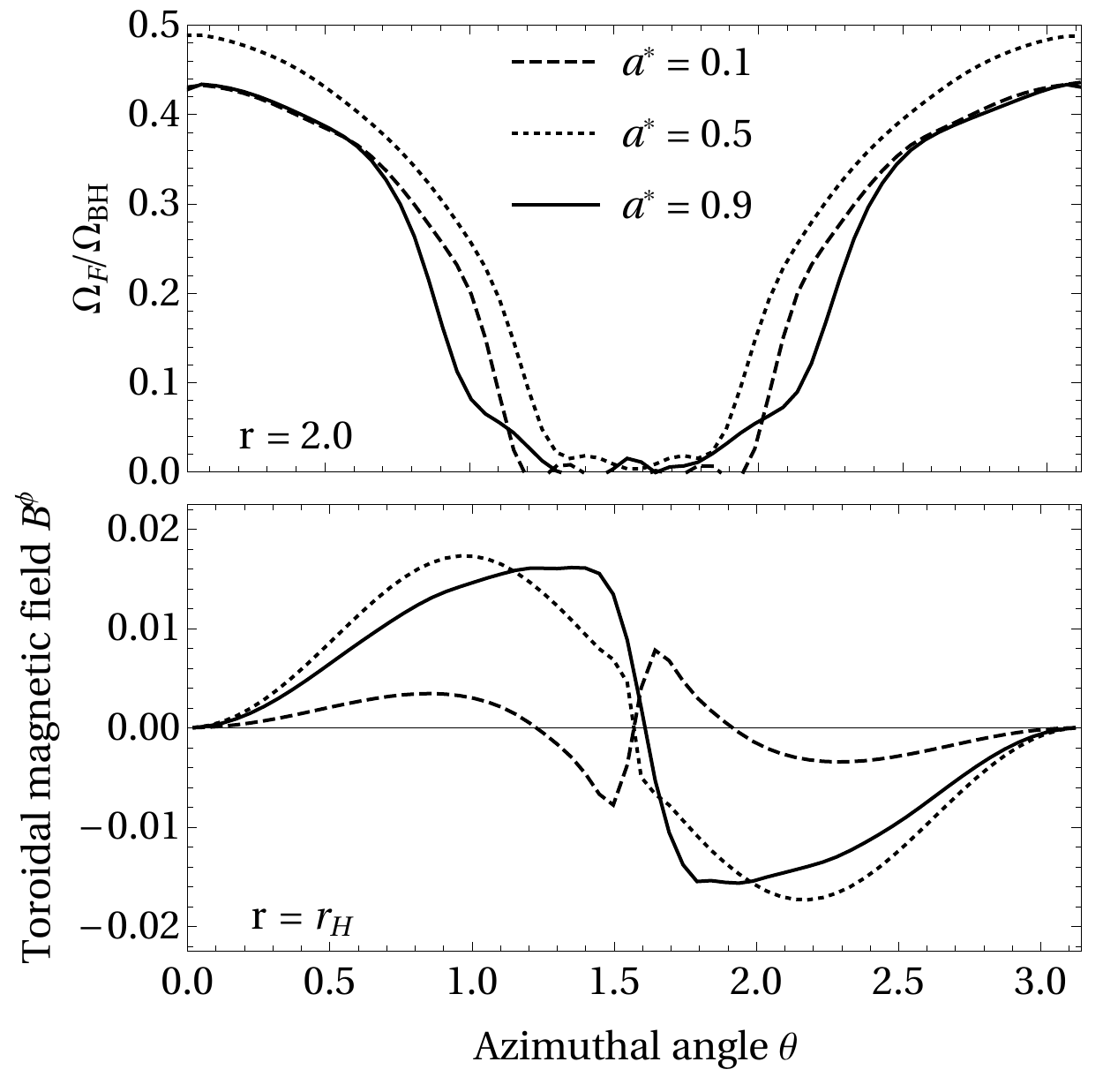}
	\caption{One-dimensional values of the field line angular velocity (\textit{top}) 
		and the toroidal magnetic field (\textit{bottom}) for the Wald test (Cartesian 
		\textsc{Carpet} grid) at $t=256M_\odot$ and using different BH dimensionless rapidities
		(see legends). The interpolation radius (for the extraction in a Cartesian grid) is 
		indicated in the respective panel, corresponding to the ergosphere radius at the equator 
		(\textit{top}) or the BH horizon radius (\textit{bottom}).}
	\label{fig:WaldOmega}
\end{figure}

\citet{Blandford1977} presented analytic equilibrium solutions of BH magnetospheres by applying 
perturbation techniques to the Grad-Shafranov equation (GSE) that match the Znajek condition \citep{Znajek1977} at 
the BH horizon and the flat space solution of \citet{Michel1973} at infinity. One of these 
results is a monopole-like magnetic field, which is often adapted to the so-called split 
monopole by mirroring the field quantities across the equatorial plane. The latter is a 
necessary step to avoid divergences of the magnetic field. In this section, however, we 
follow \citet{Komissarov2004} in considering the monopole field structure to avoid the 
challenge of resolving a current sheet at the equator. The monopole electromagnetic fields 
for slowly spinning BHs ($a^*\ll 1$) as derived in \citet{Blandford1977} can be written in 
the spatial components of vectors in Boyer-Lindquist coordinates $\left(r,\theta,\phi\right)$ 
as follows:
\begin{align}
	\begin{split}
		\mathbf{B}&=B_0\left(-\frac{\sin\theta}{2\sqrt{\gamma}},0,-\frac{a^*\sin^2\theta}{8\alpha g_{\phi\phi}}\right),\\
		\mathbf{D}&=B_0\left(0,-\frac{\Omega_{\rm F}+\beta^\phi}{2\alpha g_{\theta\theta}}\sin\theta ,0\right).
	\end{split}\label{eq:MonoData}
\end{align}
Here, $\Omega_{\rm F}$ is the field line angular velocity as defined for axially symmetric equilibrium solutions, and we employ $B_0=1$. The Cartesian simulations are conducted in a 3D domain with extensions $\left[256 M_\odot\times 256 M_\odot\times 256 M_\odot\right]$ with a grid spacing of 
$\Delta_{x,y,z}=8M_\odot$ on the coarsest grid level. We use eight additional levels of 
mesh refinement, each increasing the resolution by a factor of two and encompassing the 
central object, respectively. This means that the finest resolution of our Cartesian models 
is $\Delta_{x,y,z}^{\rm min}=0.03125 M_\odot$. The spherical simulations are conducted on a 
2D slab (axial symmetry) with extensions 
$\left[0,256M_\odot\right]\times\left[0,\pi\right]$.
In order to use a resolution that is comparable to the Cartesian setup, we employ 
$\Delta r=0.032$ and $\Delta\theta=\pi/64$. The setup is evolved for a period of 
$t=128M_\odot$. For this section we employ $f_{\rm CFL}=0.25$, the MP7 reconstruction and a 
fourth-order accurate discretization of $\mathbf{j}_{||}$.

Fig.~\ref{fig:Schwarzschildmono} summarizes the time evolution of the monopole field for 
a dynamically evolving spacetime metric for both Cartesian (3D) and spherical 
(2D, axisymmetric) meshes. As the magnetospheres considered in this section (and the following one) are idealized cases, namely a magnetic monopole and with unbound energy, we do not couple the field energy to the source terms of the BSSN 
equations. During a transient phase in which the metric terms relax to the chosen mesh and 
gauge, the electromagnetic fields can differ significantly from their initial state. This 
test demonstrates that, while the spacetime evolves, the electromagnetic fields relax toward the equilibrium 
given by Eq.~(\ref{eq:MonoData}) concurrently. Though the energy evolution shown in 
Fig.~\ref{fig:Schwarzschildmono} approaches the energy of the initial model rather well, the resulting equilibrium has to be taken with care. The evolution of dynamical 
spacetimes and corresponding GRFFE fields can be subject to the influence of small changes 
of the BH mass and spin (due to finite numerical resolution), as well as an involved array 
of geometric source terms (see Sect.~\ref{sec:MaxConservative}).

The geometric (i.e., spacetime) quantities determined by the initial data for spinning BHs 
presented in \citet{Liu2009} relax to their equilibrium state depending on the chosen 
numerical resolution of the mesh and specification of gauge quantities (i.e., the lapse 
and shift) during an initialization phase. The choice of these quantities is preferably 
done in a way that causes the least possible noise across all metric quantities during 
their evolution. As an example, instead of providing the spacetime data with the analytic 
lapse function defined in Boyer-Lindquist coordinates \citep{Liu2019}, we specify the lapse
initially as:
\begin{align}
	\tilde{\alpha}(0)=2\times\left[1+\left(1+\frac{M}{2r}\right)^4\right]^{-1}\label{eq:alphainit}.
\end{align}
With this initialization, the spacetime relaxes swiftly to its equilibrium state during 
the first $\Delta t_{\rm init}\approx 25M_\odot$. The tests presented in this section give 
some important hints on the strategies chosen to set up BH magnetospheres for our future 
research. The goal of this test was to show that the magnetospheric data are conserved 
throughout the (dynamic) relaxation of the spacetime induced, for example, by the BSSN algorithms 
of the \textsc{Einstein Toolkit}. As both the magnetospheric energy as well as the field 
line angular velocity at the equator, are recovered after $\Delta t_{\rm init}$, our GRFFE 
code passes this test of spacetime-field coupling.

\subsubsection{The Wald magnetosphere}
\label{sec:Wald_Magnetosphere}

The immersion of a BH into a magnetic field that is uniform at infinity was originally 
suggested by \citet{Wald1974} and then explored throughout the literature, both as a test 
and as a laboratory for force-free plasma \citep{Komissarov2004,Komissarov2007a,Carrasco2017,Parfrey2019}. 
In this section, we reproduce the initial data of the Wald magnetosphere of a Schwarzschild 
BH in Boyer-Lindquist coordinates \citep[rescaled according to the prescription of][]{Liu2009} 
and evolve it for different BH spins. We therefore, extend the testing 
of the GR capacities of our code to rapidly spinning BH (up to $a^*=0.9$). The Wald magnetosphere of a Schwarzschild BH in the spatial components of Boyer-Lindquist 
coordinates $\left(r,\theta,\phi\right)$ can be initialized as follows:
\begin{align}
	\begin{split}
		\mathbf{B}=B_0\left(-\sqrt{\frac{r}{2+r}}\cos\theta,\frac{2\sin\theta}{\sqrt{r+\left(2+r\right)}},0\right),\qquad\mathbf{D}=\left(0,0,0\right).\label{eq:WaldInitial}
	\end{split}
\end{align}
We employ $B_0=1$ as normalization of the magnetic field strength. The Cartesian simulations 
are conducted in a 3D domain with extensions $\left[512 M_\odot\times 512 M_\odot\times 512 M_\odot\right]$ 
with a grid spacing of $\Delta_{x,y,z}=16M_\odot$ on the coarsest grid level. We employ 
nine additional levels of mesh refinement, such that the finest resolution of our Cartesian 
models is $\Delta_{x,y,z}^{\rm min}=0.03125 M_\odot$. The spherical simulations are conducted in axial symmetry with extensions $\left[0,256M_\odot\right]\times\left[0,\pi\right]$. 
In order to use a resolution which is comparable to the Cartesian setup, we employ 
$\Delta r=0.032$ and $\Delta\theta=\pi/64$. The setup is evolved for a period of $t=128M_\odot$ 
in Cartesian coordinates and $t=256M_\odot$ in spherical coordinates. For this section we 
employ $f_{\rm CFL}=0.25$, the MP7 reconstruction and a fourth-order accurate discretization 
of $\mathbf{j}_{||}$.

Figure~\ref{fig:WaldA05} shows the results from the time evolution of these fields 
in spacetimes of rotating BHs for a selected case ($a^*=0.5$) in Cartesian and spherical 
coordinates. The magnetic field lines connecting to the BH, which are initially not rotating, 
are gradually twisted in case of a spinning central object. Also, current sheets form along 
the equatorial plane within the BH ergosphere for high dimensionless spins ($a^*=0.9$, 
Fig.~\ref{fig:WaldA09}), preventing the development of static magnetospheric conditions 
\citep[cf.][]{Komissarov2004}. The overall topology of the magnetic field throughout the 
BH ergosphere broadly coincides with respective equilibrium solutions of Kerr magnetospheres 
\citep[as derived, e.g., in][]{Nathanail2014,Mahlmann2018}. 

The simulations in spherical coordinates are significantly more expensive than in Cartesian 
coordinates, due to the severe restrictions on the timestep imposed by the converging spherical 
mesh close to the central singularity. Future code developments will include mesh-coarsening 
strategies close to $r=0$ to overcome this restriction. Alternatively, we may use other methods to exclude the BH singularity from our computational domain \citep[e.g., using shifted Kerr-Schild coordinates as in][]{Paschalidis2013}. Obtaining a stable evolution of the 
spacetime with spherical coordinates for $a^*\gtrsim 0.9$ is very challenging unless we 
employ rather fine grid spacing in $\theta$, which makes this numerical experiment 
(currently) too expensive as a validation test of our code. Even in axisymmetric simulations, 
the timestep restriction from the $\theta$ coordinate, which imposes a timestep proportional to
$r \times \Delta \theta$, is too restrictive when the coordinate origin is included in the 
computational domain, which is necessary for a spacetime evolution without excision as
the one employed here. In order to alleviate this shortcoming of doing high-resolution 
simulations in spherical coordinates that include the origin, algorithms to circumvent the 
timestep restrictions imposed by both the $\theta$ and $\phi$ coordinates are currently being 
developed \citep{Zlochower2020}.

In Fig.~\ref{fig:WaldOmega} we extract the field line angular velocity and toroidal magnetic 
field at different locations for Cartesian coordinates for comparison with Fig.~5 in 
\citet{Komissarov2004} (computed for a BH with $a^*=0.9)$. The chosen extraction location is 
slightly different from the literature in order to represent the complete range of BH spins. 
We find that our GRFFE code qualitatively reproduces the results in the literature, though 
some differences remain to be mentioned. \citet{Komissarov2004} uses spherical coordinates 
and axial symmetry, as opposed to our 3D simulations with mesh refinement. More even, the 
angular resolution of 800 cells in the $\theta$-direction is almost ten times the resolution 
that we used on our finest refinement level (the resolution limit is simply imposed by 
the aim of running numerical tests that do not consume disproportionate computational resources).
Without evolving the spacetime \citep[as in][]{Komissarov2004} the numerical grid may extend outward in the radial direction from the event horizon as a boundary, in practice, excising the central singularity and allowing for significantly larger timesteps. A similar effect may have the usage of shifted Kerr-Schild coordinates \citep[as in][]{Paschalidis2013}. The quantitative difference in the shape of the angular velocity distribution 
\citep[V-shape in Fig.~5 of][vs. U-shape in Fig.~\ref{fig:WaldOmega}]{Komissarov2004} may, 
hence, be significantly improved with increasing resolution or by resorting to a GRFFE code 
in spherical coordinates (as we plan to do once the timestep restrictions are overcome). 
Also, we point out that we show the toroidal component of the magnetic field $\mathbf{B}$ 
rather than $\mathbf{H}$. The overall form of the toroidal field for the rapidly rotating 
case ($a^*=0.9$) corresponds well (up to a difference in sign) with Fig.~5 of \citet{Komissarov2004}. 
The BH magnetosphere simulations in this section have been repeated on a fixed Kerr-Schild background metric, effectively confirming the results presented in this section. For a direct comparison of numerical data, comparable resolutions and exact convergence 
(longer simulations) are required; this is beyond the scope of the test presented here.

We finally note that the small numerical resistivity of our code \citepalias[prompted by the spatial high-order of our MP reconstruction methods; see][]{Mahlmann2020c} inhibits, in part, the formation of a stationary current-sheet in the solution. For that to happen, one needs to add extra dissipation at current sheets, for example by modifying Ohm's law \citep[as in, e.g.,][]{McKinney2006} or artificially decreasing the conductivity perpendicular to field lines \citep{Paschalidis2013}. The ideal FFE aligned rotator test presented in Sec.~\ref{sec:ffaligned} lucidly illustrates this statement.

In conclusion, especially field lines threading the ergosphere are gradually twisted by the 
rotating BH. Due to the broad coincidence with \citet{Komissarov2004}, and the reproduction 
of magnetospheres which resemble respective equilibrium solutions of the 
\citep{Nathanail2014,Mahlmann2018}, the Wald magnetosphere test is passed.

\section{Conclusions}
\label{sec:conclusions}

We have developed a new GRFFE code that models magnetically dominated plasma in
dynamical spacetimes with support for both Cartesian and spherical coodinates provided by the 
\textsc{Carpet} grid of the \textsc{Einstein Toolkit}. Our simulation tool combines techniques 
from an array of literature \citep[especially][]{Komissarov2002,McKinney2004,Palenzuela2009,Paschalidis2013,Parfrey2017} 
and improves further on numerical strategies as well as the understanding of their limits:
\begin{itemize}
	\item We explicitly couple the continuity equation of charge to our conservative scheme 
	(Sect.~\ref{sec:GRFFE}) and, thus, ensure a consistent modeling of the force-free current 
	(Eq.~\ref{eq:FFCurrent}).
	\item We employ a hyperbolic/parabolic cleaning of errors \citep[extending the 
	techniques in][to general relativity]{Palenzuela2009,Mignone2010}. Allowing for arbitrary 
	advection speeds for the cleaning of divergence errors significantly improves the conservation 
	of total charge in spacetimes of spinning BHs (see Appendix~\ref{sec:Numerical_Errors}).
	\item The current parallel to the magnetic field $\mathbf{j}_\parallel$ is the dominant 
	driver of resistivity in GRFFE schemes. In case of the force-free current (Eq.~\ref{eq:CurrentParallel}), 
	high-order discretization allows us to model (smooth) force-free plasma waves with nearly 
	theoretical order (Sect.~2.1, Paper II), and diffusing only due to numerical resistivity. 
	\item Current sheets are genuine regions of significant physical resistivity. Conventional 
	GRFFE methods (i.e., schemes that do not include phenomenological models of artificial 
	physical resistivity) are unable to properly resolve such resistive layers, especially for 
	highly accurate reconstruction methods (Paper II). Thus, at discontinuities, the order of 
	convergence of GRFFE is significantly reduced; a true limit of applicability of GRFFE is reached.
\end{itemize}

Writing our \textsc{Einstein Toolkit} thorn from scratch enabled us to implement suitable 
finite volume integrators for both Cartesian and spherical coordinates. Spherical 
coordinate systems prove exceptionally valuable for the highly accurate modeling of magnetar 
magnetospheres (Sect.~\ref{sec:Magnetar_Magnetospheres}). For the simulation of BH 
magnetospheres on dynamically evolving spacetimes, our GRFFE method will benefit from future 
updates to the support of spherical coordinates in the \textsc{Einstein Toolkit} 
\citep{Mewes2018,Mewes2020}. With the presented numerical code we broadly exploit the 
modular nature of the \textsc{Einstein Toolkit} and implement a cutting-edge tool for 
the simulation of GRFFE on dynamical spacetimes in Cartesian and spherical coordinates.

\section{Acknowledgments}

We appreciate the helpful comments and perspectives contributed by the anonymous referee. JM acknowledges a Ph.D. grant of the \textit{Studienstiftung des Deutschen Volkes}. 
We acknowledge the support from the grants AYA2015-66899-C2-1-P, PGC2018-095984-B-I00, PROMETEO-II-2014-069, and PROMETEU/2019/071. We acknowledge the partial support of the PHAROS COST 
Action CA16214 and GWverse COST Action CA16104. 
PCD acknowledges the Ramon y Cajal funding (RYC-2015-19074) supporting his research. VM is supported by the Exascale Computing 
Project (17-SC-20-SC), a collaborative effort of the U.S. Department of Energy 
(DOE) Office of Science and the National Nuclear Security Administration. Work at
Oak Ridge National Laboratory is supported under contract DE-AC05-00OR22725 with the
U.S. Department of Energy. VM also acknowledges partial support from the National 
Science Foundation (NSF) from Grant Nos.\ OAC-1550436, AST-1516150, PHY-1607520, 
PHY-1305730, PHY-1707946, and PHY-1726215 to Rochester Institute of Technology (RIT). 
The shown numerical simulations have been conducted on infrastructure of the 
\textit{Red Espa\~{n}ola de Supercomputación} (AECT-2020-1-0014) as well as of the 
\textit{University of Valencia} Tirant and LluisVives supercomputers.

\bibliographystyle{aa}
\bibliography{literature.bib}

\appendix
\section{Code performance (3D Cartesian): Cleaning of errors}
\label{sec:Numerical_Errors}

\begin{figure}
	\centering
	\includegraphics[width=0.45\textwidth]{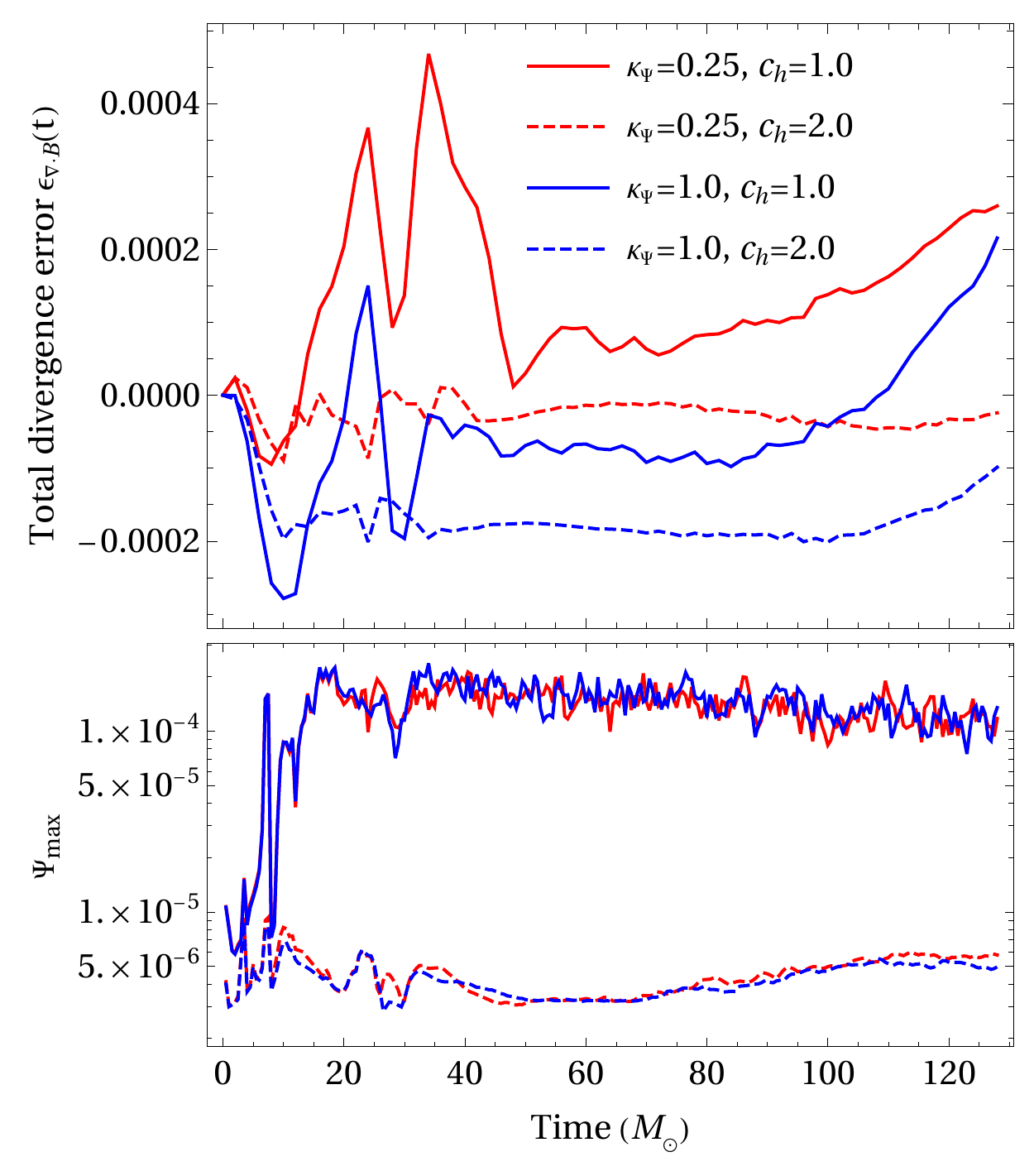}
	\caption{Time evolution of numerical errors ($\text{div}\mathbf{B}=0$, \textit{top} panel) 
		and the corresponding maximum cleaning potential $\Psi$ (\textit{bottom} panel). We present 
		combinations of different $\kappa_{\Psi}$ and $c_h$ for a fixed $\kappa_{\Phi}=1.0$.}
	\label{fig:PsiCleaning}
\end{figure}

We describe the implementation of the generalized Lagrange multiplier method employed to 
preserve the electromagnetic differential constraints $\text{div}\mathbf{B}=0$ and 
$\text{div}\mathbf{D}=\rho$ in Sect.~\ref{sec:CleaningErrors}. This section explores the 
code performance for the cleaning of divergence errors in the 3D Cartesian version of the BH 
monopole test (with the setup from Sect.~\ref{sec:BH_Magnetospheres}) for different 
choices of the parameters governing the numerical cleaning of errors. We measure the numerical 
errors to the aforementioned conditions by considering the global measures:
\begin{align}
	\varepsilon_{\nabla\cdot\mathbf{B}}\left(t\right)&=\int \left[\nabla\cdot\mathbf{B}\left(t\right)\right]\text{d}V-\int \left[\nabla\cdot\mathbf{B}\left(t=0\right)\right]\text{d}V\label{eq:divBerror}.
\end{align}
Here, we employ the 3D region outside of the BH horizon as an integration region, and 
subtract the initially present errors due to the discretization of the magnetic field. 
Figure~\ref{fig:PsiCleaning} shows the evolution of numerical errors and the corresponding 
cleaning potentials for different combinations of the parameters $\kappa_{\Psi}$, and $c_h$. 
The optimization of these parameters may differ for different applications and can be critical 
in highly dynamical processes where strong numerical violations of the divergence constraints 
occur \citep[e.g., by strong violations of the force-free conditions, see also the discussion 
in][]{Mahlmann2019}. For the tests at hand, the exact calibration of the parameters of the 
cleaning method may have very small effects (the total magnetospheric energy presented in 
Fig.~\ref{fig:Schwarzschildmono} is not notably changed by any of the different combinations 
shown in Fig.~\ref{fig:PsiCleaning}). However, their analysis provides crucial information 
about the code's performance, and in other applications the proper calibration of the cleaning 
routines has a significant impact \citep[][]{Mahlmann2019}.

As is lucidly shown in Fig.~\ref{fig:PsiCleaning}, the introduction of the superluminal 
advection velocity $c_h$ into the augmented system of equations (Eq.~\ref{eq:MaxAugII}) for 
divergence cleaning reduces the error $\varepsilon_{\nabla\cdot\mathbf{B}}$ (especially in 
the early and late phase of the evolution) significantly. Furthermore, the maximum magnitude 
of the cleaning potential $\Psi$ decreases by two orders of magnitude. Small variations in 
$\varepsilon_{\nabla\cdot\mathbf{B}}$ are also observed for stronger damping of errors by 
greater values for $\kappa_{\Psi}$. Though the presented tests for flat background geometries 
employ $c_h=1$, we conclude from the results in Fig.~\ref{fig:PsiCleaning} that $c_h=2$ 
improves the code performance (i.e., reducing the arising numerical errors) for general 
relativistic spacetimes.

This comparison of parameters responsible for the cleaning of numerical errors emphasizes 
the strong need for a diligent calibration for each setup (i.e., boundary conditions, geometry, 
etc.) at hand. The standard configurations employed for the hyperbolic/parabolic cleaning 
of numerical errors should and will be readjusted in the light of future applications of our GRFFE method.

\section{Conservation at refinement boundaries}
\label{sec:Refluxing}

\begin{figure}
	\centering
	\includegraphics[width=0.48\textwidth]{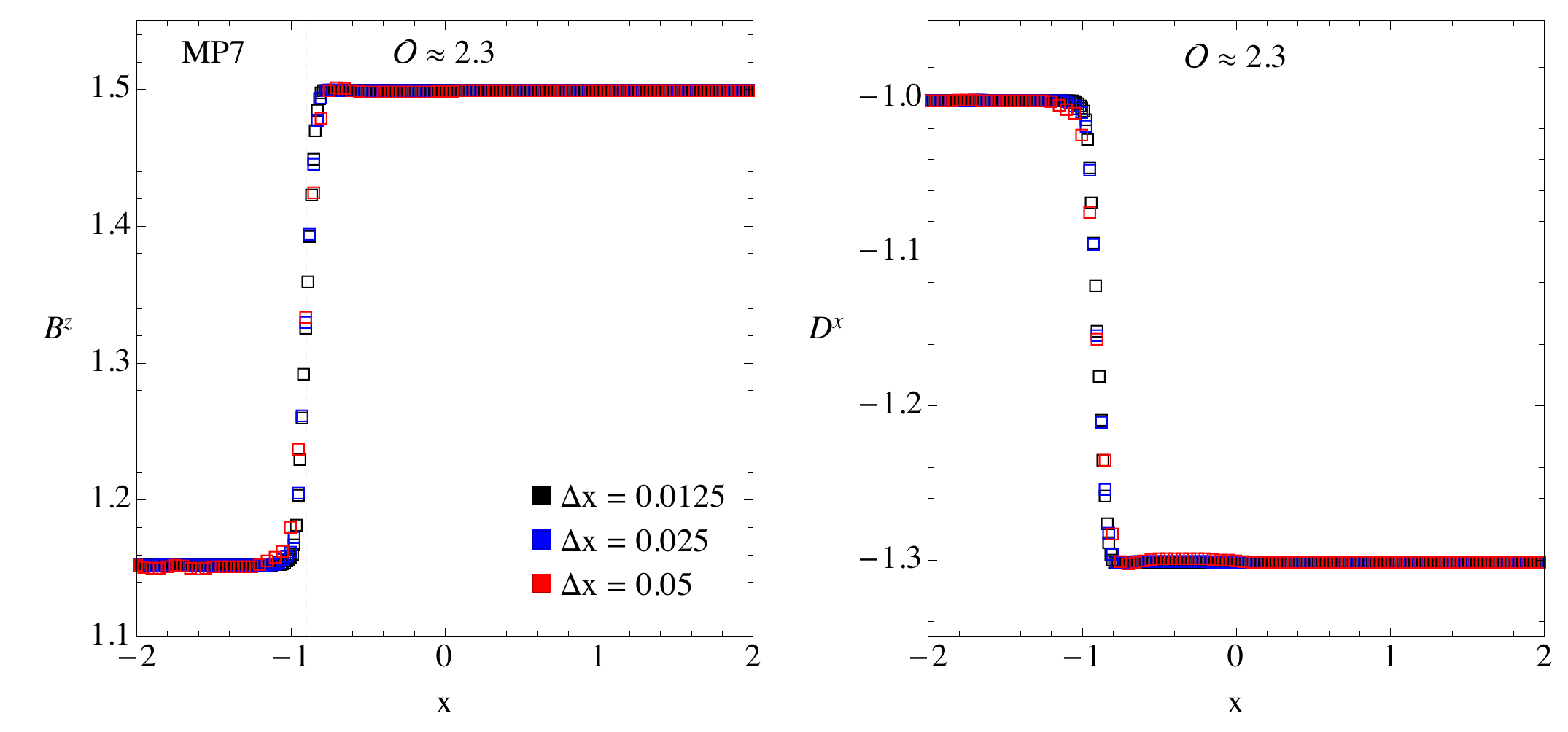}
	\caption{Degenerate Alfvén wave test on a $x\in\left[-2,2\right]$ 
			grid ($f_\textsc{cfl}=0.25$) at $t=2.0$ and for different resolutions. The theoretically expected position of the Alfven wave is indicated by a gray dashed line.}
	\label{fig:DegAlfvenBD}
\end{figure}
\begin{figure}
	\centering
	\includegraphics[width=0.48\textwidth]{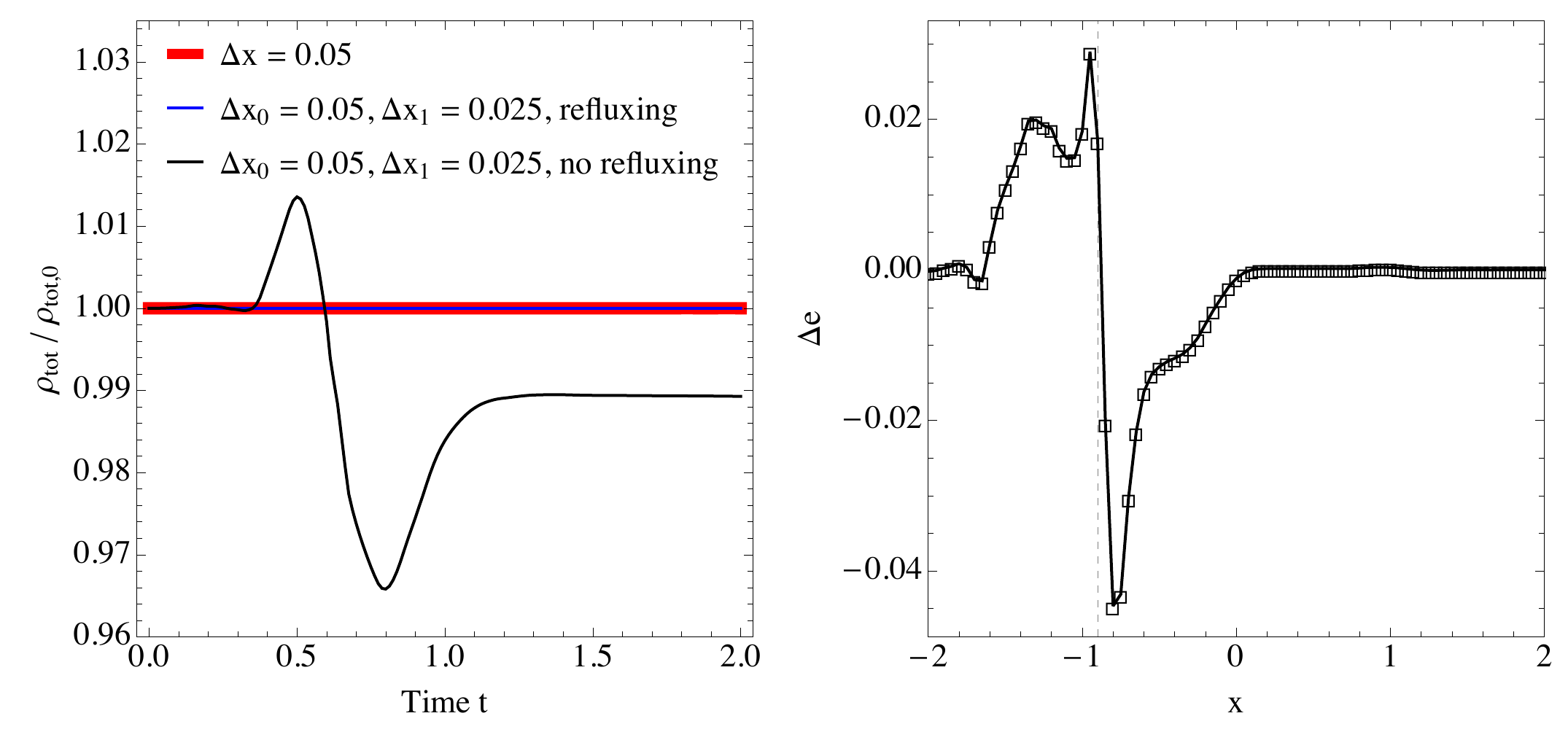}
	\caption{Same test as in Fig.~\ref{fig:DegAlfvenBD}, but with an additional level of mesh refinement for $x\leq -0.25$. \textit{Left}: Comparison of global charge conservation between a uniform grid (red line), a refined mesh with refluxing (blue line), and a refined grid without refluxing (black line). We display the relative deviation of the total charge from its initial value for different setups. \textit{Right}: Difference between the electromagnetic energy density for the refined mesh with and without refluxing at $t=2.0$.}
	\label{fig:DegAlfvenReflux}
\end{figure}

Alfvén waves carry charge (and current). They are an excellent testbed for probing the conservative properties of a numerical method \citep[cf.][]{Mahlmann2020c}. In this section, we examine the quality of our conservative method at mesh refinement boundaries. We consider a (smoother) variation the degenerate Alfvén wave test from \citet{Komissarov2002} and \citet{Yu2011}. For this, we boost the initial data given by Eq.~(\ref{eq:standingAlfven}) into a frame with $\beta=0.5$:

	\begin{align}
		\begin{split}
			\mathbf{B}&=\left(1,\sqrt{3},2B^z/\sqrt{3}\right),\qquad \mathbf{D}=\left(-B^z,-B^z/\sqrt{3},\sqrt{3}\right),\\
			B^z&=\left\{\begin{array}{cc}
				1& x\leq 0\\
				1+0.15\left[1+\sin\left[5\pi\left(x-0.1\right)\right]\right]& 0<x\leq 0.2\\
				1.3 & x>0.2
			\end{array}\right. .
		\end{split} \label{eq:standingAlfvenBoost}
	\end{align}

Fig.~\ref{fig:DegAlfvenBD} shows components of the electric and magnetic fields of the solution developed up to $t=2$ on a uniform mesh \citep[comparable to the respective tests in][]{Komissarov2002,Yu2011}. We then repeat this test with an additional level of refinement for $x\leq -0.25$ and display the main results in Fig.~\ref{fig:DegAlfvenReflux}. The degenerate Alfvén wave thus crosses one mesh refinement boundary during its evolution.

The quality of our conservative method for global quantities - such as the electric charge - is impaired on the level of an error of several percent if no refluxing is used (black line in the left panel of Fig.~\ref{fig:DegAlfvenReflux}). The same statement (though on a smaller scale) is also true for the energy, as we show in the right panel of Fig.~\ref{fig:DegAlfvenReflux}. If refluxing is used, on the other hand, energy is conserved up to machine precision or violation of the force-free conditions. Such levels of nonconservation in simple 1D tests justify the use of refluxing techniques (cf. Sec.~\ref{sec:Code_Implementation}), especially in highly dynamic 3D simulations.

\end{document}